\documentclass[a4paper,fleqn]{cas-dc}

\usepackage[numbers]{natbib}
\usepackage{afterpage}
\usepackage{cuted}
\usepackage{caption}
\usepackage{adjustbox}
\usepackage{capt-of}
\usepackage{placeins}
\usepackage{threeparttable}

\usepackage{booktabs}
\usepackage{tabularx}
\usepackage{array}

\usepackage{graphicx}
\usepackage{subcaption}

\newif\ifmarkedversion

\markedversionfalse

\usepackage{xcolor}

\ifmarkedversion
    \usepackage[
        draft,
        authormarkup=none,
        commandnameprefix=ifneeded
    ]{changes}
\else
    \usepackage[
        final,
        commandnameprefix=ifneeded
    ]{changes}
\fi

\definechangesauthor[
    name={Revision},
    color=blue
]{rev}

\newcommand{\deletedtablerow}[5]{%
    \ifmarkedversion
        \deleted[id=rev]{#1} &
        \deleted[id=rev]{#2} &
        \deleted[id=rev]{#3} &
        \deleted[id=rev]{#4} &
        \deleted[id=rev]{#5} \\
    \fi
}

\DeclareRobustCommand{\revadded}[1]{%
    \ifmarkedversion
        {\color{blue}#1}%
    \else
        #1%
    \fi
}

\newcolumntype{Y}{>{\raggedright\arraybackslash}X}
\newcommand{\rowrule}{\specialrule{0.25pt}{0pt}{0pt}}

\def\tsc#1{\csdef{#1}{\textsc{\lowercase{#1}}\xspace}}
\tsc{WGM}
\tsc{QE}
\tsc{EP}
\tsc{PMS}
\tsc{BEC}
\tsc{DE}
\begin{document}
\let\WriteBookmarks\relax
\def\floatpagepagefraction{1}
\def\textpagefraction{.001}
\shorttitle{PRIN for Low-Light Image Enhancement}
\shortauthors{Q. Xie et al.}

\title [mode = title]{PRIN: A Perceptual-Structural Retinex-Inspired Network for Low-Light Image Enhancement}                      
%\tnotemark[1,2]

%\tnotetext[1]{This document is the results of the research
%   project funded by the National Science Foundation.}

%\tnotetext[2]{The second title footnote which is a longer text matter
%   to fill through the whole text width and overflow into
%   another line in the footnotes area of the first page.}

\author[1,3]{Qiyi Xie}[orcid=0009-0006-9506-8359]
\fnmark[1]
\ead{xieqiyi@seas.upenn.edu}
\credit{Conceptualization, Methodology, Software, Validation, Formal analysis, Visualization, Writing - original draft, Writing - review \& editing}

\author[1]{Wei Chen Ng}[orcid=0009-0003-7777-6341]
\fnmark[1]
\ead{mcs2509006@xmu.edu.my}
\credit{Software, Validation, Formal analysis, Visualization, Writing - original draft}

\author[1]{Ziyao Li}[orcid=0009-0001-3064-7695]
\ead{EEE2409020@xmu.edu.my}
\credit{Visualization, Formal analysis, Writing - original draft}

\author[1]{Yuning Wu}[orcid=0009-0001-1500-8158]
\ead{EEE2309312@xmu.edu.my}
\credit{Visualization, Formal analysis, Writing - original draft}

\author[3]{Huasen Wu}[orcid=0009-0000-4293-2058]
\ead{wu77@seas.upenn.edu}
\credit{Resources, Writing - review \& editing}

\author[3]{Tianshi Miao}[]
\ead{miaots@seas.upenn.edu}
\credit{Resources, Writing - review \& editing}

\author[1,2]{Jian Ding Tan}[orcid=0000-0001-5169-9470]
\cormark[1]
\ead{jianding.tan@xmu.edu.my}
\credit{Supervision, Project administration, Writing - review \& editing}

% Affiliation 1: XMUM
\affiliation[1]{organization={School of Artificial Intelligence and Robotics, Xiamen University Malaysia},
                addressline={Jalan Sunsuria, Bandar Sunsuria}, 
                city={Sepang},
                postcode={43900}, 
                state={Selangor},
                country={Malaysia}}
                
% Affiliation 2: S.M.A.R.T.
\affiliation[2]{organization={S.M.A.R.T. Nexus Center of Excellence, Xiamen University Malaysia},
                addressline={Jalan Sunsuria, Bandar Sunsuria}, 
                city={Sepang},
                postcode={43900}, 
                state={Selangor},
                country={Malaysia}}
                
% Affiliation 3: UPenn
\affiliation[3]{organization={School of Engineering and Applied Science, University of Pennsylvania},
                addressline={220 South 33rd Street}, 
                city={Philadelphia},
                postcode={19104}, 
                state={PA},
                country={United States}}
                
\fntext[fn1]{These authors contributed equally to this work.}

\cortext[cor1]{Corresponding author}

\begin{abstract}
Low-light image enhancement (LLIE) focuses not only on visibility recovery but also on preserving structurally meaningful content and perceptual naturalness. Existing learning-based methods may achieve favorable pixel-wise fidelity while still producing over-smoothed textures, color deviations, or spatially incoherent illumination. Retinex-inspired methods provide interpretable decomposition into reflectance and illumination components, however, the asymmetric restoration roles of these components are rarely encoded through branch-specific frequency-domain refinement. To address this issue, we propose a Perceptual-Structural Retinex-Inspired Network (PRIN), which integrates Retinex priors with restoration-role-aware enhancement. 
PRIN employs a shared encoder and asymmetric decoders to estimate reflectance and illumination components.
\replaced[id=rev]{%
The reflectance branch performs attention-guided and edge-aware wavelet refinement to preserve textures and boundaries, whereas the illumination branch promotes spatially coherent brightness adjustment.%
}{%
In the reflectance branch, a Wavelet Detail-Guided Reflectance Refiner (WDGRR) performs attention-guided and edge-aware wavelet refinement to preserve textures and boundaries. In the illumination branch, a Wavelet Structure-Aware Illumination Refiner (WSAIR) performs spatially coherent brightness adjustment.%
}
After the recomposition of reflectance and illumination branches, a curve-based post-composition enhancement head calibrates residual tone and exposure deviations. 
Experiments on LOL-v1, LOL-v2-Syn, and LOL-v2-Real show that 
\replaced[id=rev]{%
the proposed network ties for the best structural similarity index
(SSIM) and obtains the second-lowest learned perceptual image patch
similarity (LPIPS) on LOL-v1, achieves the best SSIM and LPIPS on
LOL-v2-Syn, and obtains the highest peak signal-to-noise ratio (PSNR)
and the second-lowest LPIPS on LOL-v2-Real. The source code, pretrained models, supplementary results, and subsequent updates are available at: https://github.com/LucasXQY/Perceptual-Structural-Retinex-Inspired-Network-PRIN-for-Low-Light-Image-Enhancement%
}{%
PRIN achieves a favorable perception-distortion tradeoff and the lowest LPIPS values across all three benchmarks while securing the top-tier SSIM and competitive PSNR among the existing LLIE methods.%
}
\end{abstract}

\begin{highlights}
\item PRIN performs restoration-role-aware Retinex low-light enhancement.
\item Asymmetric decoders separately model reflectance and illumination restoration.
\item Branch-specific wavelet refiners preserve details and coherent illumination.
\item Post-composition curve refinement calibrates residual tone and exposure. \\

\deleted[id=rev]{%
PRIN achieves the lowest LPIPS across three evaluated LLIE benchmarks.%
}
\end{highlights}


\maketitle

\section{Introduction}
In practice, low-light images typically suffer from various coupled degradations, including insufficient brightness, weak contrast, color bias, attenuation of fine details, and noise amplification \cite{9088214}\deleted[id=rev]{[62]}. The degradations do not occur independently: when brightness is boosted, noise and color distortion tend to become more apparent \cite{10.1145/3689642}; moreover, when denoising is too strong, textures and boundaries may be over-smoothed \cite{Panagiotou2023Denoising}. Therefore, the goal of LLIE is not just limited to visibility recovery. The quality of an enhanced image should be judged by whether it preserves perceptually meaningful information. In other words, a result may have sufficient brightness or achieve good pixel-wise fidelity scores while still being visually unreliable if fine textures are over-smoothed, object boundaries become blurred, colors deviate from their natural appearance, or illumination transitions become spatially incoherent. An effective LLIE method should preserve structural details and perceptual naturalness, rather than pursuing luminance amplification or distortion reduction alone.

Recently, due to the strong ability to learn complex mappings between estimations and desired outputs, many deep learning methods have been proposed to improve LLIE. Existing deep approaches include direct enhancement methods and physically inspired decomposition-based methods. For the first type, direct enhancement methods infer enhanced images or transformation parameters without explicitly separating intrinsic appearance from illumination variation. For instance, Zero-DCE \cite{Guo2020} models enhancement as image-specific curve estimation without requiring paired or unpaired supervision. EnlightenGAN \cite{Jiang2021} performs low-light enhancement through unpaired adversarial learning. Moreover, LLFlow \cite{Wang2022} utilizes a normalizing flow framework to model the one-to-many mapping from low-light inputs to normal-light outputs. Benefiting from the powerful representational capabilities of deep networks, these methods are able to capture complex mapping relationships between inputs and outputs and typically achieve competitive quantitative and visual results. These approaches demonstrate strong representation capability, yet their restoration behavior is not necessarily associated with explicitly interpretable image components.

For the second type, take Retinex-inspired methods as examples, which provide a more structured modeling basis for LLIE. According to Retinex theory, which is derived from human visual perception, an observed image can be represented through the interaction between reflectance and illumination. The reflectance component is closely related to intrinsic appearance and local structure, and the illumination component describes spatial light distribution \cite{Land:71}. For example, Retinex-Net combines a Decomposition Network (Decom-Net) with an Illumination Adjustment Network (Enhance-Net) to enhance low-light images \cite{Wei2018}. Retinexformer proposes a single-stage framework, which utilizes an illumination-guided transformer to model long-range dependencies and suppress noise introduced during illumination enhancement \cite{cai2023retinexformer}. Moreover, RetinexMamba combines the Mamba state-space model and the Retinex framework. It also replaces the illumination-guided self-attention module with a fused attention module to achieve a balance between efficiency and visual quality \cite{Bai2025}. 

These methods \replaced[id=rev]{%
provide a structured representation of image formation, which can improve the interpretability of the enhancement process.%
}{%
aim to preserve clear physical meanings so as to offer better interpretability and generalization.%
} 
\added[id=rev]{%
However, Retinex decomposition is non-unique, in that a given image may be decomposed into infinitely many valid reflectance and illumination pairs. Consequently, architectural priors and loss-based regularization are required to restrict the feasible decomposition space.%
}
\replaced[id=rev]{%
Moreover, imposing a Retinex decomposition does not by itself ensure that the learned components follow their intended restoration roles.%
}{%
Despite this, Retinex decomposition only separates an observed image into reflectance and illumination components; it does not automatically make sure that the subsequent restoration process is consistent with the distinct roles of these components.%
}
Some existing Retinex-based methods\deleted[id=rev]{, such as KinD and Diff-Retinex,} have introduced separate processing strategies for reflectance and illumination. \added[id=rev]{KinD introduces separate restoration mechanisms for reflectance and illumination, including reflectance denoising and controllable illumination adjustment \cite{Zhang2019KinD}. Diff-Retinex instead adopts component-specific diffusion processes for reflectance and illumination restoration \cite{yi2023diffretinex}.} However, their branch difference is mainly realized through component-specific network modules, enhancement controls, or generative restoration strategies, rather than through an explicit frequency-domain encoding of component restoration roles.

This distinction is important because reflectance restoration and illumination adjustment are inherently asymmetric tasks. Reflectance is closely associated with intrinsic appearance, object boundaries, texture patterns, and local color structures, and therefore requires careful preservation and recovery of detail-related information. Illumination mainly characterizes spatial light distribution and should be adjusted with an emphasis on smooth exposure regulation and coherent brightness transitions across regions. When these two components are refined without sufficiently differentiated frequency-aware operations, the enhanced results may suffer from over-smoothed textures, unstable local contrast, color deviation, or unnatural illumination transitions.

Frequency-domain refinement provides an interpretable mechanism for encoding these different restoration roles, as the different frequency representations contain different information, and different enhancing methods could be applied. The Haar wavelet transform is applied in this research; it decomposes an image into 4 subbands (one approximation band (LL) and three directional detail bands (LH, HL, and HH)). The low-frequency approximation subband contains spatial organization and broad illumination variation, and three high-frequency subbands contain boundary and texture cues. To visually illustrate the difference, Fig.~\ref{fig:haar_visualization} 
shows the Haar wavelet decomposition of an image, including its low-frequency subband and combined three high-frequency responses. Therefore, the key idea of this work is to perform branch-specific wavelet refinement according to the restoration role of each Retinex component.

\par\medskip
\noindent
\begin{minipage}{\columnwidth}
    \centering
    \setlength{\tabcolsep}{2pt}
    \renewcommand{\arraystretch}{0.95}
    \begin{tabular}{@{}cc@{}}
        \includegraphics[width=0.47\linewidth]{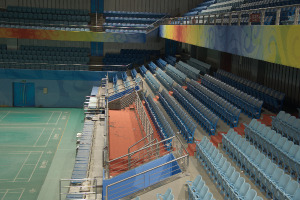} &
        \includegraphics[width=0.47\linewidth]{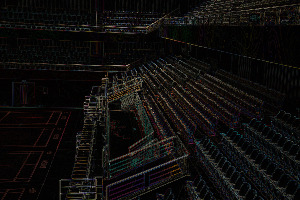} \\
        {\scriptsize LL subband} &
        {\scriptsize Combined $(LH, HL, HH)$ subband}
    \end{tabular}
    \captionof{figure}{The decomposition result of Haar wavelet 
    transform.}
    \label{fig:haar_visualization}
\end{minipage}
\par\medskip

Another issue is that the network performance is commonly evaluated with emphasis on distortion-oriented metrics, such as mean squared error (MSE) and peak signal-to-noise ratio (PSNR), on paired datasets. Although these metrics are useful for measuring pixel-wise agreement with reference images, higher distortion scores may not indicate better visual quality. 
\added[id=rev]{%
One reason is that the reference image itself may carry mild softness (e.g., from capture conditions or in-camera denoising), so an output that smooths away fine details can move closer to the reference in the squared-error sense without any gain in perception quality. %
}
As illustrated in 
Fig.~\ref{fig:blur_psnr_comparison}, Gaussian-smoothed variants of an enhanced result achieve progressively higher PSNR values, increasing from 29.480 dB to 30.637 dB, while their boundaries and textures become more blurred. 
\replaced[id=rev]{This observation reflects a documented limitation of squared-error fidelity measures, which weight all pixel-wise errors equally regardless of their perceptual role \cite{wang2009mse}.}{This observation is consistent with  the ideas from previous studies on the difference between distortion reduction and perceptual quality [2,21]. }
Therefore, this work reports PSNR as a complementary pixel-fidelity measure, while placing greater emphasis on SSIM for spatial structural consistency and LPIPS for perceptual similarity.

Motivated by the asymmetric restoration roles of Retinex components and the need for perceptually reliable enhancement beyond distortion-oriented optimization, we propose the Perceptual-Structural Retinex-Inspired Network (PRIN) to achieve a better balance between structural fidelity and perceptual quality. PRIN first uses a shared encoder to extract low-level features from the input image. Then, two asymmetric decoders are applied to reconstruct the reflectance and illumination (branch) components to achieve a balance between detail recovery and illumination correction. Specifically, the reflectance decoder focuses on detail recovery by enhancing salient structures and edge-related information. The illumination decoder adopts extra downsampling stages to capture broader contextual dependencies for smoother and more globally consistent illumination estimation. After the element-wise multiplication of the two branch components, we also introduced a curve-based post-composition enhancement module to calibrate residual tone and exposure imbalance.

\begin{center}
\begin{minipage}{\columnwidth}
    \centering

    \begin{minipage}[t]{0.47\linewidth}
        \centering
        \includegraphics[width=\linewidth]{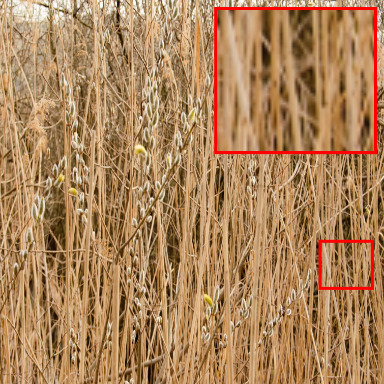}
        {\scriptsize Original, PSNR: 29.480 dB\par}
    \end{minipage}
    \hfill
    \begin{minipage}[t]{0.47\linewidth}
        \centering
        \includegraphics[width=\linewidth]{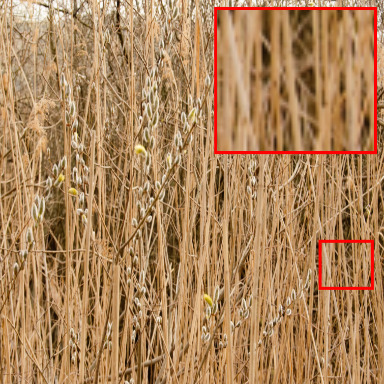}
        {\scriptsize $r=0.5$, $\alpha=0.2$, PSNR: 30.140 dB\par}
    \end{minipage}

    \vspace{3pt}

    \begin{minipage}[t]{0.47\linewidth}
        \centering
        \includegraphics[width=\linewidth]{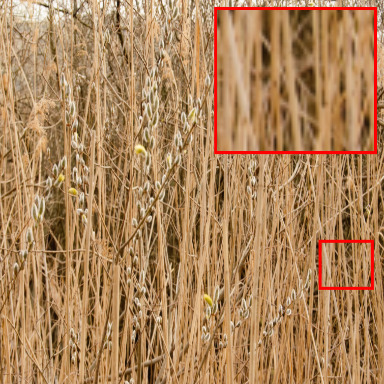}
        {\scriptsize $r=0.8$, $\alpha=0.2$, PSNR: 30.515 dB\par}
    \end{minipage}
    \hfill
    \begin{minipage}[t]{0.47\linewidth}
        \centering
        \includegraphics[width=\linewidth]{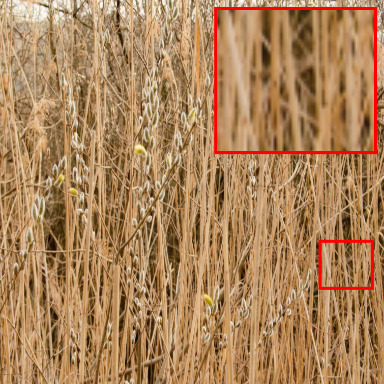}
        {\scriptsize $r=1.0$, $\alpha=0.2$, PSNR: 30.637 dB\par}
    \end{minipage}

    \captionof{figure}{
    Visual comparison between the original enhanced result and its blur-based post-processing variants. The blur variants are generated by applying Gaussian smoothing with different radii $r$ to the original output and then blending the smoothed image with the original result using a fixed weight $\alpha=0.2$.
    }
    \label{fig:blur_psnr_comparison}
\end{minipage}
\end{center}

The main contributions of this work are summarized as follows.
\begin{itemize}
    \item \replaced[id=rev]{%
We propose PRIN, a regularized restoration-role-aware Retinex-inspired framework for low-light image enhancement. The ambiguous Retinex decomposition is constrained through asymmetric decoder topologies, branch-specific frequency refinement, bounded outputs, and loss-based regularization.%
}{%
We propose PRIN, a restoration-role-aware Retinex-inspired framework for low-light image enhancement, which employs asymmetric decoder topologies that the reflectance branch preserves spatial detail through same-scale refinement, while the illumination branch expands contextual modeling through additional downsampling.%
}

    \item We design two branch-specific wavelet refinement modules: WDGRR 
    enhances reflectance representations for texture and boundary preservation, and WSAIR refines illumination representations for spatially coherent brightness adjustment. This design translates the component roles motivated by Retinex decomposition into differentiated frequency-domain restoration operations.

    \item We integrate a curve-based post-composition enhancement head and a task-aligned composite objective to coordinate decomposed restoration, final tone regulation, structural preservation, color stability, and exposure control.

    \item We conduct comprehensive experiments including benchmark comparisons, qualitative visual analysis, network variant comparison, module ablation, Gaussian-blur metric analysis, GT-mean analysis, and SSIM-LPIPS distribution analysis to examine not only numerical performance but also the perceptual-structural behavior of the proposed method.
\end{itemize}

The rest of this paper is organized as follows. Section 2 reviews the most relevant studies on LLIE. Section 3 presents the proposed PRIN in detail. Section 4 presents the experimental settings and comprehensive results, including quantitative and qualitative comparisons, variant and ablation studies, GT-mean analysis, and perceptual-structural performance evaluation. Section 5 discusses the limitations and future directions of this work. Finally, Section 6 concludes the paper.

\section{Related Work}
In this section, we review the studies most relevant to PRIN from three perspectives: Retinex-based decomposition, frequency-aware enhancement, and attention-driven perceptual adaptation in order to clarify the motivation and technical positioning of our method.

\subsection{Retinex-Based LLIE Methods}
Classical Retinex theory is proposed based on the human visual system, and the key idea is that an observed image can be decomposed as the product of its reflectance and illumination \cite{Land:71}. Modern computer vision has adopted this physical model, decoupling an image into these two independent components to enable the explicit manipulation of illumination while preserving the scene's intrinsic properties. Since degradation under low-light conditions can be explained as an image with insufficient illumination and relatively constant reflectance, the Retinex decomposition aligns perfectly with the task of LLIE. Retinex-Net was the first to integrate this principle into a deep learning architecture, which also established the paired LOL-v1 benchmark dataset. This thereby laid the foundation for a subsequent series of physics-guided LLIE methods \cite{Wei2018}.

Subsequent research on Retinex-based LLIE can be mainly categorized into two types. The first one is to introduce physical priors or restoration modules for illumination and reflectance components, thereby enhancing model robustness, respectively. For instance, Zhang et al. \cite{Zhang2019KinD} highlights the issue that direct brightness amplification will also magnify hidden artifacts. Thus, after the Retinex decomposition, KinD introduces a noise suppression module in the reflectance branch and a ratio parameter in the illuminance branch for manual illumination control. Moreover, Yang et al. \cite{yang2021sgmnet} proposed SGM-Net (Sparse Gradient Minimization Sub-Network), which employs sparse gradient regularization to low-amplitude structures while preserving major edges to obtain robust illumination estimation. SGM-Net also deploys specific enhancement sub-networks for each respective component to improve image contrast while suppressing severe noise.

Another type of research focuses on more tightly integrating model-based interpretability with data-driven adaptiveness. For example, URetinex-Net formulates Retinex optimization as a learnable network that replaces hand-crafted priors with data-driven learning while maintaining a structured iterative process \cite{Wu2022URetinexNet}. More recently, some researchers attempted to integrate Retinex with global context learners. For instance, Cai et al. \cite{cai2023retinexformer} propose a Retinex-based single-stage framework (Retinexformer) that integrates illumination-guided self-attention to model long-range dependencies under illumination cues, and it achieves excellent results.

Despite recent progress, Retinex-based methods still face a major challenge: the decomposition process is underdetermined. As a result, the network often learns inaccurate reflectance and illumination, leading to color shifts and texture loss \cite{Wei2018}. To address this issue, an architecture must preserve the distinct properties of each component. This is also the core motivation behind PRIN.

\subsection{Multi-Scale and Frequency-Aware LLIE Methods}
The visual information in an image is spread out over different spatial frequencies or scales. For example, the low-frequency component mainly contains global exposure and contrast information, while the high-frequency component mainly contains edge and texture information \replaced[id=rev]{\cite{zhou2026dlfn}}{[26]]}. The different image information contained by different frequency components or scales suggests a possible strategy that we could enhance or extract information from different frequencies or scales for precise and effective image enhancement. Based on this, various multi-scale (or multi-frequency) methods have been proposed in recent years. For instance, K. Chen et al. \cite{CHEN2025109689} proposed the Refined Mixed Attention Network (RMANet), a multi-scale method, which uses a progressive architecture ranging from coarse-grained to fine-grained. At the coarser scale, the network focuses on recovering global illumination, while at a finer scale, it focuses on extracting and preserving textures. Moreover, to prevent amplified noise during the enhancement, it also integrates a fine-grained blended attention module during the multi-scale fusion stage. Furthermore, the Diffusion-based Low-Frequency Enhancement Network (DLFN) uses a Laplacian pyramid to decompose images into multi-scale components. Additionally, the network employs a dual-branch strategy that leverages the generative priors of diffusion models to reconstruct denoised low-frequency features while employing a high-frequency module to enhance multi-sub-band textures, followed by progressive fusion \cite{zhou2026dlfn}. We can easily observe that these studies attempt to enhance different scales or frequency components separately to achieve more controllable image enhancement.

Additionally, due to the outstanding denoising and generative abilities, diffusion models are increasingly being applied to LLIE. However, directly applying this technology may result in severe loss of details \cite{10758698}. To solve the problem, recent research has begun integrating diffusion models with some decoupling and enhancing techniques. For instance, Diff-Retinex employs Retinex theory as a prior to explicitly decouple an image into illumination and reflectance components. Then, it applies two independent diffusion modules: one for noise \replaced[id=rev]{reduction}{deduction} and detail generation on the reflectance component and another for illumination enhancement on the illumination component \cite{yi2023diffretinex}. PyDiff first introduces an Inverse Pyramid Resolution Process to accelerate diffusion sampling, addressing the issues of slow computation speed and global color shift in the diffusion models. It decouples the degradation restoration process across the scale space \deleted[id=rev]{that} in the coarse-scale stage, it focuses on the illumination and color recovery, and in the fine-scale stage, it focuses on generating details \cite{10.24963/ijcai.2023/199}. H. Jiang et al. \cite{jiang2023diffll} proposed wavelet-based diffusion models (WCDM), which operate directly within the frequency domain. It first decomposes the image into four subbands (LL, LH, HL, HH) using the 2D-DWT. Then, the diffusion model is applied to the LL subband, and the other three (LH, HL, HH) subbands are enhanced by a high-frequency restoration module to restore illumination without affecting the details.

Inspired by the multi-scale and frequency-aware LLIE methods, we introduce different wavelet refinement modules to separately optimize the illumination and reflectance components.
\replaced[id=rev]{
These modules provide frequency-specific regularization that encourages the learned representations to follow their intended restoration roles. The detailed principles and architectural implementations will be introduced in the Methodology section.
}{%
to strictly preserve the physical priors of Retinex theory. We will introduce the detailed principles and architectural implementations in the Methodology section.
}

\subsection{Attention and Perceptual Adaptation}
To better address structural variations, spatially oriented attention mechanisms are increasingly being developed and integrated into vision tasks. For instance, coordinate attention combines positional information with channel attention to make it possible to model long-range dependencies that are sensitive to location \cite{Hou2021}. Triplet Attention rotates the input tensor to compute attention weights directly across three dimension pairs: height-channel, width-channel, and height-width to capture cross-dimensional interactions using a lightweight, three-branch architecture that avoids the information loss caused by dimensionality reduction \cite{Misra2021}.

Recently, attention mechanisms are increasingly used in LLIE to achieve effective spatial-channel fusion. For example, RMANet applies mixed attention across multiple scales to progressively recover structural details \cite{CHEN2025109689}. DLFN uses spatial-channel co-attention specifically to guide its high-frequency enhancement module \cite{zhou2026dlfn}. Furthermore, Retinexformer directly uses the estimated illumination map to guide self-attention computations, effectively modeling long-range dependencies across regions with different lighting conditions \cite{cai2023retinexformer}.

Moreover, global tonal adjustment is important, as precise spatial and structural restoration does not guarantee a visually natural result. In other words, during the image enhancement process, global illumination and color distribution may drift; consequently, the output may preserve structures well but have unnatural tones \cite{9879647}. Recently, an increasing number of methods have begun to integrate global adjustment modules. For instance, EnlightenGAN introduces a global-local discriminator architecture, where the global discriminator explicitly guides the network to align the overall illumination and chromatic distributions with unpaired, high-quality natural images \cite{Jiang2021}. Moreover, DRBN employs a specific perceptual alignment stage that uses unpaired high-quality priors to globally refine the contrast and color distributions of the initially restored signal \cite{Yang2021}. Furthermore, Zero-DCE predicts image-specific, higher-order tone-mapping curves to enable smooth, global dynamic range adjustment \cite{Guo2020}. 

Inspired by the methods, PRIN uses attention mechanisms and a perceptual adaptation module together with its physics-based framework to achieve low-light image enhancement. The proposed Wavelet-Detail-Guided Reflectance Refiner (WDGRR) employs gated channel attention, triplet attention, and explicit edge injection to enhance reflectance representations related to structural features and fine details. The Wavelet-Structure-Aware Illumination Refiner (WSAIR) employs coordinate attention to achieve smoother and more spatially coherent illumination refinement. PRIN also introduces a lightweight curve-based post-composition enhancement head applied after reflectance-illumination recomposition to calibrate residual tone and exposure imbalance.

\subsection{Comparison between PRIN and other methods}
To further clarify the technical positioning of the proposed method, 
Table~\ref{tab:technical_positioning} compares representative Retinex-inspired and frequency-aware LLIE methods with proposed PRIN. 

\begin{table*}[!t]
    \centering
    \caption{Technical comparison of representative low-light image enhancement 
    methods and the proposed PRIN.}
    \label{tab:technical_positioning}
    \setlength{\tabcolsep}{4pt}
    \renewcommand{\arraystretch}{1.18}
    \footnotesize
    \begin{tabularx}{\textwidth}{
        @{}p{2.15cm}
        p{3.20cm}
        Y
        Y@{}
    }
        \toprule
        \textbf{Method} &
        \textbf{Modeling Principle} &
        \textbf{Restoration Focus} &
        \textbf{Frequency / Calibration Design} \\
        \midrule

        Retinex-Net \cite{Wei2018} &
        Learned Retinex decomposition &
        Reflectance-illumination decomposition and illumination enhancement &
        No explicit frequency refinement \\
        \rowrule

        KinD \cite{Zhang2019KinD} &
        Retinex restoration with degradation modeling &
        Reflectance restoration and adjustable illumination enhancement &
        No explicit frequency refinement \\
        \rowrule

        SGM-Net \cite{yang2021sgmnet} &
        Sparse-gradient Retinex restoration &
        Gradient-regularized illumination estimation and component enhancement &
        Structure regularization without wavelet refinement \\
        \rowrule

        URetinex-Net \cite{Wu2022URetinexNet} &
        Unfolded Retinex optimization &
        Iterative decomposition and restoration learning &
        No explicit frequency refinement \\
        \rowrule

        Retinexformer \cite{cai2023retinexformer} &
        Illumination-guided Retinex Transformer &
        Illumination-guided global feature restoration &
        Attention-based long-range modeling \\
        \rowrule

        Diff-Retinex \cite{yi2023diffretinex} &
        Diffusion-based Retinex restoration &
        Component-wise generative restoration &
        Diffusion refinement for reflectance and illumination \\
        \rowrule

        DiffLL \cite{jiang2023diffll} &
        Wavelet-domain diffusion restoration &
        Low-frequency generation and high-frequency detail recovery &
        Subband-specific wavelet processing \\
        \rowrule

        DLFN \cite{zhou2026dlfn} &
        Laplacian frequency-aware diffusion &
        Low-frequency restoration and high-frequency detail enhancement &
        Laplacian multi-frequency modeling \\
        \midrule

        \textbf{PRIN (Ours)} &
        \textbf{Role-aware Retinex wavelet restoration} &
        \textbf{Detail-oriented reflectance and structure-aware illumination refinement} &
        \textbf{Branch-specific Haar refinement and post-composition calibration} \\

        \bottomrule
    \end{tabularx}
\end{table*}

Existing Retinex-based methods provide interpretable decomposition or 
illumination-guided restoration, whereas frequency-aware methods exploit 
multi-scale or subband representations to improve global enhancement and local detail recovery. In contrast, PRIN focuses on associating frequency-domain refinement with the distinct restoration roles of Retinex components.

\section{Proposed Method}
The proposed PRIN is designed based on a simple but consequential idea: the reflectance component and the illumination component of a low-light image should be enhanced in different ways.  
\replaced[id=rev]{Inspired by the Retinex theory, the reflectance component is expected to emphasize intrinsic information, e.g., the textures, edges, etc.; while the illumination component is expected to preserve extrinsic light information, which is smoother across the space. The expectations are treated as regularizing priors rather than as a unique physical decomposition.
}{According to the Retinex theory, the reflectance component mainly contains the intrinsic information, e.g., the textures, edges, etc.; the illumination component mainly contains the extrinsic light information, which is smoother across the space. Our solution is to preserve an explicit Retinex prior and tailor the enhancement strategy to the specific semantics of individual components.}

\added[id=rev]{%
We use the term restoration-role-aware in a specific and verifiable sense. A decomposition network is considered role-aware when its two branches are not just identical operators with independently learned weights, but are structurally differentiated according to the distinct restoration roles of the factors they estimate. Three aspects are selected to make the differentiation to be visible, which are: the effective receptive field, the frequency-domain operator, and the training objective. When this structural differentiation aligns the inductive bias of each branch with the role of the factor it estimates, the network is role-aware in this sense. Section 4.8 provides quantitative evidence for this distinction and further shows that the observed role-specific behavior remains consistent across different choices of the refinement coefficients, rather than being tied to a particular coefficient setting. %
}

Given a low-light input image  $I \in [0,1]^{H \times W \times 3}$, the whole process \replaced[id=rev]{%
is represented using the following Retinex-inspired factorization:%
}{%
could be decomposed according to the Retinex theory and represented as the following equation:%
}
\begin{equation}
    I = R \odot L
\end{equation}

where R denotes the reflectance component, L denotes the illumination component, and $\odot$ denotes element-wise multiplication. 

\added[id=rev]{%
This decomposition is ill-posed as a given observation $I$ may correspond to infinite pairs of reflectance and illumination. For example, for a positive scaling factor $\alpha$ satisfying the feasible component ranges,
\begin{equation}
    I=R\odot L=(\alpha R)\odot(L/\alpha)
\end{equation}
}
\added[id=rev]{%
PRIN restricts this ambiguous solution space through four mechanisms, each of which encodes one role prior. Table \ref{tab:role_aware_mechanisms} summarizes the correspondence between the priors, the mechanisms that realize them, and the sections in which they are described. As a result, the network can predict regularized reflectance and illumination estimations $(\hat{R},\hat{L})$, which is supervised by the network design and losses.%
}

\begin{table*}[!t]
    \centering

    % Only mark the table caption in blue
    \captionsetup{
        labelfont={color=blue},
        textfont={color=blue}
    }
    \caption{Mechanisms used to encode restoration-role awareness in the proposed decomposition framework.}
    \label{tab:role_aware_mechanisms}

    \setlength{\tabcolsep}{4pt}
    \renewcommand{\arraystretch}{1.18}
    \footnotesize

    \begin{tabularx}{\textwidth}{
        @{}p{2.65cm}
        p{5.10cm}
        Y@{}
    }
        \toprule
        \textbf{Mechanism} &
        \textbf{Role prior it encodes} &
        \textbf{Realisation} \\
        \midrule

        Asymmetric branch design &
        Illumination is globally consistent and low-frequency; reflectance is not. & The illumination decoder applies three additional down-sampling stages, reaching a spatial resolution of $1/32$ of the input and thereby providing a larger effective receptive field; the reflectance decoder does not. \\
        \rowrule

        Component-specific wavelet refinement & The two factors require different sub-band treatment. & WDGRR refines reflectance sub-bands with a detail-oriented path; WSAIR refines illumination sub-bands with a structure-oriented path. \\
        \rowrule

        Bounded output projections & The scaling freedom of Eq.~(2) must be removed. & Sigmoid-bounded branch outputs constrain $\hat{R}$ and $\hat{L}$ to a feasible range. \\
        \rowrule

        Loss-based regularization &
        Illumination should be piecewise smooth; the product must stay faithful; sub-bands are constrained band-wise. & $\mathcal{L}_{\text{illum}}$, an edge-aware weighted total variation (TV) term applied to $\hat{L}$ only (\S3.7.6); $\mathcal{L}_{\text{base}}$, product consistency on $\hat{I}_{\text{pre}}$ (\S3.7.5); and $\mathcal{L}_{\text{wav}}$, sub-band weighting. \\

        \bottomrule
    \end{tabularx}
\end{table*}

The network \deleted[id=rev]{predicts $\hat{R}$ and $\hat{L}$ respectively,} then recomposes 
\replaced[id=rev]{the estimations }{them} into an intermediate result, which can be represented as
\begin{equation}
    \hat{I}_{pre} = \hat{R} \odot \hat{L}
\end{equation}

The $\hat{I}_{pre}$ will then be fed to a lightweight curve enhancement head to obtain the final enhanced image $\hat{I}$:
\begin{equation}
    \hat{I} = C (\hat{I}_{pre})
\end{equation}

where $C(\cdot)$ denotes the iterative curve-based refinement module.

\begin{figure*}
	\centering
	\includegraphics[width=1\textwidth]{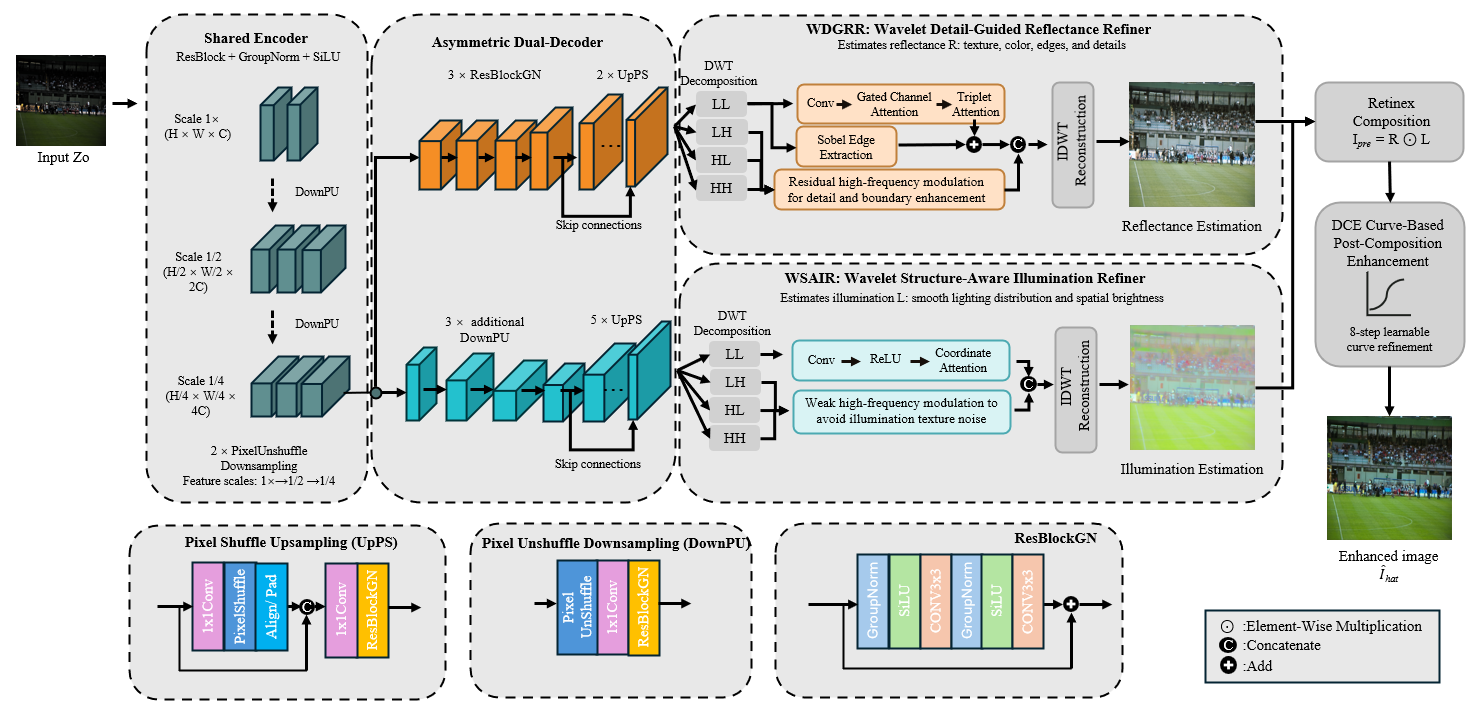}
	\caption{The overall architecture of the proposed PRIN}
	\label{Fig: 3.1}
\end{figure*}

As illustrated in Figure \ref{Fig: 3.1}, the proposed architecture can be mainly divided into four stages: (1) shared feature extraction, (2) asymmetric dual-decoder decoding, (3) wavelet-domain refinements, and (4) curve-based post-composition enhancement. Firstly, a shared encoder extracts shallow-to-mid-level features by two downsampling stages from the input low-light image. The features are then sent to two decoders. The reflectance decoder maintains the latent spatial resolution through three same-scale refinement and fusion blocks before applying two upsampling stages to recover the full-resolution reflectance representation. This design avoids unnecessary spatial compression and helps retain texture- and boundary-related information. In contrast, the illumination decoder introduces three additional downsampling stages followed by five upsampling stages, allowing it to capture broader contextual dependencies for smoother and more spatially coherent illumination estimation. Next, the two branch components \replaced[id=rev]{are}{would be} sent to the Wavelet Detail-Guided Reflectance Refiner and Wavelet Structure-Aware Illumination Refiner, respectively. Both refiners use a discrete wavelet transform (DWT) to separate the low-frequency subband (LL) and high-frequency subbands (LH, HL, and HH). Then we apply different enhancement strategies on the different subbands and use inverse DWT to reconstruct the pre-enhanced results. Finally, a post-composition curve enhancer will be applied on the pre-enhanced results to get the final enhanced results.

\subsection{Shared encoder}
The encoder first pads the input image so that the spatial dimensions are multiples of 32 to ensure the normal pixel unshuffle operations and DWT operations. Then, the padded input will be fed into a residual block equipped with group normalization to extract a 64-channel shallow feature map. After this, two downsampling stages are applied. Each stage consists of a pixel unshuffle operation, a $1 \times 1$ convolution for channel mixing, and a residual block for feature refinement. Through this process, the encoder maps the input to a 256-channel latent representation while preserving complex local details. 

Standard downsampling techniques, such as max-pooling or strided convolutions, unavoidably lead to the loss of spatial information \cite{10.1007/978-3-031-26409-2_27}. Thus, we choose the pixel unshuffle, which performs a space-to-depth transformation, packing spatial elements into the channel dimension prior to feature mixing. Thus, it can help prevent the loss of structured local details\deleted[id=rev]{[45]}, which is crucial for the downstream reflectance branch to reconstruct high-fidelity textures. 

\subsection{Asymmetric dual-decoder design}
The output of the shared encoder is then sent to two asymmetric decoders. The two decoders employ different topologies, since reflectance recovery and illumination adjustment require different types of contextual information.

The reflectance branch applies two residual blocks, followed by a bottleneck block. It then performs two upsampling stages with pixel shuffle and skip fusion. In this branch, we did not add extra downsampling steps to reduce the loss of spatial information. The illumination branch is deeper \added[id=rev]{in} that we add three additional pixel unshuffle downsampling operations interleaved with residual blocks before entering its bottleneck to enlarge the receptive field. After these, five upsampling stages are performed. Thus, smoother and more globally coherent illumination fields can be gained.

The pixel shuffle upsampling is performed by first expanding channels with a $1 \times 1$ convolution and then spatially upsampling using a pixel shuffle operation. Next, the resulting feature map is concatenated with the corresponding encoder skip feature, fused by another $1 \times 1$ convolution, refined with coordinate attention, and finally processed by a residual block. The asymmetric design can help preserve both local details and directional spatial information.

\subsection{Haar DWT and IDWT}
The pair of Haar Discrete Wavelet Transform (Haar DWT) and Haar Inverse Discrete Wavelet Transform (Haar IDWT) is used in both branch refiners for feature decomposition and reconstruction. It could be described as follows:

Let the $F\in\mathbb{R}^{C\times H\times W}$ be a decoded feature map input to the Haar DWT. C denotes the channels, H denotes the height of the feature map, and W denotes the width of the feature map. The Haar DWT decomposes it into four subbands:
\begin{equation}
    \mathcal{W}(F)=\{F_{LL},F_{LH},F_{HL},F_{HH}\}
\end{equation}

where $F_{LL}$ is the low-frequency/approximation band, $F_{LH}$ is the horizontal high-frequency band, $F_{HL}$ is the vertical high-frequency band, and $F_{HH}$ is the diagonal high-frequency band. The Haar DWT is implemented by applying fixed-weight $2 \times 2$ filters across the spatial dimensions with a stride of 2. And the dimension will be changed as follows:
Given a decoded branch feature 
$F \in \mathbb{R}^{C \times H \times W}$, 
the Haar DWT produces four subbands:
$F_{LL}$, $F_{LH}$, $F_{HL}$, and $F_{HH}$, 
each belonging to 
$\mathbb{R}^{C \times \frac{H}{2} \times \frac{W}{2}}$. 
After concatenation along the channel dimension, the complete wavelet-domain representation is 
$F_{\mathcal{W}} \in \mathbb{R}^{4C \times \frac{H}{2} \times \frac{W}{2}}$.

After the enhancement from the branch-specific refiners, the inverse transform (IDWT) will then be applied to reconstruct the spatial feature map. The dimension will be changed from $\mathbb{R}^{4C \times \frac{H}{2} \times \frac{W}{2}}$ (4 subbands) to $\mathbb{R}^{C \times H \times W}$ (reconstructed image). The process can be expressed as:
\begin{equation}
    F_{recon}=\mathcal{W}^{-1}(F_{LL},F_{LH},F_{HL},F_{HH})
\end{equation}

\added[id=rev]{The Haar DWT is adopted mainly because it matches the structure of the proposed refiners. First, it decomposes decoder features into spatially localized subbands, allowing subband-specific convolutions, attention mechanisms, and Sobel-based edge refinement to operate directly in the spatial domain. In contrast, Fourier- or cosine-domain representations do not provide the same localized subband structure required by our design. Second, the Haar basis has very short spatial support and low computational overhead, making it suitable for relatively small latent feature maps. Its 2×2 decomposition is also naturally compatible with the space-to-channel operations used in PRIN. We use only one decomposition level because additional coarse-scale information is already captured by the encoder-decoder hierarchy. Therefore, Haar DWT is selected for its structural compatibility with the proposed role-aware refinement.}

\subsection{WDGRR: Wavelet Detail-Guided Reflectance Refiner}
The WDGRR is used in the reflectance branch in order to preserve detail, texture, and edge while allowing controlled modification of high-frequency components.

\subsubsection{Low-frequency refinement}
WDGRR first extracts the low-frequency component $R_{LL}$ and processes it with a $3 \times 3$ convolution:
\begin{equation}
    \tilde{R}_{LL} = \phi_{3 \times 3}(R_{LL})
\end{equation}

Then we integrate a gated channel attention block and a triplet attention block with the resulting output $\tilde{R}_{LL}$ to encourage cross-dimensional spatial sensitivity and help preserve directional structure:
\begin{equation}
    R_{LL}^{att} = \mathcal{T}(\mathcal{G}(\tilde{R}_{LL}))
\end{equation}

where $\mathcal{G}(\cdot)$ is the operation of global average pooling followed by a channel-wise gating operation, and $\mathcal{T}(\cdot)$ computes attention across multiple tensor orientations to capture channel width, channel height, and spatial interactions.

After the two attention blocks, we apply the explicit edge injection block. The original low-frequency subband is first processed with Sobel filters to obtain an edge map $E(R_{LL})$, which is then added back to the $R_{LL}^{att}$, and we could get the final enhanced LL component of the reflectance branch:
\begin{equation}
R_{LL}^{ref} = R_{LL}^{att} + 0.1 E(R_{LL})
\end{equation}

Thus, the whole enhancing process of the LL subband of the reflectance can be expressed as:
\begin{equation}
R^{\mathrm{ref}}_{LL} = \mathcal{T} \left(\mathcal{G} \left(\phi_{3\times3}(R_{LL})\right) \right) + 0.1\,E(R_{LL}).
\end{equation}

\subsubsection{High-frequency refinement}
The three high-frequency subbands are concatenated first and processed jointly:
\begin{equation}
    R_H = [R_{LH}; R_{HL}; R_{HH}]
\end{equation}

Then we add a small residual modulation network to predict a correction term $\Delta R_H$:
\begin{equation}
    \Delta R_H = \psi(R_H)
\end{equation}

where $\psi(\cdot)$ \replaced[id=rev]{is}{s} composed of a $3 \times 3$ convolution, ReLU, a $1 \times 1$ convolution, batch normalization, and a tanh activation. The correction is then added back with a scale factor of 0.1:
\begin{equation}
    R_H^{ref} = R_H + 0.1 \Delta R_H
\end{equation}

\subsubsection{Reconstruction}
After the refinement of the 4 subbands, the refined reflectance feature is reconstructed as:
\begin{equation}
    R_{ref}^{feat} = \mathcal{W}^{-1}(R_{LL}^{ref}, R_{LH}^{ref}, R_{HL}^{ref}, R_{HH}^{ref})
\end{equation}

and then projected to the reflectance output by a $1 \times 1$ convolution and sigmoid activation.
\added[id=rev]{%
The sigmoid projection bounds the reflectance estimate to $[0,1]$, thereby restricting the scale ambiguity of the learned decomposition.%
}

\subsection{WSAIR: Wavelet Structure-Aware Illumination Refiner}
The WSAIR is used in the illumination branch, focusing on the spatial consistency.

\subsubsection{Low-frequency refinement}
Let the $L_{LL}$ be the low-frequency subband of the illumination branch. WSAIR applies a $3 \times 3$ convolution, ReLU activation, and coordinate attention on this subband, which can be expressed as:
\begin{equation}
    L_{LL}^{ref} = \mathcal{A}_{coord}(\sigma(\phi_{3 \times 3}(L_{LL})))
\end{equation}

\subsubsection{High-frequency refinement}
Similar to the reflectance part, three high-frequency subbands of the illumination branch are first concatenated and processed jointly:
\begin{equation}
    L_H = [L_{LH}; L_{HL}; L_{HH}]
\end{equation}

Then we process the $L_H$ with a grouped $3 \times 3$ convolution, ReLU, a $1 \times 1$ convolution, batch normalization, and tanh to achieve the residual modulation network, which can be expressed as:
\begin{equation}
    \Delta L_H = \eta(L_H)
\end{equation}

Next, we add it back with a scale factor of 0.05:
\begin{equation}
    L_H^{ref} = L_H + 0.05 \Delta L_H
\end{equation}

\subsubsection{Reconstruction}
After the refinement of the 4 subbands, the refined illumination feature is reconstructed:
\begin{equation}
    L_{ref}^{feat} = \mathcal{W}^{-1}(L_{LL}^{ref}, L_{LH}^{ref}, L_{HL}^{ref}, L_{HH}^{ref})
\end{equation}

and then the reconstructed feature will pass through a $1 \times 1$ convolution and sigmoid activation to produce the illumination output.
\added[id=rev]{%
The sigmoid projection similarly bounds the illumination estimate to $[0,1]$ and further restricts the feasible decomposition space.%
}

\subsection{Curve-Based Post-Composition Enhancement}
After branch-specific wavelet refinement and output projection, the predicted reflectance map $\hat{R}$ and illumination map $\hat{L}$ are recomposed to form a preliminary enhanced image:
\begin{equation}
\hat{I}_{\mathrm{pre}} = \hat{R} \odot \hat{L}.
\end{equation}

For the curve-based post-composition enhancement module, the idea is inspired by the Zero-DCE network, but the function and learning setting are different. Rather than directly enhancing the input without reference, we apply this module only after reflectance–illumination recomposition, serving as a lightweight post-composition refinement head. The design allows the network to first establish a Retinex decomposition and then use adaptive curve mapping to compensate for residual tone imbalance, insufficient local contrast, and exposure deviation in the recomposed output.

The module first predicts a stack of curve parameters through seven convolutional layers with internal feature concatenation. The final layer outputs $3T$ channels, where we choose $T = 8$ as the number of curve iterations, which can be expressed as:
\begin{equation}
    \mathbf{F}^{(k)} = \begin{cases}
    \hat{I}_{pre} & k = 0 \\
    \phi(\text{Conv}_k(\mathbf{F}^{(k-1)})), & k = 1, 2, 3, 4 \\
    \phi(\text{Conv}_5([\mathbf{F}^{(4)}, \mathbf{F}^{(3)}])), & k = 5 \\
    \phi(\text{Conv}_6([\mathbf{F}^{(5)}, \mathbf{F}^{(2)}])), & k = 6
    \end{cases}
\end{equation}

After this, we apply the $tanh(\cdot)$ activation on the last layer. The process can be expressed as:
\begin{equation}
    \mathbf{P} = \tanh\left(\text{Conv}_7([\mathbf{F}^{(6)}, \mathbf{F}^{(1)}])\right)
\end{equation}

Next, the parameters are split into $T=8$ groups:
\begin{equation}
    \mathbf{P} = [\mathbf{P}^{(1)}, \mathbf{P}^{(2)}, \dots, \mathbf{P}^{(T)}], \quad T = 8
\end{equation}

where $\mathbf{P}^{(t)} \in \mathbb{R}^{H \times W \times 3}$. The image is then enhanced iteratively:
\begin{equation}
    \hat{I}^{(t+1)} = \hat{I}^{(t)} + P_t \odot \hat{I}^{(t)} \odot (1 - \hat{I}^{(t)}), \quad t = 0, \dots, T - 1
\end{equation}

where the $\hat{I}^{(0)} = \hat{I}_{pre}$ and the final output of the whole PRIN is the \deleted[id=rev]{final} output of the curve-based post-composition enhancement module:
\begin{equation}
    \hat{I}_{enhanced} = \hat{I}^{T=8}
\end{equation}

\subsection{Loss Design}
\replaced[id=rev]{%
The training objective combines final-output data fidelity with structure-, decomposition-, and appearance-oriented regularization. In addition to supervising the process and the output, the loss terms
restrict the ambiguous Retinex-inspired decomposition and discourage solutions that violate the intended restoration roles of the two components.
}{%
The loss design closely follows the architecture, and it mainly includes the decomposition-aware and reconstruction-aware terms.} 
The final objective is a weighted sum of eight terms:

\begin{equation}
\begin{aligned}
    \mathcal{L}_{total} &= \lambda_{rec}\mathcal{L}_{rec} + \lambda_{ssim}\mathcal{L}_{ssim} + \lambda_{edge}\mathcal{L}_{edge} + \lambda_{wav}\mathcal{L}_{wav} \\ &+ \lambda_{base}\mathcal{L}_{base} + \lambda_{illum}\mathcal{L}_{illum} + \lambda_{col}\mathcal{L}_{col} + \lambda_{oe}\mathcal{L}_{oe}
\end{aligned}
\end{equation}

The default weights of the training are:
\begin{equation}
\begin{aligned}
\lambda_{rec} = 1.0, \quad \lambda_{ssim} = 0.15, \quad \lambda_{edge} = 0.05, \quad \lambda_{wav} = 0.08, \\
\quad \lambda_{base} = 0.20, \quad \lambda_{illum} = 0.08, \quad \lambda_{col} = 0.02, \quad \lambda_{oe} = 0.01
\end{aligned}
\end{equation}

\added[id=rev]{%
The reconstruction and structural terms supervise the final enhanced image, while the remaining terms provide regularization. In particular, the intermediate reconstruction loss constrains the recomposition $\hat{R}\odot\hat{L}$, the illumination smoothness loss discourages texture leakage into $\hat{L}$, and the edge- and wavelet-domain loss terms encourage the preservation of structural information. The color and over-exposure terms further restrict undesirable outputs after post-composition refinement.%
}

\added[id=rev]{%
It should be noted that, these loss terms restrict the feasible learned decomposition but do not guarantee a unique physical reflectance-illumination solution.%
}

\subsubsection{Robust reconstruction loss}
The main reconstruction term is the Charbonnier loss between the final output and ground truth:
\begin{equation}
    \mathcal{L}_{rec} = \frac{1}{N} \sum \sqrt{(\hat{I} - I^{gt})^2 + \epsilon^2}
\end{equation}

This penalty is less sensitive to outliers than a plain $l2$ loss and is suitable for enhancement tasks with local intensity discrepancies.

\subsubsection{Structural similarity loss}
To encourage perceptual structural alignment, we apply the structural similarity index measure loss:
\begin{equation}
    \mathcal{L}_{ssim} = 1 - SSIM(\hat{I}, I^{gt})
\end{equation}

This term complements pixel-wise reconstruction by enforcing local luminance, contrast, and structure consistency.

\subsubsection{Edge preservation loss}
We also introduce the following loss to encourage edge preservation, which compares Sobel gradient magnitudes of the output and reference:
\begin{equation}
    \mathcal{L}_{edge} = \| \nabla_s \hat{I} - \nabla_s I^{gt} \|_1
\end{equation}

This term is consistent with the detail-preserving role of WDGRR and helps protect boundaries from over-smoothing.

\subsubsection{Wavelet-domain fidelity loss}
Moreover, we add a wavelet-domain loss on the final enhanced image. Both prediction and reference are decomposed by the Haar DWT. The low-frequency subband is weighted by 0.5, while the three high-frequency bands are weighted higher:
\begin{equation}
    \mathcal{L}_{wav} = 0.5 \| \hat{I}_{LL} - I_{LL}^{gt} \|_1 + \sum_{b \in \{LH, HL, HH\}} \| \hat{I}_b - I_b^{gt} \|_1
\end{equation}

This loss directly encourages frequency-consistent reconstruction and aligns with the architecture’s wavelet-domain design.

\subsubsection{Pre-curve reconstruction supervision}
The network also supervises the intermediate recomposed image $\hat{I}_{pre} = \hat{R} \cdot \hat{L}$:
\begin{equation}
    \mathcal{L}_{base} = \text{Charbonnier}(\hat{I}_{pre}, I^{gt})
\end{equation}

This term supervises the preliminary recomposed output before curve-based 
calibration, preventing the post-composition head from dominating the entire 
enhancement process.

\subsubsection{Edge-aware illumination smoothness}
Next, we apply an edge-aware regularizer to constrain the smoothness of the illumination branch output. Let L denote the illumination estimate and I the low-light input. The gradients of L are weighted by exponentials derived from the gradients of I:
\begin{equation}
    \mathcal{L}_{illum} = \mathbb{E} \left[ e^{-\alpha|\partial_x I|} |\partial_x L| + e^{-\alpha|\partial_y I|} |\partial_y L| \right],
\end{equation}

where $\alpha = 10$. This encourages smooth illumination in homogeneous regions while permitting sharper changes near meaningful image boundaries.

\subsubsection{Color consistency loss}
To constrain the color consistency, we introduce two terms that one is hue alignment in HSV space, and the other \deleted[id=rev]{one} is the image-wise mean color alignment in RGB space:
\begin{equation}
    \mathcal{L}_{col} = \| H(\hat{I}) - H(I^{gt}) \|_1 + 0.5 \| \mu(\hat{I}) - \mu(I^{gt}) \|_1
\end{equation}

where the $H(\cdot)$ denotes the hue channel is extracted after converting an RGB image into HSV space and $\mu(\cdot)$ denotes the image-wise mean RGB vector computed over all pixels. This helps reduce color drift during enhancement, especially after curve-based post-composition enhancement.

\subsubsection{Over-exposure suppression Loss}
Finally, to prevent blown-out highlights, the model penalizes intensities above a threshold $\tau = 0.95$:
\begin{equation}
    \mathcal{L}_{oe} = \mathbb{E} \left[ \max(\hat{I} - \tau, 0)^2 \right].
\end{equation}

This term is lightweight and useful, especially to prevent the curve enhancement head from increasing brightness aggressively.

\section{Experiments}
In this section, we evaluate the performance of our proposed method. First, we introduce the benchmark datasets and the experimental settings used in our evaluation. Then, we conduct comprehensive quantitative comparisons and qualitative visual assessments with recent state-of-the-art competitors to demonstrate the performance of PRIN. Finally, we perform a series of ablation studies to validate the effectiveness and individual contributions of our core architectural designs.

\subsection{Experiments Settings}
\subsubsection{Benchmark Databases}
To demonstrate the effectiveness of the proposed PRIN for low-light image enhancement, we select three widely used benchmark databases: LOL \cite{Wei2018}, LOL-v2-Syn, and LOL-v2-Real \cite{yang2021sgmnet} for performance evaluation. The LOL dataset consists of 500 low-light image pairs that we would divide into 485 paired images for training and 15 pairs for testing. The LOL-v2-Syn dataset consists of 1,000 paired low-light images that we would divide into 900 pairs for training and 100 for testing. The LOL-v2-Real consists of 789 pairs of low-light images that we would divide into 689 pairs for training and 100 pairs for testing.

\begin{table}[H]
\centering
\caption{Quantitative comparison on the LOL dataset. The best and second-best results are highlighted in \textbf{bold} and \underline{underlined}, respectively.}
\label{tab:lol_comparison}
\setlength{\tabcolsep}{5pt}
\renewcommand{\arraystretch}{1.15}

\begin{tabular}{lcccc}
\toprule
Method & Reference & SSIM$\uparrow$ & LPIPS$\downarrow$ & PSNR$\uparrow$ \\
\midrule

\deletedtablerow
{MBLLEN [31]}
{BMVC'18}
{0.724}
{0.225}
{18.040}

\deletedtablerow
{DeepUPE [41]}
{CVPR'19}
{0.636}
{0.357}
{19.160}

\deletedtablerow
{DeepLPF [35]}
{CVPR'20}
{0.737}
{0.366}
{20.660}

MIRNet \cite{zamir2020mirnet}
& ECCV'20
& 0.820
& 0.160
& 23.480 \\

EnlightenGAN \cite{Jiang2021}
& TIP'21
& 0.651
& 0.322
& 17.600 \\

\deletedtablerow
{KinD+ [59]}
{IJCV'21}
{0.832}
{0.158}
{21.800}

\deletedtablerow
{RetinexDIP [63]}
{TCSVT'21}
{0.531}
{0.388}
{19.290}

SGM-Net \cite{yang2021sgmnet}
& TIP'21
& 0.816
& 0.169
& 19.370 \\

CDEF \cite{lei2023cdef}
& TMM'22
& 0.585
& 0.407
& 16.335 \\

\deletedtablerow
{LVENet [11]}
{JON'22}
{0.822}
{0.186}
{23.210}

SCI \cite{ma2022sci}
& CVPR'22
& 0.526
& 0.392
& 14.784 \\

SNR-Net \cite{xu2022snr}
& CVPR'22
& 0.842
& 0.233
& 24.613 \\

Uformer \cite{wang2022uformer}
& CVPR'22
& 0.741
& 0.354
& 19.001 \\

\deletedtablerow
{Bread [10]}
{IJCV'23}
{0.823}
{0.166}
{21.960}

DiffLL \cite{jiang2023diffll}
& TOG'23
& \underline{0.844}
& 0.219
& \textbf{26.336} \\

Diff-Retinex \cite{yi2023diffretinex}
& ICCV'23
& 0.812
& 0.372
& 21.751 \\

GDP \cite{fei2023gdp}
& CVPR'23
& 0.546
& 0.431
& 15.904 \\

LLFormer \cite{wang2023llformer}
& AAAI'23
& 0.816
& 0.169
& 23.650 \\

\deletedtablerow
{MBPNet [58]}
{TIP'23}
{0.834}
{0.128}
{22.880}

Retinexformer \cite{cai2023retinexformer}
& ICCV'23
& 0.843
& 0.121
& 25.152 \\

UHDF \cite{li2023uhdfour}
& ICLR'23
& 0.820
& 0.146
& 23.310 \\

WeatherDiff \cite{ozdenizci2023weatherdiff}
& TPAMI'23
& 0.811
& 0.272
& 17.913 \\

\added[id=rev]{FourierDiff \cite{fourierdiff2024}}
& \added[id=rev]{CVPR'24}
& \added[id=rev]{0.612}
& \added[id=rev]{0.287}
& \added[id=rev]{17.561} \\

LightenDiff \cite{jiang2024lightendiffusion}
& ECCV'24
& 0.803
& 0.192
& 20.453 \\

QuadPrior \cite{wang2024quadprior}
& CVPR'24
& 0.808
& 0.232
& 20.301 \\

UHDFormer \cite{wang2024uhdformer}
& AAAI'24
& 0.821
& 0.259
& 23.093 \\

\added[id=rev]{LYT-Net \cite{brateanu2025lyt}}
& \added[id=rev]{SPL'25}
& \added[id=rev]{0.824}
& \added[id=rev]{\textbf{0.118}}
& \added[id=rev]{22.371} \\

IGDFormer \cite{wen2025igdformer}
& PR'25
& 0.821
& 0.235
& 24.112 \\

\added[id=rev]{UJDE \cite{li2025ujde}}
& \added[id=rev]{ICLR'25}
& \added[id=rev]{0.746}
& \added[id=rev]{0.254}
& \added[id=rev]{19.819} \\

DLFN \cite{zhou2026dlfn}
& PR'26
& \textbf{0.853}
& 0.216
& \underline{25.632} \\

\midrule

\textbf{PRIN (Ours)}
& /
& \textbf{0.853}
& \underline{0.119}
& 23.390 \\

\bottomrule
\end{tabular}
\end{table}

\subsubsection{Implementation Details}
We implement the proposed network based on the PyTorch framework on one NVIDIA GeForce RTX 4090 GPU. The ADAM optimizer \cite{kingma2015adam} with an initial learning rate of $1 \times 10^{-4}$ is employed to optimize the model parameters. The batch size is set to 2, and the model is trained for 2500 epochs until the loss converges. Moreover, all three datasets are used without any resizing that LOL-v1 and LOL-v2-Real are processed at $600 \times 400$, while LOL-v2-Syn is processed at $384 \times 384$. To ensure compatibility with the operations of pixel unshuffle/shuffle and wavelet-based transforms, each input image is padded by reflection to the nearest spatial size divisible by 32 before being fed into the network, and the output is then cropped back to the original image size after inference. In the testing, the original image resolution is unchanged.

\begin{table}[H]
\centering
\caption{Quantitative comparison on the LOL-v2-Syn dataset. The best and second-best results are highlighted in \textbf{bold} and \underline{underlined}, respectively.}
\label{tab:lolv2syn_compare}
\setlength{\tabcolsep}{5pt}
\renewcommand{\arraystretch}{1.15}
\begin{tabular}{lcccc}
\hline
\textbf{Methods} & \textbf{Reference} & \textbf{SSIM}$\uparrow$ & \textbf{LPIPS}$\downarrow$ & \textbf{PSNR}$\uparrow$ \\
\hline
Zero-DCE \cite{Guo2020}        & CVPR'20  & 0.815 & 0.140 & 17.710 \\
MIRNet \cite{zamir2020mirnet}          & ECCV'20  & 0.876 & 0.122 & 21.940 \\
RUAS \cite{liu2021ruas}            & CVPR'21  & 0.652 & 0.335 & 16.550 \\
SCI \cite{ma2022sci}             & CVPR'22  & 0.801 & 0.301 & 22.203 \\
SNR-Net \cite{xu2022snr}         & CVPR'22  & 0.912 & 0.187 & 24.142 \\
Uformer \cite{wang2022uformer}         & CVPR'22  & 0.871 & \underline{0.060} & 22.037 \\
CDEF \cite{lei2023cdef}            & TMM'22   & 0.865 & 0.298 & 23.067 \\
LLFormer \cite{wang2023llformer}        & AAAI'23  & 0.918 & 0.070 & 24.195 \\
PyDiff \cite{10.24963/ijcai.2023/199}          & IJCAI'23 & 0.904 & 0.193 & \underline{25.334} \\
GDP \cite{fei2023gdp}             & CVPR'23  & 0.652 & 0.400 & 18.550 \\
WeatherDiff \cite{ozdenizci2023weatherdiff}     & TPAMI'23 & 0.902 & 0.214 & 23.796 \\
Retinexformer \cite{cai2023retinexformer}   & ICCV'23  & \underline{0.930} & 0.064 & \textbf{25.670} \\
\added[id=rev]{FourierDiff \cite{fourierdiff2024}}
& \added[id=rev]{CVPR'24}
& \added[id=rev]{0.651}
& \added[id=rev]{0.284}
& \added[id=rev]{14.199}
\\
QuadPrior \cite{wang2024quadprior}       & CVPR'24  & 0.758 & 0.306 & 16.112 \\
LightenDiff \cite{jiang2024lightendiffusion}     & ECCV'24  & 0.707 & 0.334 & 18.550 \\
AGLLDiff \cite{lin2025aglldiff}        & AAAI'25  & 0.870 & 0.256 & 21.112 \\
\added[id=rev]{LYT-Net \cite{brateanu2025lyt}}
& \added[id=rev]{SPL'25}
& \added[id=rev]{0.919}
& \added[id=rev]{0.064}
& \added[id=rev]{23.764}
\\
\added[id=rev]{UJDE \cite{li2025ujde}}
& \added[id=rev]{ICLR'25}
& \added[id=rev]{0.795}
& \added[id=rev]{0.268}
& \added[id=rev]{18.210}
\\
DLFN \cite{zhou2026dlfn}            & PR'26    & 0.920 & 0.198 & 25.132 \\
\hline
\textbf{PRIN (Ours)}     & /    & \textbf{0.931} & \textbf{0.056} & 24.596 \\
\hline
\end{tabular}
\end{table}

\subsubsection{Metrics}
For comprehensive performance evaluation, we adopt three widely used metrics: Structural Similarity Index (SSIM), Learned Perceptual Image Patch Similarity (LPIPS), and Peak Signal-to-Noise Ratio (PSNR). SSIM and LPIPS are the primary evaluation criteria in this research, as they better capture structural integrity and perceptual quality, respectively. PSNR is reported as a supplementary reference for pixel-wise reconstruction fidelity. The evaluation protocol enables analysis on structural preservation, perceptual consistency, and conventional distortion-oriented performance while highlighting the performance of human perception quality.

Moreover, to further evaluate perceptual naturalness without relying on paired references, we report three more non-reference image quality metrics: MUSIQ, LIQE, and Q-Align. These metrics are used to reflect learned aesthetic or quality preferences that do not always coincide with paired reconstruction fidelity.

\subsection{Results and Analysis}
\subsubsection{Quantitative Results Evaluation}
For quantitative evaluation, PSNR and SSIM were computed on RGB images with pixel values normalized to [0, 1]. LPIPS was calculated using the AlexNet-based implementation, where both the enhanced image and the ground truth image were linearly mapped to [-1, 1] before evaluation. As shown in Table \ref{tab:lol_comparison}, Table \ref{tab:lolv2syn_compare} and \ref{tab:lolv2real_compare}, the proposed PRIN achieves competitive performance across the LOL, LOL-v2-Syn, and LOL-v2-Real benchmarks when compared with recent state-of-the-art low-light image enhancement methods. On the LOL dataset, our method \replaced[id=rev]{%
ties for the best SSIM at 0.853 and obtains the second-lowest LPIPS of
0.119, closely following LYT-Net.%
}{%
achieves the best SSIM of 0.853 and the best LPIPS of 0.119.%
} On LOL-v2-Syn, PRIN ranks first in both SSIM and LPIPS, reaching 0.931 and 0.056, respectively. On the LOL-v2-Real dataset, PRIN \replaced[id=rev]{%
achieves the highest PSNR of 22.428 and the second-lowest LPIPS of
0.181.%
}{%
attains the best LPIPS of 0.181 and the best PSNR of 22.428.%
} The results indicate that the proposed method is effective in preserving structural consistency and perceptual similarity while maintaining reasonable reconstruction fidelity.

\begin{table}[H]
\centering
\caption{Quantitative comparisons on LOL-v2-Real. The best and second-best results are highlighted in \textbf{bold} and \underline{underlined}, respectively.}
\label{tab:lolv2real_compare}
\setlength{\tabcolsep}{5pt}
\renewcommand{\arraystretch}{1.15}
\begin{tabular}{lcccc}
\hline
\textbf{Methods} & \textbf{Reference} & \textbf{SSIM}$\uparrow$ & \textbf{LPIPS$\downarrow$} & \textbf{PSNR$\uparrow$} \\
\hline
EnlightenGAN \cite{Jiang2021}     & TIP'21       & 0.678     & 0.364     & 18.676 \\
RUAS \cite{liu2021ruas}             & CVPR'21      & 0.626     & 0.395     & 15.351 \\
SNR-Net \cite{xu2022snr}          & CVPR'22      & 0.849     & 0.237     & 21.482 \\
URetinex-Net \cite{Wu2022URetinexNet}     & CVPR'22      & 0.853     & 0.208     & 21.093 \\
CDEF \cite{lei2023cdef}             & TMM'22       & 0.630     & 0.349     & 19.757 \\
SCI \cite{ma2022sci}              & CVPR'22      & 0.540     & 0.345     & 17.304 \\
Uformer \cite{wang2022uformer}          & CVPR'22      & 0.759     & 0.347     & 18.442 \\
LLFormer \cite{wang2023llformer}         & AAAI'23      & 0.792   & 0.192     & 20.485 \\
UHDF \cite{li2023uhdfour}          & ICLR'23      & \underline{0.854}     & 0.292     & 21.785 \\
GDP \cite{fei2023gdp}              & CVPR'23      & 0.493     & 0.435     & 14.294 \\
WaveNet \cite{dang2023wavenet}          & PG'23       & 0.829     & 0.253     & 20.009 \\
DevigNet \cite{luo2024devignet}         & AAAI'24      & 0.706     & 0.256     & 19.813 \\
\added[id=rev]{FourierDiff \cite{fourierdiff2024}}
& \added[id=rev]{CVPR'24}
& \added[id=rev]{0.607}
& \added[id=rev]{0.294}
& \added[id=rev]{16.857}
\\
SGF \cite{li2024sgf}              & TMM'24       & 0.850     & 0.228     & 22.052 \\
QuadPrior \cite{wang2024quadprior}        & CVPR'24      & 0.811     & 0.302     & 20.483 \\
AGLLDiff \cite{lin2025aglldiff}          & AAAI'25      & 0.783     & 0.291     & 19.306 \\
\added[id=rev]{LYT-Net \cite{brateanu2025lyt}}
& \added[id=rev]{SPL'25}
& \added[id=rev]{0.848}
& \added[id=rev]{\textbf{0.135}}
& \added[id=rev]{21.924}
\\
\added[id=rev]{UJDE \cite{li2025ujde}}
& \added[id=rev]{ICLR'25}
& \added[id=rev]{0.792}
& \added[id=rev]{0.266}
& \added[id=rev]{20.349}
\\
DLFN \cite{zhou2026dlfn}             & PR'26        & \textbf{0.859} & 0.201     & \underline{22.412} \\
\hline
\textbf{PRIN (Ours)}     & /     & 0.852     & \underline{0.181}     & \textbf{22.428} \\
\hline
\end{tabular}
\end{table}

\begin{figure*}[H]
    \centering
    \setlength{\tabcolsep}{1pt}
    \renewcommand{\arraystretch}{0.92}

    \begin{tabular}{@{}ccccc@{}}
        %==================== Top block: folder 669 ====================
        \includegraphics[width=0.19\textwidth]{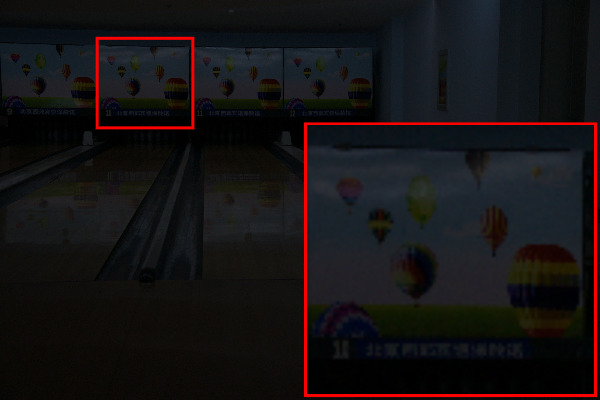} &
        \includegraphics[width=0.19\textwidth]{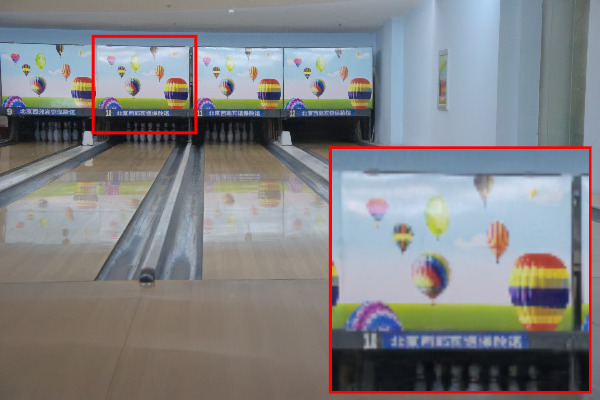} &
        \includegraphics[width=0.19\textwidth]{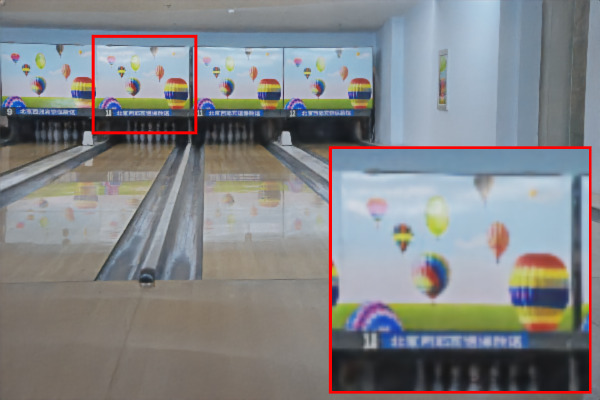} &
        \includegraphics[width=0.19\textwidth]{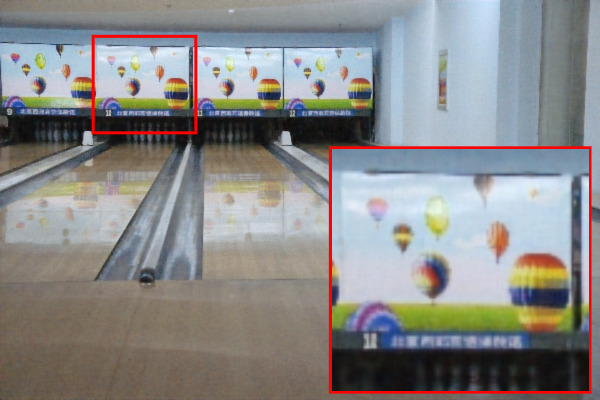} &
        \includegraphics[width=0.19\textwidth]{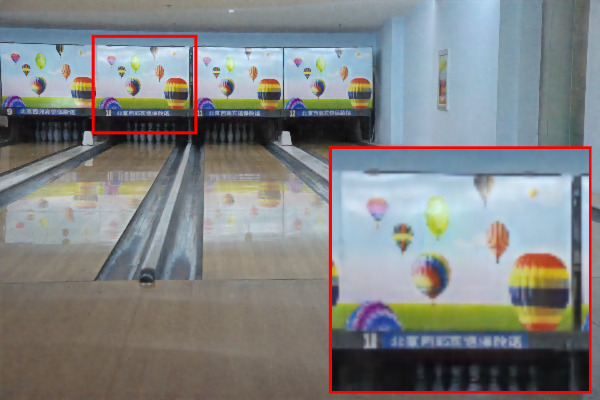} \\

        \includegraphics[width=0.19\textwidth]{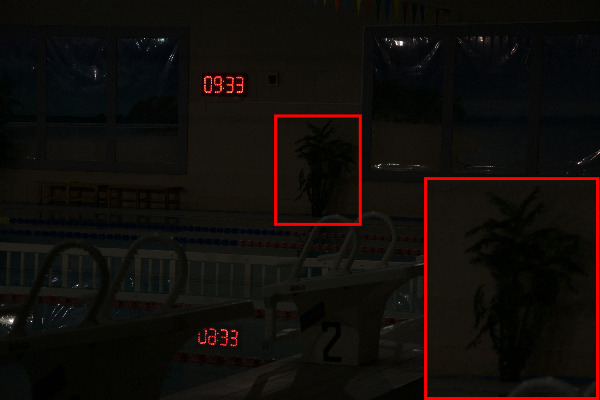} &
        \includegraphics[width=0.19\textwidth]{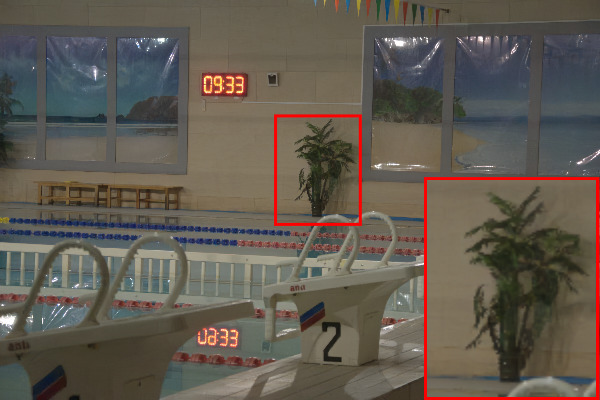} &
        \includegraphics[width=0.19\textwidth]{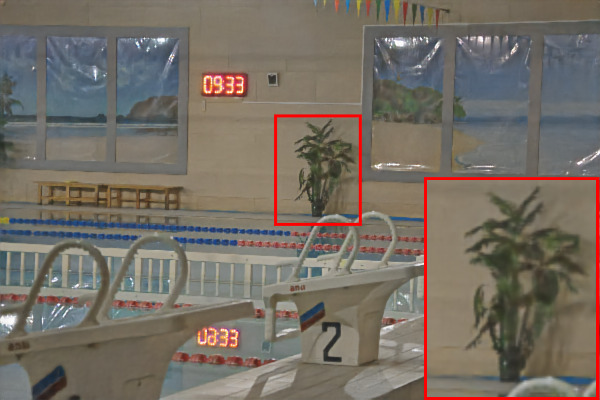} &
        \includegraphics[width=0.19\textwidth]{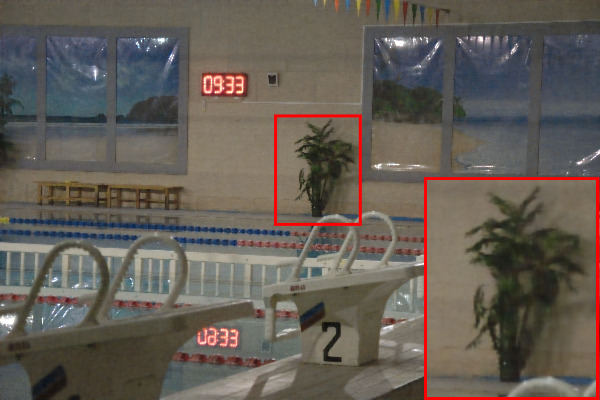} &
        \includegraphics[width=0.19\textwidth]{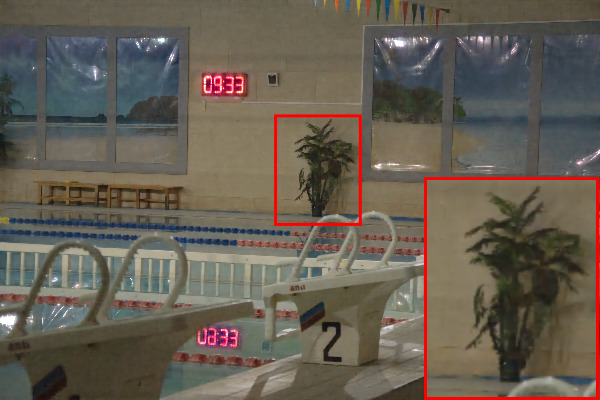} \\

        \includegraphics[width=0.19\textwidth]{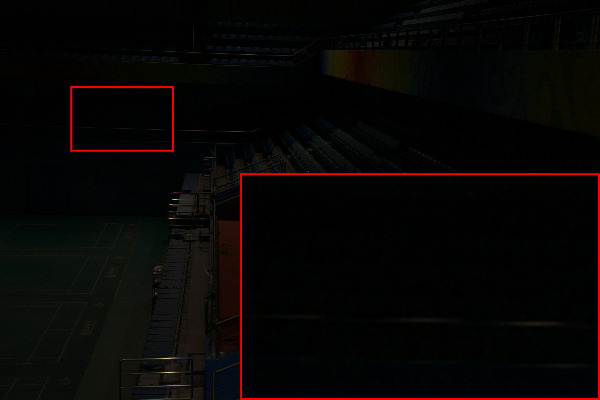} &
        \includegraphics[width=0.19\textwidth]{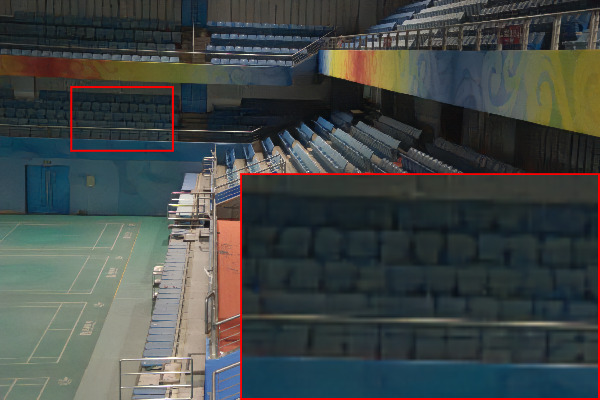} &
        \includegraphics[width=0.19\textwidth]{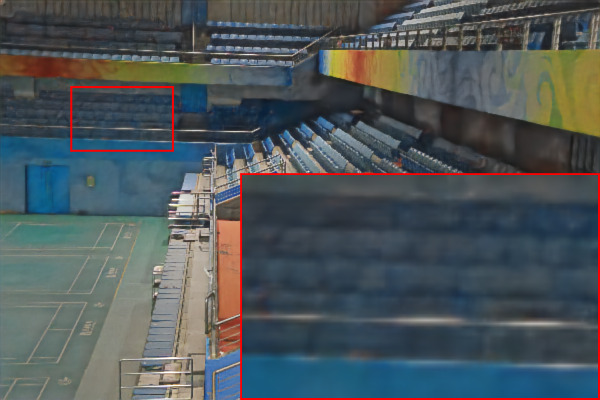} &
        \includegraphics[width=0.19\textwidth]{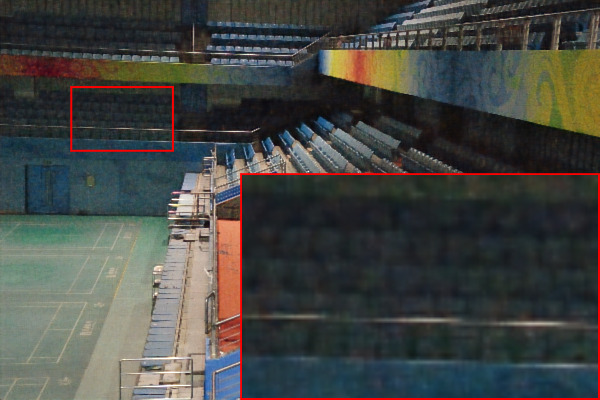} &
        \includegraphics[width=0.19\textwidth]{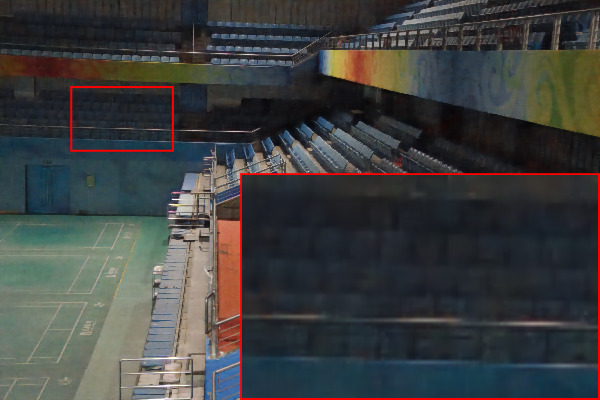} \\

        \textbf{Input} & \textbf{BEM} & \textbf{KinD} & \textbf{LLFormer} & \textbf{MIRNet} \\[2pt]

        %==================== Bottom block: same three folders ====================
        \includegraphics[width=0.19\textwidth]{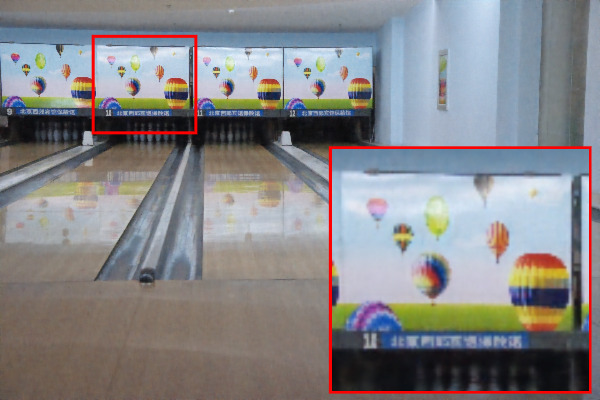} &
        \includegraphics[width=0.19\textwidth]{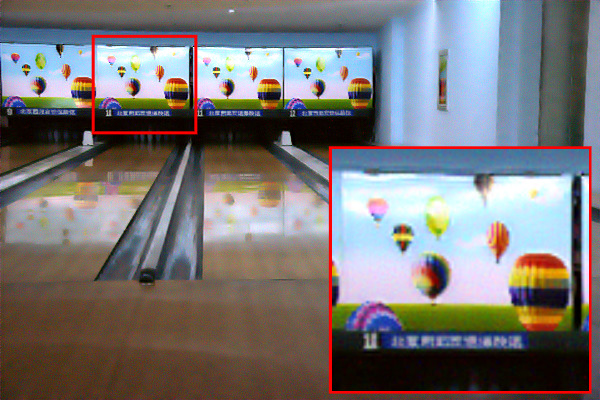} &
        \includegraphics[width=0.19\textwidth]{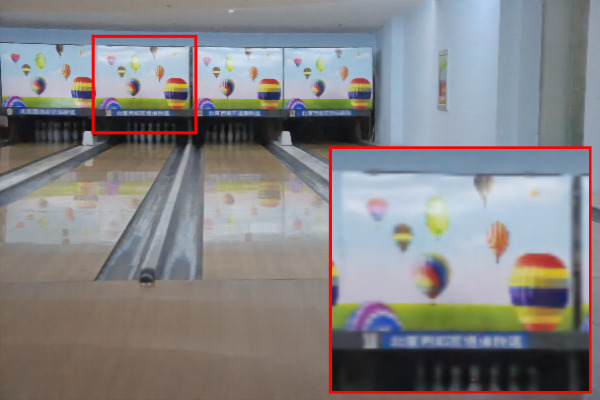} &
        \includegraphics[width=0.19\textwidth]{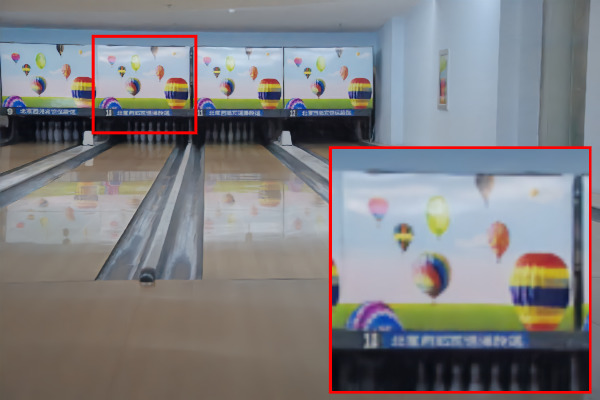} &
        \includegraphics[width=0.19\textwidth]{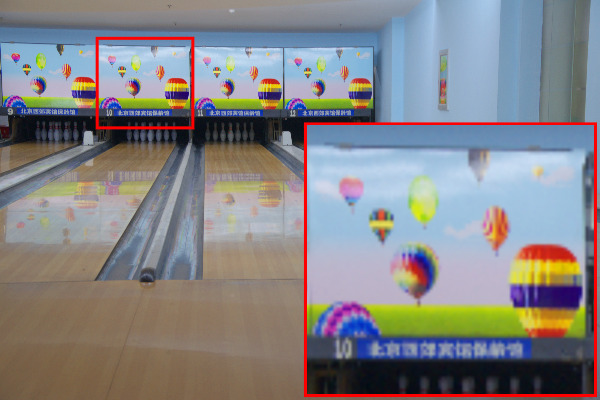} \\

        \includegraphics[width=0.19\textwidth]{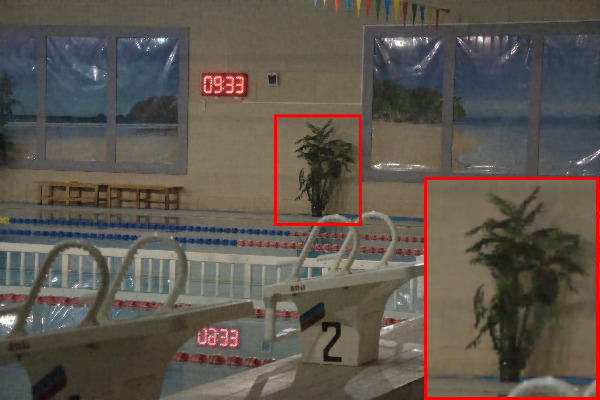} &
        \includegraphics[width=0.19\textwidth]{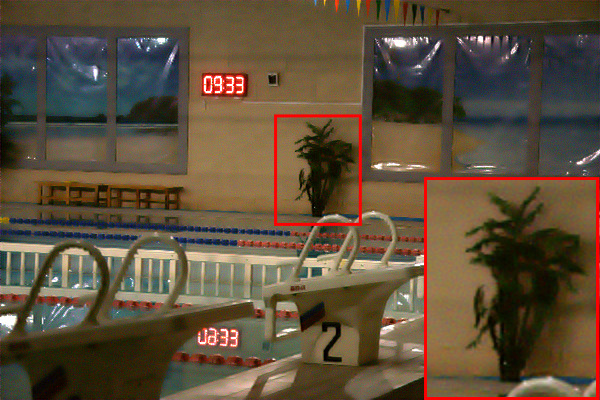} &
        \includegraphics[width=0.19\textwidth]{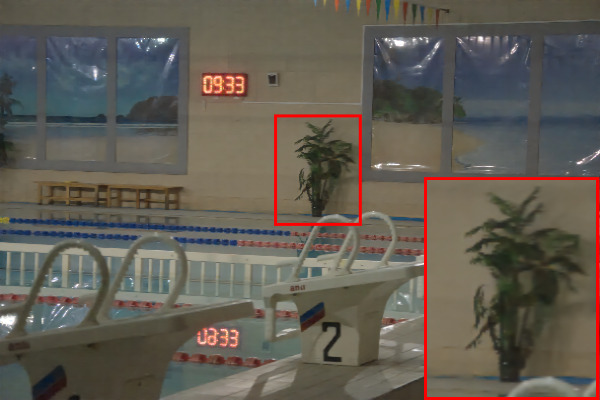} &
        \includegraphics[width=0.19\textwidth]{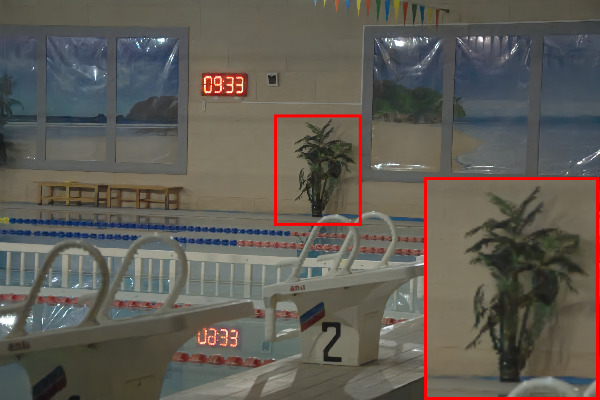} &
        \includegraphics[width=0.19\textwidth]{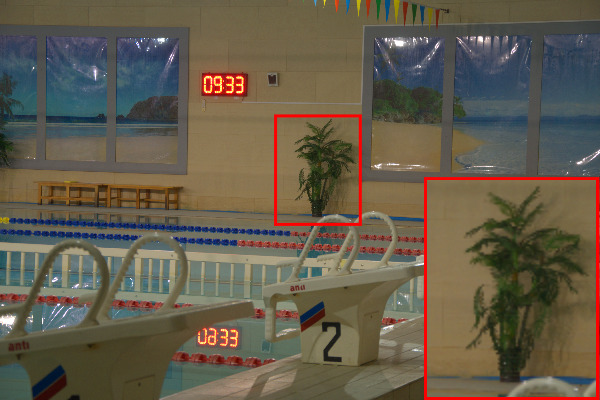} \\

        \includegraphics[width=0.19\textwidth]{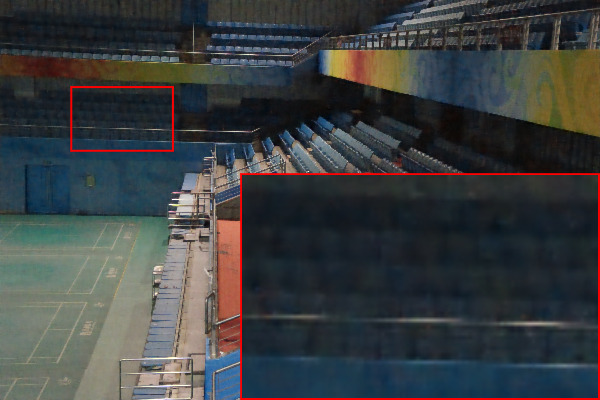} &
        \includegraphics[width=0.19\textwidth]{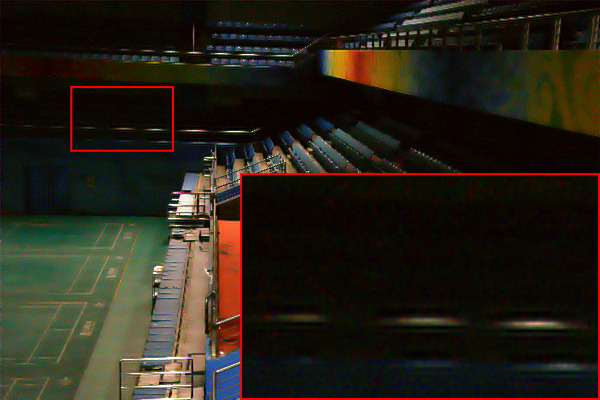} &
        \includegraphics[width=0.19\textwidth]{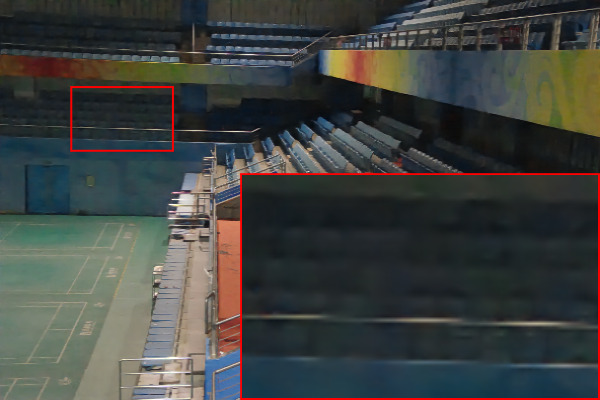} &
        \includegraphics[width=0.19\textwidth]{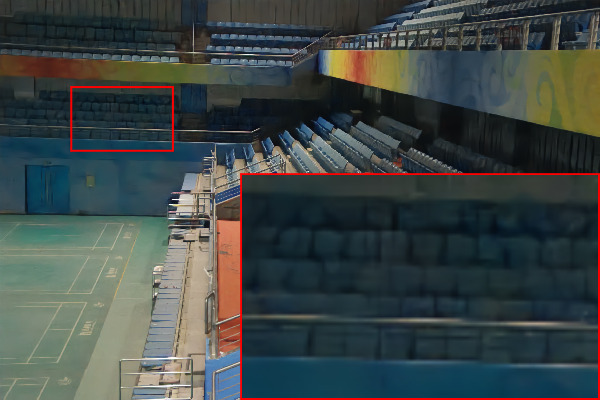} &
        \includegraphics[width=0.19\textwidth]{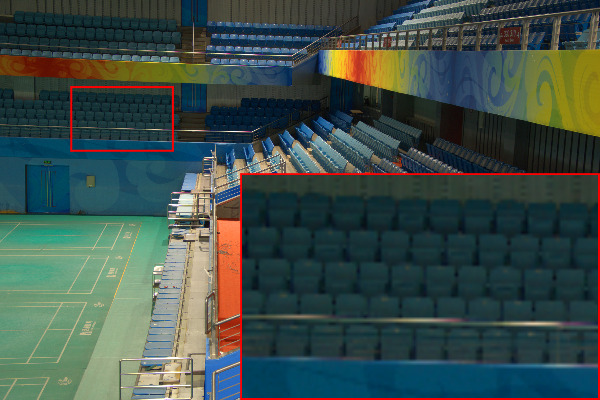} \\

        \textbf{Retinexformer} & \textbf{RUAS} & \textbf{SNR} & \textbf{Ours} & \textbf{Ground Truth} \\
    \end{tabular}

    \caption{Visual comparison of different low-light image enhancement methods on representative examples from the LOL-v1 dataset.}
    \label{fig:visual_comparison_lolv1}
\end{figure*}

\subsubsection{Qualitative Results Evaluation}
In this section, we conduct experiments to compare our proposed PRIN with several representative LLIE methods in LOL-v1, LOL-v2-Syn, and LOL-v2-Real datasets, including BEM \cite{Huang_Yang_Lin_Qi_Bull_Anantrasirichai_2026}, KinD \cite{Zhang2019KinD}, LLFormer \cite{wang2023llformer}, MIRNet \cite{zamir2020mirnet}, Retinexformer \cite{cai2023retinexformer}, RUAS \cite{liu2021ruas}, Restormer \cite{Restormer}, and SNR-Net \cite{xu2022snr}.

Figure \ref{fig:visual_comparison_lolv1} shows how PRIN compares visually to other methods on the LOL-v1 dataset. Most current methods show four types of degradation: uneven lighting, chromatic aberration, loss of structural detail, and edge blurring. For example, the illumination estimation of RUAS is unstable that there exists severe highlight clipping (e.g., the bowling lane) while the shadow regions are not enhanced enough. KinD and LLFormer improve global luminance but still suffer from high-frequency detail attenuation, causing fine topological structures (e.g., stadium seating) to collapse. Moreover, there is a bluish-gray color shift in the results of LLFormer, and MIRNet exhibits a reddish chromatic bias and suppressed mid-tones. Furthermore, while SNR-Net and Retinexformer achieve more acceptable luminance, they still have issues with edge preservation and localized detail recovery. The recent method, BEM, provides similar visual-enhancing results to PRIN, but it has problems of diminished local contrast and desaturated mid-tones. The results of PRIN demonstrate excellent exposure balancing and color constancy while maintaining edges and fine textures and better perceptual quality.

\begin{figure*}[!t]
\centering
\setlength{\tabcolsep}{1pt}
\renewcommand{\arraystretch}{0.85}

\begin{adjustbox}{max width=\textwidth,max height=0.72\textheight,center}
\begin{tabular}{ccccc}
    %==================== Top block ====================
    \includegraphics[width=0.19\textwidth]{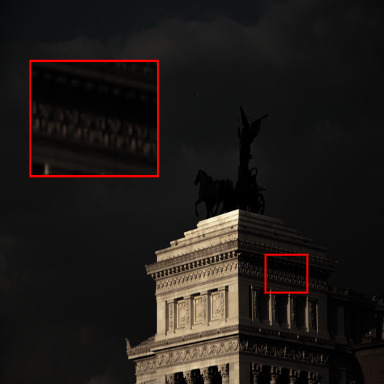} &
    \includegraphics[width=0.19\textwidth]{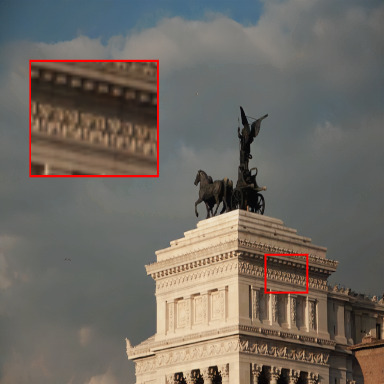} &
    \includegraphics[width=0.19\textwidth]{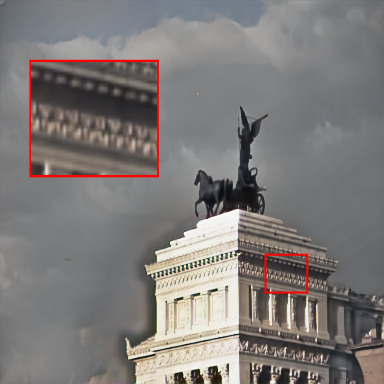} &
    \includegraphics[width=0.19\textwidth]{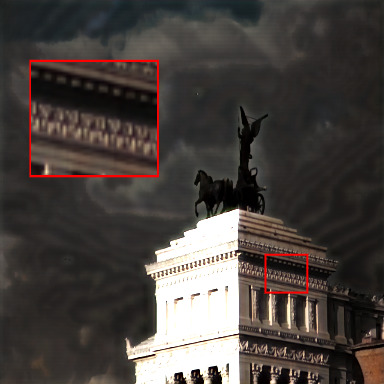} &
    \includegraphics[width=0.19\textwidth]{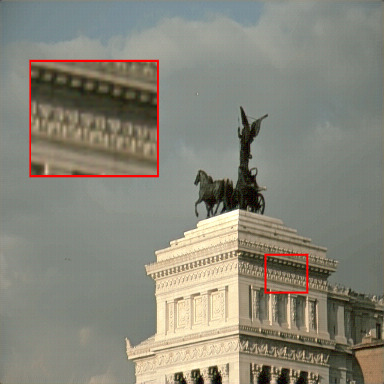} \\

    \includegraphics[width=0.19\textwidth]{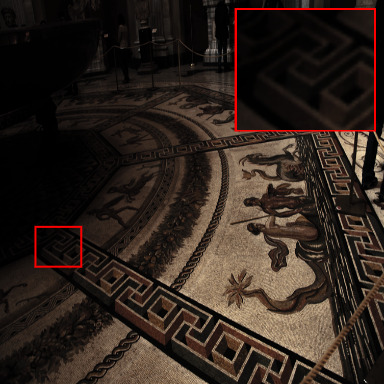} &
    \includegraphics[width=0.19\textwidth]{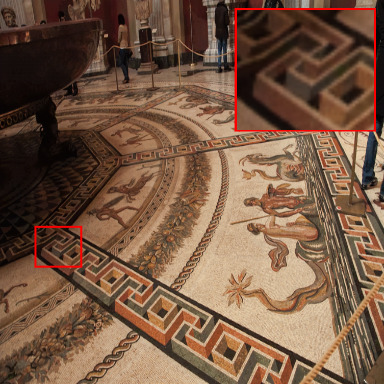} &
    \includegraphics[width=0.19\textwidth]{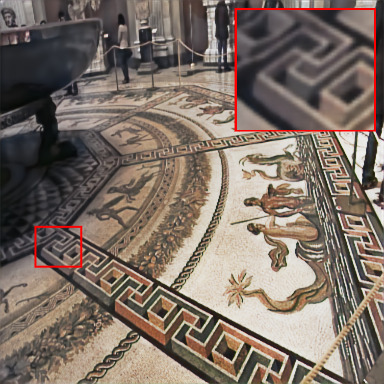} &
    \includegraphics[width=0.19\textwidth]{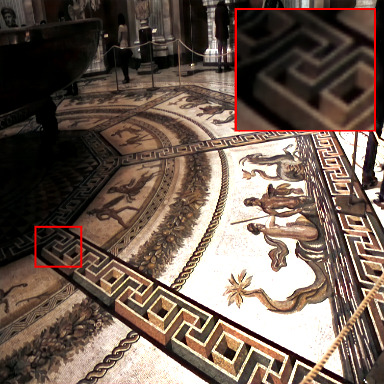} &
    \includegraphics[width=0.19\textwidth]{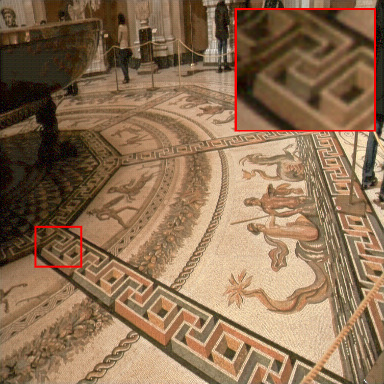} \\

    \includegraphics[width=0.19\textwidth]{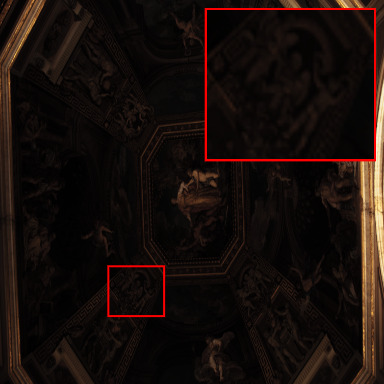} &
    \includegraphics[width=0.19\textwidth]{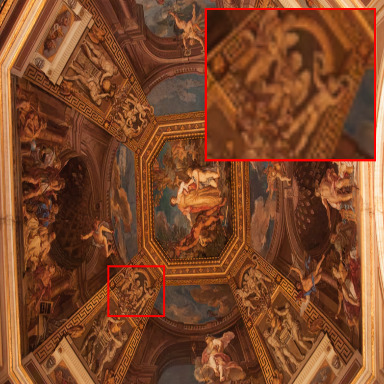} &
    \includegraphics[width=0.19\textwidth]{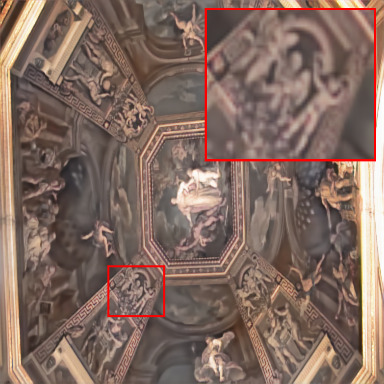} &
    \includegraphics[width=0.19\textwidth]{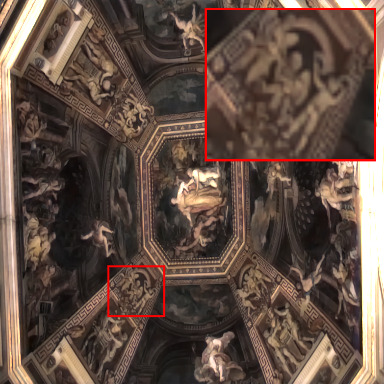} &
    \includegraphics[width=0.19\textwidth]{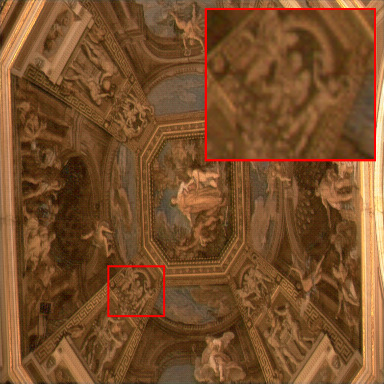} \\

    \textbf{Input} & \textbf{BEM} & \textbf{KinD} & \textbf{MIR} & \textbf{Restormer} \\[2pt]

    %==================== Bottom block ====================
    \includegraphics[width=0.19\textwidth]{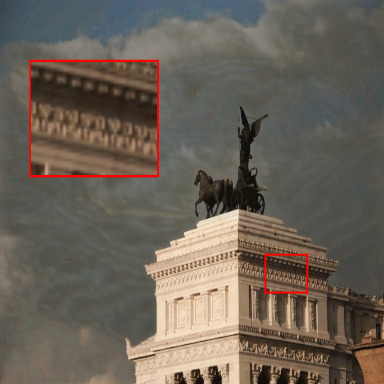} &
    \includegraphics[width=0.19\textwidth]{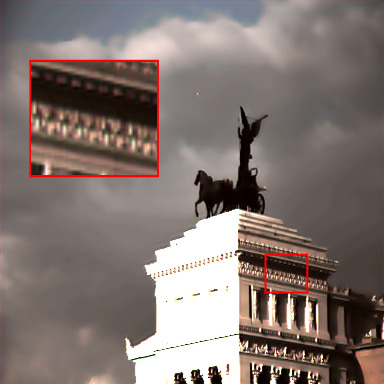} &
    \includegraphics[width=0.19\textwidth]{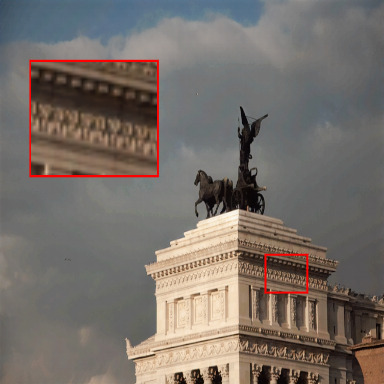} &
    \includegraphics[width=0.19\textwidth]{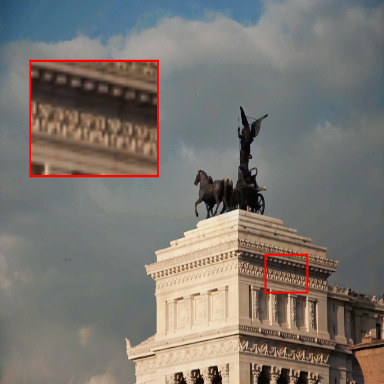} &
    \includegraphics[width=0.19\textwidth]{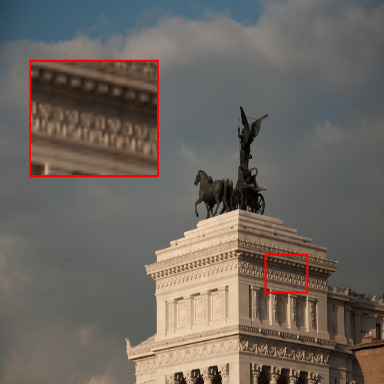} \\

    \includegraphics[width=0.19\textwidth]{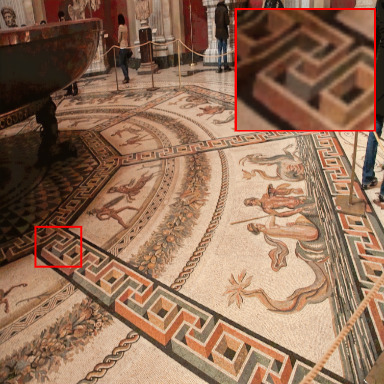} &
    \includegraphics[width=0.19\textwidth]{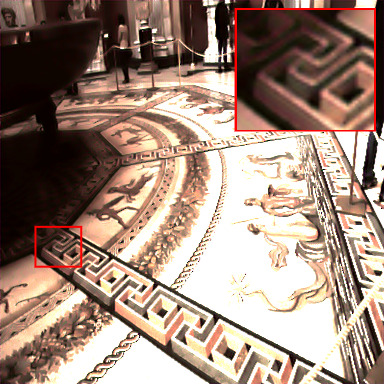} &
    \includegraphics[width=0.19\textwidth]{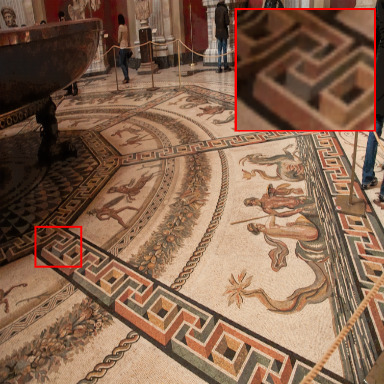} &
    \includegraphics[width=0.19\textwidth]{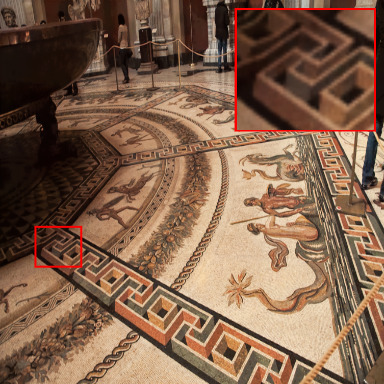} &
    \includegraphics[width=0.19\textwidth]{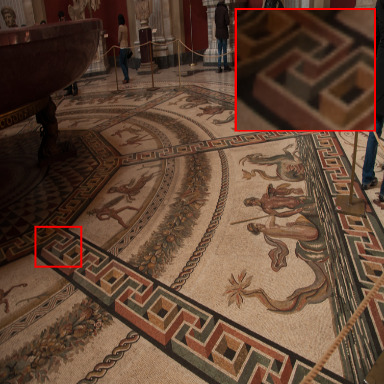} \\

    \includegraphics[width=0.19\textwidth]{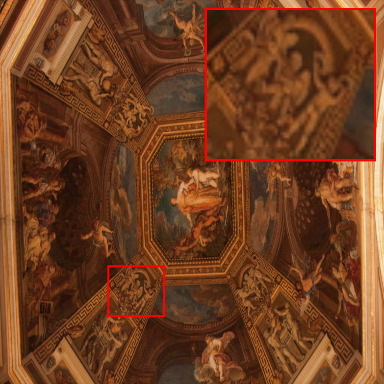} &
    \includegraphics[width=0.19\textwidth]{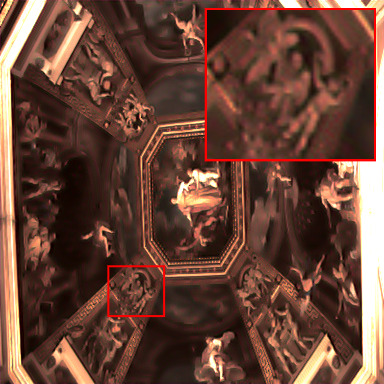} &
    \includegraphics[width=0.19\textwidth]{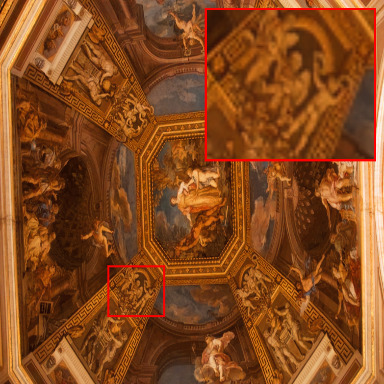} &
    \includegraphics[width=0.19\textwidth]{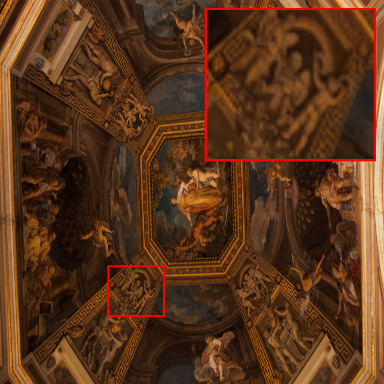} &
    \includegraphics[width=0.19\textwidth]{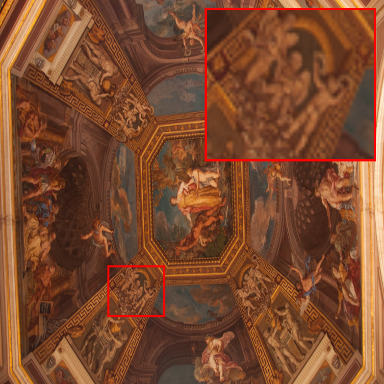} \\

    \textbf{Retinexformer} & \textbf{RUAS} & \textbf{SNR} & \textbf{PRIN (Ours)} & \textbf{Ground Truth} \\
\end{tabular}
\end{adjustbox}

\caption{Visual comparison of different low-light image enhancement methods on representative examples from the LOL-v2-Syn dataset.}
\label{fig:visual_comparison_lolv2syn}
\end{figure*}

Figure \ref{fig:visual_comparison_lolv2syn} shows the comparison results on the LOL-v2-Syn dataset, which offers degradation-free references for evaluating illumination correction and texture fidelity. As shown in the figure, BEM and SNR-Net can preserve visually coherent global illumination, but the high-frequency details are smoothed, and the local contrast in complex patterns is reduced (e.g., the ceiling motifs). KinD and Retinexformer can boost luminance levels but cause a desaturated, muted effect, and they fail to recover structural elements. Restormer is effective at restoring global illumination, but it makes the colors look warmer (brings color shifts). Moreover, MIRNet has relatively conservative enhancement, resulting in underexposed mid-tones, while RUAS is more aggressive, 

\FloatBarrier

causing unnatural highlight clipping in high-reflectance regions (e.g., the monument facade). The proposed method, PRIN, demonstrates robust restoration capabilities, yielding accurate color rendition, enhanced local contrast, and superior structural preservation across varying scenes without introducing chromatic or exposure artifacts.

\begin{figure*}[t]
    \centering
    \setlength{\tabcolsep}{1pt}
    \renewcommand{\arraystretch}{0.95}

    \begin{tabular}{ccccc}
        %==================== Top block: LOLv2-Real ====================
        \includegraphics[width=0.195\textwidth]{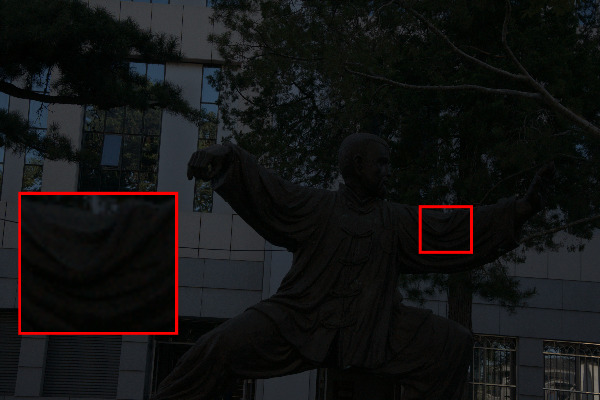} &
        \includegraphics[width=0.195\textwidth]{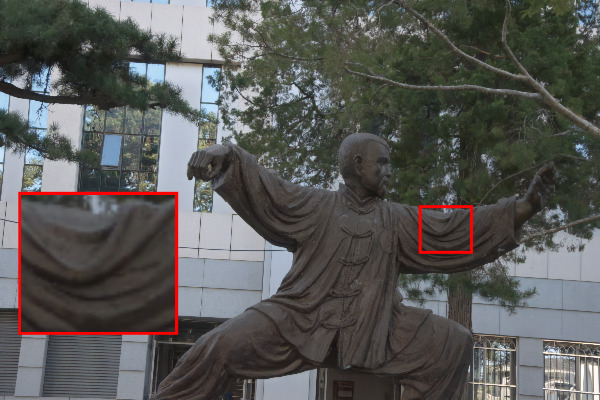} &
        \includegraphics[width=0.195\textwidth]{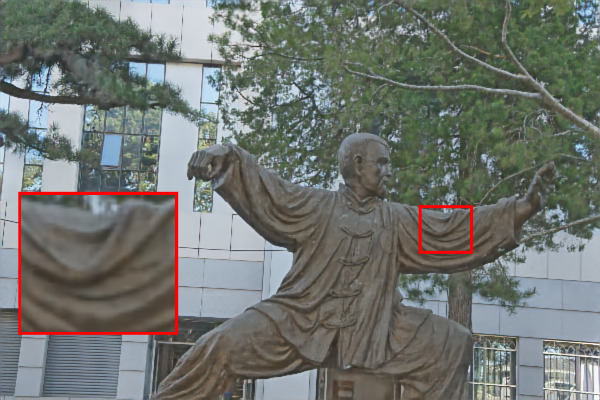} &
        \includegraphics[width=0.195\textwidth]{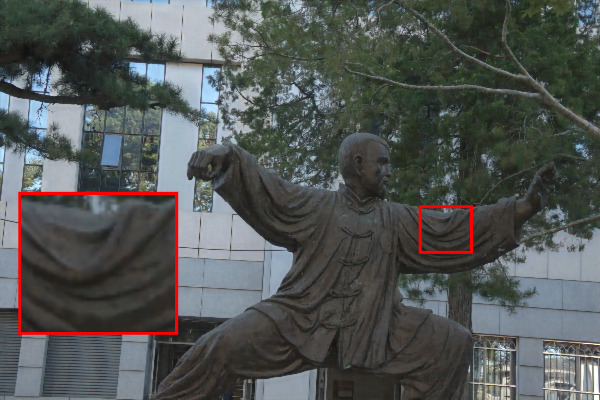} &
        \includegraphics[width=0.195\textwidth]{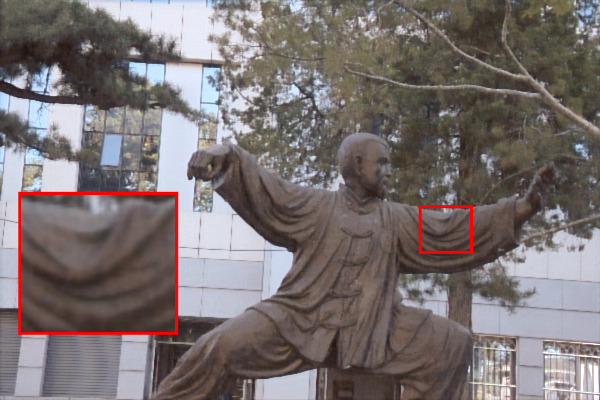} \\

        \includegraphics[width=0.195\textwidth]{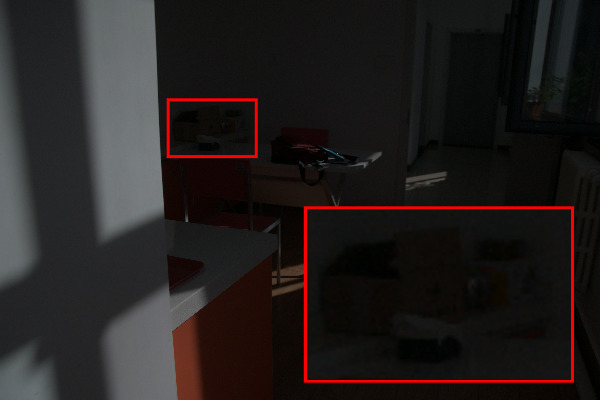} &
        \includegraphics[width=0.195\textwidth]{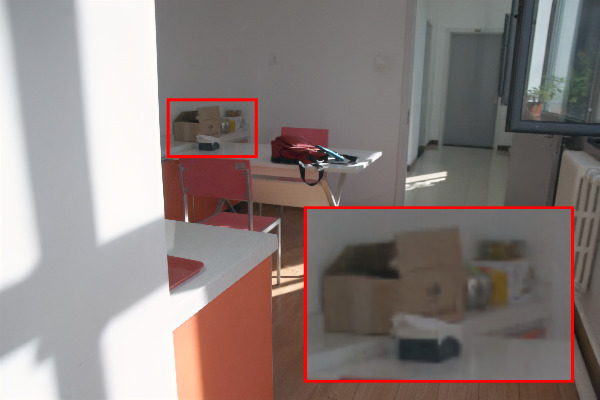} &
        \includegraphics[width=0.195\textwidth]{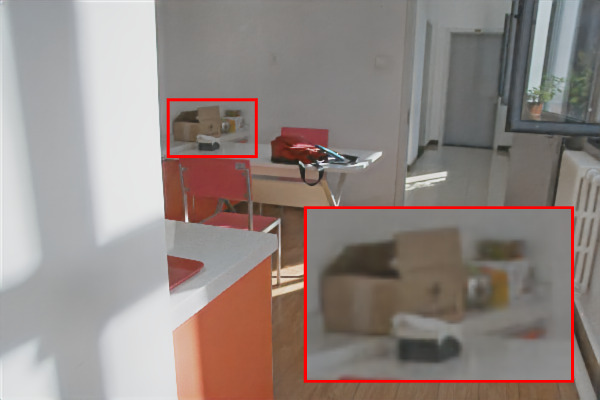} &
        \includegraphics[width=0.195\textwidth]{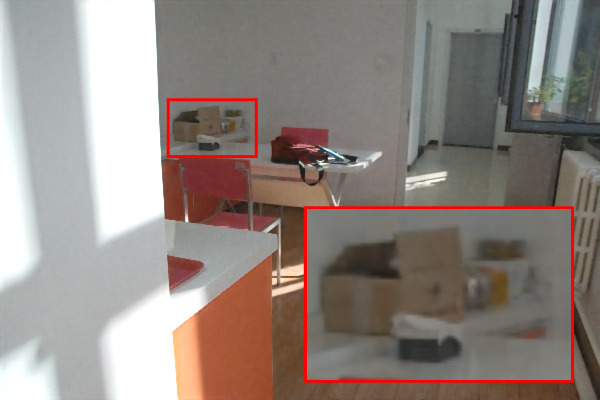} &
        \includegraphics[width=0.195\textwidth]{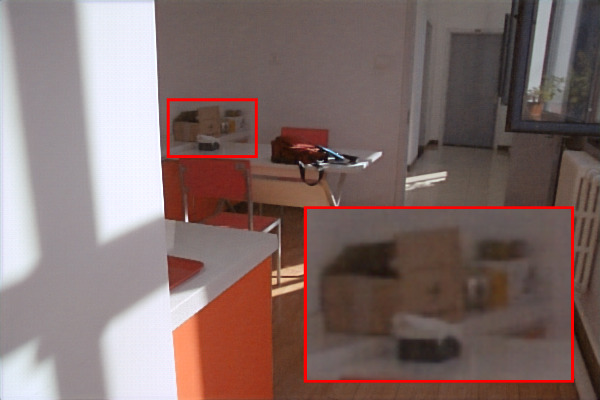} \\

        \includegraphics[width=0.195\textwidth]{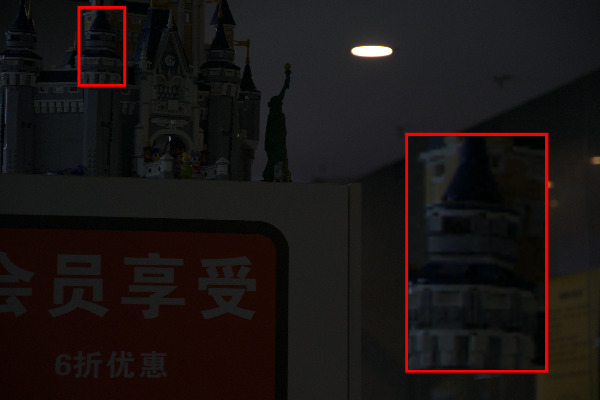} &
        \includegraphics[width=0.195\textwidth]{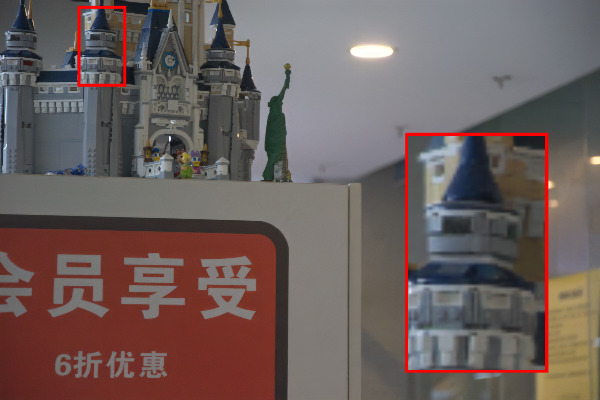} &
        \includegraphics[width=0.195\textwidth]{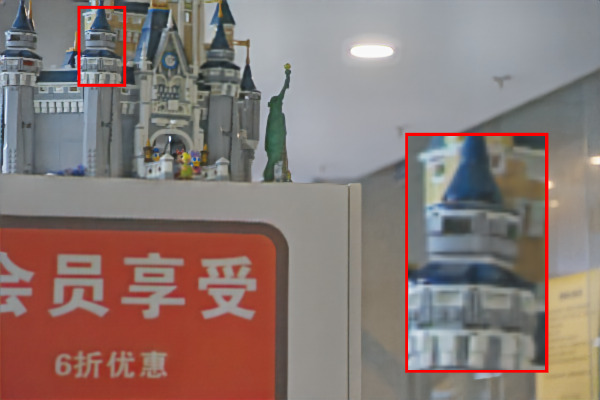} &
        \includegraphics[width=0.195\textwidth]{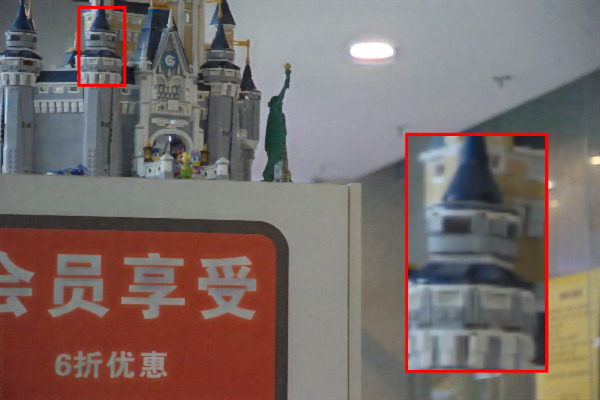} &
        \includegraphics[width=0.195\textwidth]{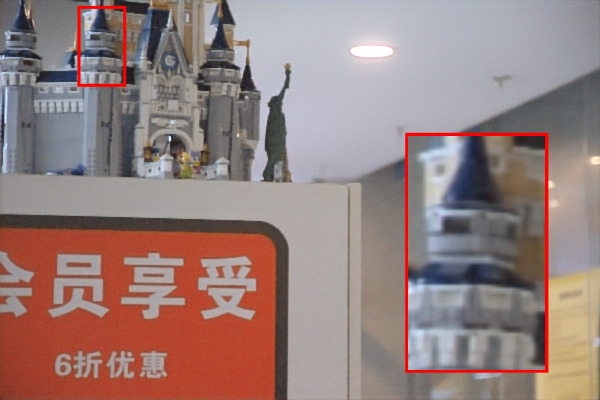} \\

        \textbf{Input} & \textbf{BEM} & \textbf{KinD} & \textbf{MIRNet} & \textbf{Restormer} \\[2pt]

        %==================== Bottom block: LOLv2-Real ====================
        \includegraphics[width=0.195\textwidth]{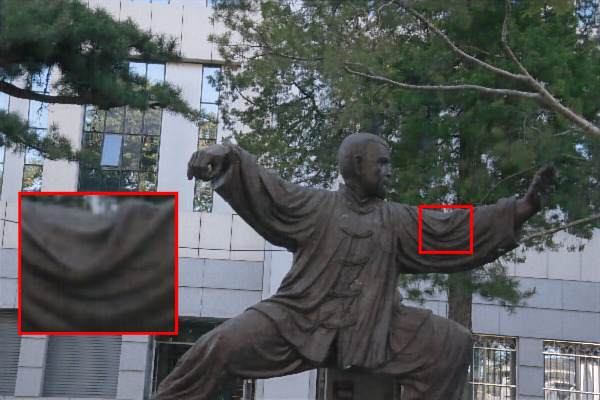} &
        \includegraphics[width=0.195\textwidth]{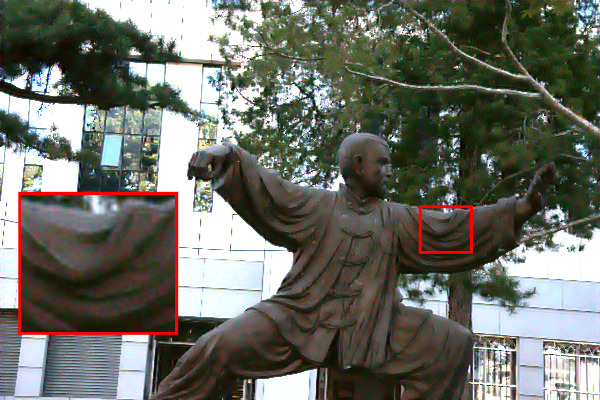} &
        \includegraphics[width=0.195\textwidth]{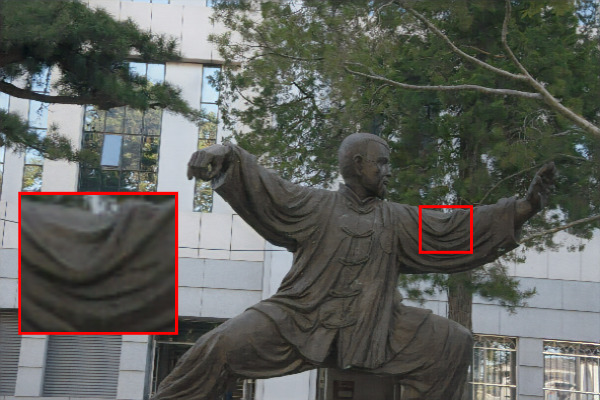} &
        \includegraphics[width=0.195\textwidth]{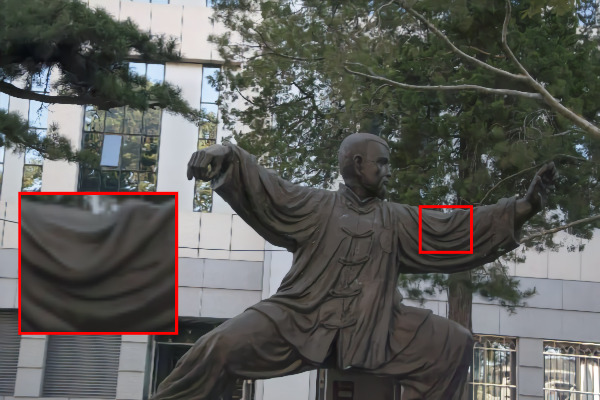} &
        \includegraphics[width=0.195\textwidth]{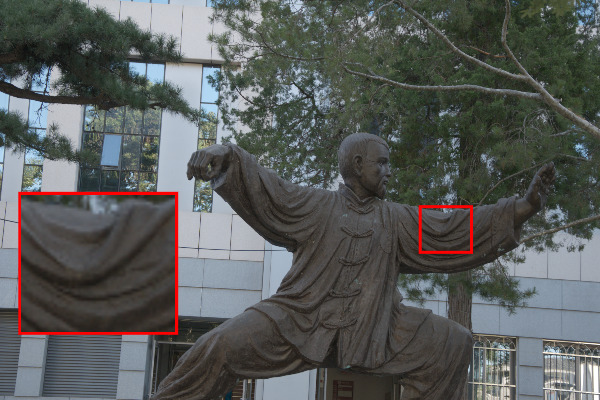} \\

        \includegraphics[width=0.195\textwidth]{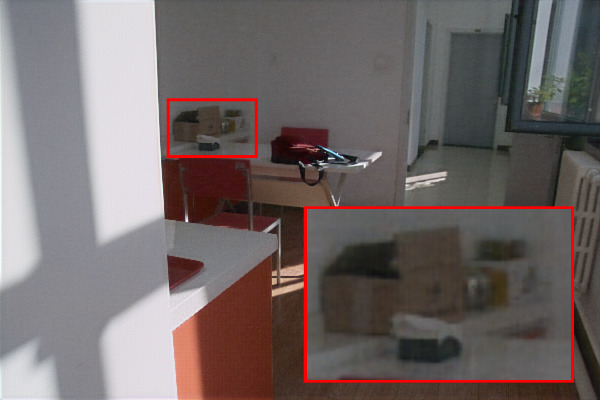} &
        \includegraphics[width=0.195\textwidth]{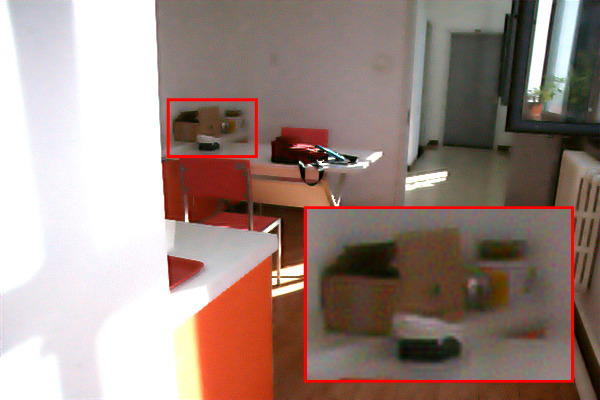} &
        \includegraphics[width=0.195\textwidth]{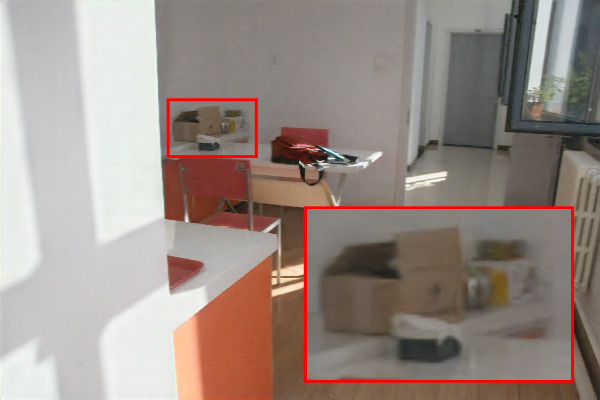} &
        \includegraphics[width=0.195\textwidth]{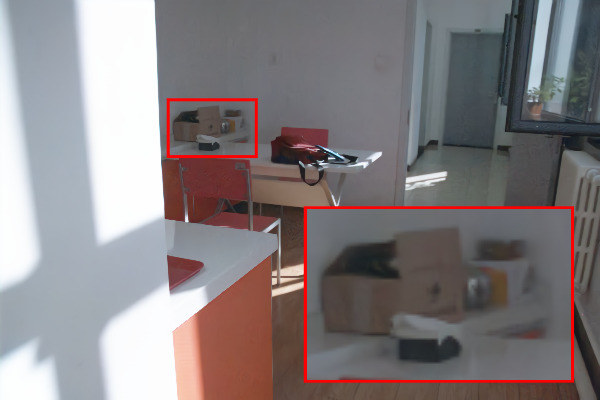} &
        \includegraphics[width=0.195\textwidth]{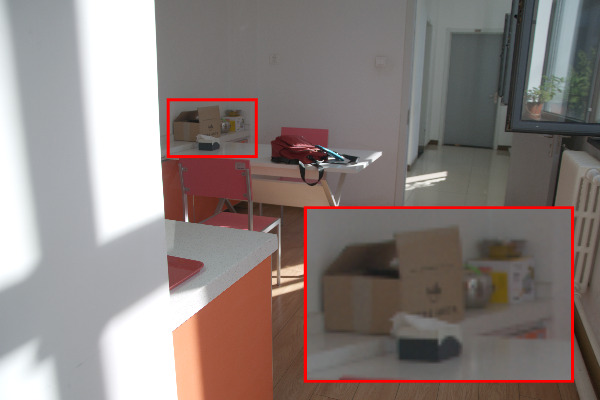} \\

        \includegraphics[width=0.195\textwidth]{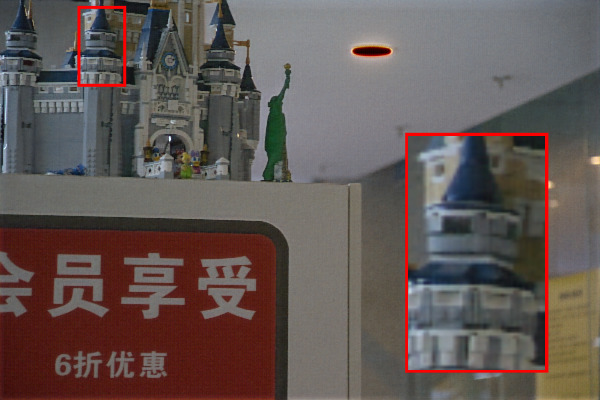} &
        \includegraphics[width=0.195\textwidth]{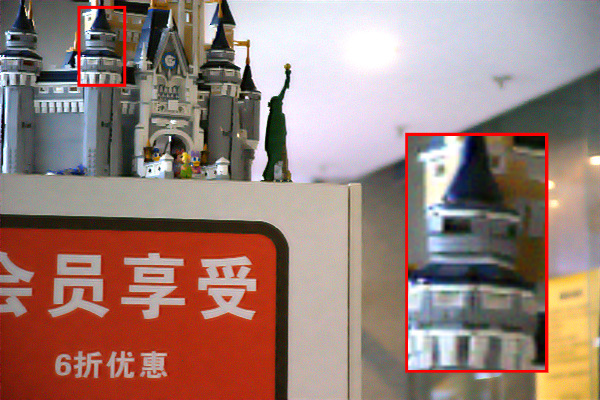} &
        \includegraphics[width=0.195\textwidth]{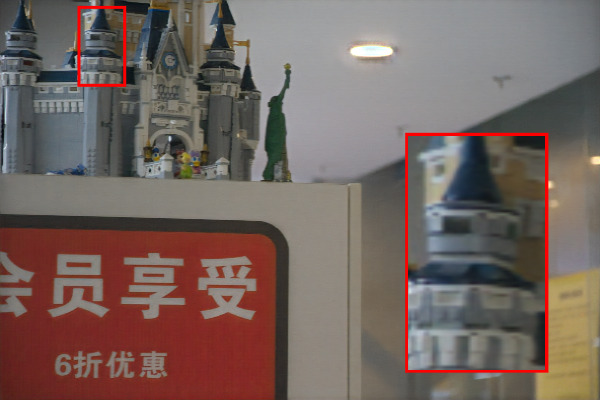} &
        \includegraphics[width=0.195\textwidth]{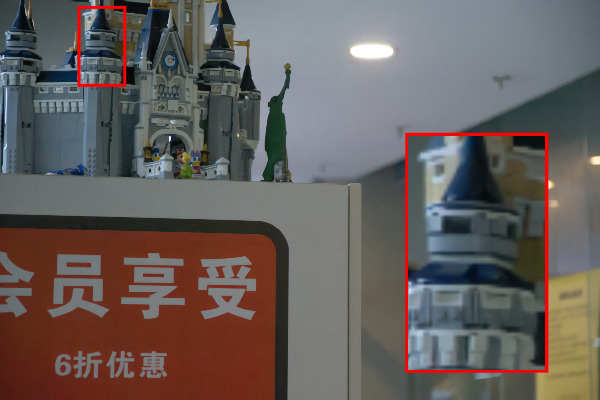} &
        \includegraphics[width=0.195\textwidth]{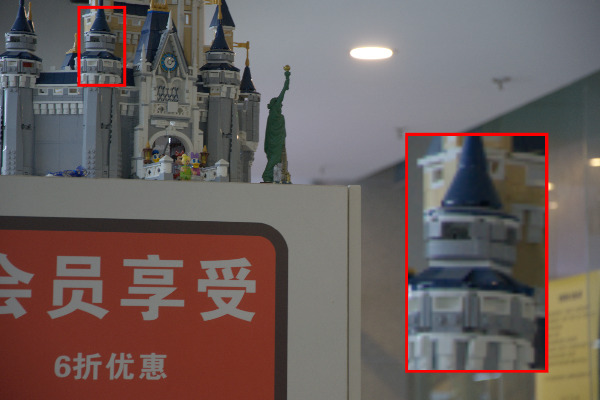} \\

        \textbf{Retinexformer} & \textbf{RUAS} & \textbf{SNR} & \textbf{PRIN (Ours)} & \textbf{Ground Truth} \\
    \end{tabular}

    \caption{Visual comparison of different low-light image enhancement methods on representative examples from the LOL-v2-Real dataset.}
    \label{fig:visual_comparison_lolv2real}
\end{figure*}

Figure \ref{fig:visual_comparison_lolv2real} shows the visual comparison results of how well the algorithms worked on the LOL-v2-Real dataset. This dataset has complex, real-world degradations like severe underexposure, illumination that changes with space, and severe texture degradation. In the figure, BEM and KinD mitigate global darkness, but they have problems with local contrast degradation and structural flattening that the textures are washed out. There's a distinct warm bias in the results of Restormer and local color saturation artifacts in indoor scenarios. MIRNet and Retinexformer exhibit constrained localized detail recovery, struggling to delineate complex object boundaries (e.g., bottle labels). The exposure control of RUAS is not stable, which highlights clipping in mixed-illumination scenes. SNR-Net recovers color successfully, but the textures are smoothed out. The proposed method, PRIN, produces perceptually natural enhancements. It improves the visibility of shadow regions while alleviating highlight saturation in high-dynamic-range scenes and preserving object contours and color representations with reasonable fidelity.

Notably, since the paired ground-truth images in LOL-v2-Real are captured empirically, they inherently contain mild degradations such as slight motion blur or suboptimal local contrast. In several displayed high-frequency regions, such as text boundaries and surface textures, PRIN produces sharper visible boundaries than the paired reference.

\subsection{Variant Study}
To better understand the progressive evolution of the proposed framework, we conducted the variant study by comparing four representative network variants. V1a and V1b are two classical Retinex-based baselines, which follow a conventional encoder-decoder design with wavelet-domain processing. V1a adopts a lighter Conv-BN-ReLU design with single-channel illumination prediction, while V1b uses BatchNorm-PReLU pre-activation blocks with a wider backbone and three-channel illumination estimation. V2 introduces the main architectural upgrade by replacing conventional pooling-based sampling with pixel unshuffle/pixel shuffle, reformulating the backbone with GroupNorm-SiLU blocks, enabling decoder attention, and adopting a curve-based post-enhancement head. V3 is built based on V2, in which the wavelet module is shifted to an earlier stage, decoder attention is made selective rather than uniform, and the post-enhancement head is replaced by a global affine-gamma tone module. In this work, V2 is retained as the final model. The detailed comparison of the four variants is provided in Table~\ref{tab:variant_study}. The quantitative comparison of the final results of four variants is provided in Table~\ref{tab:variant_results}.

\begin{table*}[t]
\centering
\caption{Comparison of different network variants.}
\label{tab:variant_study}
\setlength{\tabcolsep}{4pt}
\begin{tabular}{l l l l l l}
\toprule
Variant & Block Design & Down/Up sampling & Illumination & Wavelet Position & Post-Enhancement \\
\midrule
V1a & Conv-BN-ReLU & MaxPool / ConvTranspose & 1-channel & 64-channel & GammaHead \\
V1b & BatchNorm+PReLU (PreAct) & MaxPool / ConvTranspose & 3-channel & 64-channel & GammaHead \\
V2  & GroupNorm + SiLU & PixelUnshuffle / PixelShuffle & 3-channel & 64-channel & DCECurveEnhancer \\
V3  & GroupNorm + SiLU & PixelUnshuffle / PixelShuffle & 3-channel & 128-channel & GlobalAffineGammaHead \\
\bottomrule
\end{tabular}
\end{table*}

\begin{table*}[t]
\centering
\caption{Quantitative comparison of different network variants on three datasets. The best and second-best results are highlighted in \textbf{bold} and \underline{underlined}, respectively.}
\label{tab:variant_results}
\setlength{\tabcolsep}{4pt}
\begin{tabular}{l c ccc ccc ccc}
\toprule
\multirow{2}{*}{Variant} & \multirow{2}{*}{Params (M)} 
& \multicolumn{3}{c}{LOL-v1} 
& \multicolumn{3}{c}{LOL-v2-Real} 
& \multicolumn{3}{c}{LOL-v2-Syn} \\
\cmidrule(lr){3-5} \cmidrule(lr){6-8} \cmidrule(lr){9-11}
& 
& LPIPS$\downarrow$ & SSIM$\uparrow$ & PSNR$\uparrow$
& LPIPS$\downarrow$ & SSIM$\uparrow$ & PSNR$\uparrow$
& LPIPS$\downarrow$ & SSIM$\uparrow$ & PSNR$\uparrow$ \\
\midrule
V1a & \textbf{5.210}& 0.219  & 0.803  & 20.55  & -  & -  & -  & -  & -  & -  \\
V1b & \underline{23.394}  & 0.155  & 0.830  & 22.02  & \underline{0.269}  & \underline{0.809}  & 18.870  & -  & -  & -  \\
\textbf{V2 (current)}  & 24.412  & \underline{0.119}  & \textbf{0.853}  & \textbf{23.390}  & \textbf{0.181}  & \textbf{0.852}  & \textbf{22.428}  & \textbf{0.056}  & \textbf{0.931}  & \underline{24.596}  \\
V3  & 26.247  & \textbf{0.110}  & \underline{0.852}  & \underline{23.306}  & 0.427  & 0.760  & \underline{20.176}  & \underline{0.058}  & \underline{0.929}  & \textbf{24.661}  \\
\bottomrule
\end{tabular}
\end{table*}

\textbf{Quantitative Results Evaluation.} Among the \replaced[id=rev]{four}{three} variants, V2 provides the most balanced performance across the three datasets. Compared with the classical baselines V1a and V1b, V2 improves LPIPS, SSIM, and PSNR, which verifies the effectiveness of the optimized backbone design, decoder attention, and post-enhancement strategy. Although V3 achieves better LPIPS and PSNR on LOL-v1 and LOL-v2-Syn datasets, the performance drops significantly on LOL-v2-Real, especially in terms of LPIPS and SSIM. Thus, V2 is selected as the final adopted version, since it achieves the best trade-off between perceptual quality, structural fidelity, and cross-dataset generalization.

\begin{figure*}[t]
    \centering
    \setlength{\tabcolsep}{1.5pt}
    \renewcommand{\arraystretch}{0.95}

    % ===== Left: LOL-v1 (2x3) =====
    \begin{minipage}[t]{0.58\textwidth}
        \centering
        {\scriptsize \textbf{LOL-v1}}\par
        \vspace{1mm}
        \begin{tabular}{ccc}
            \scriptsize Input & \scriptsize V1a & \scriptsize V1b \\
            \includegraphics[width=0.36\linewidth]{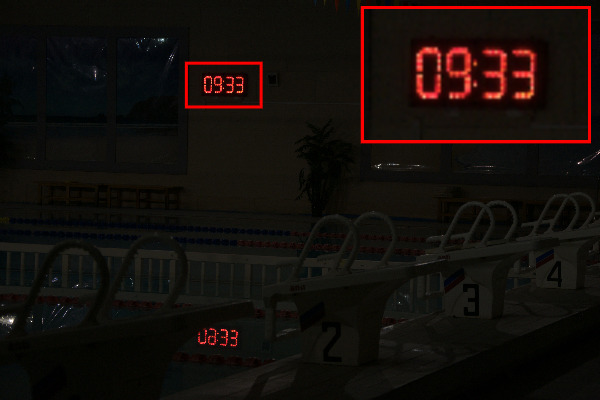} &
            \includegraphics[width=0.36\linewidth]{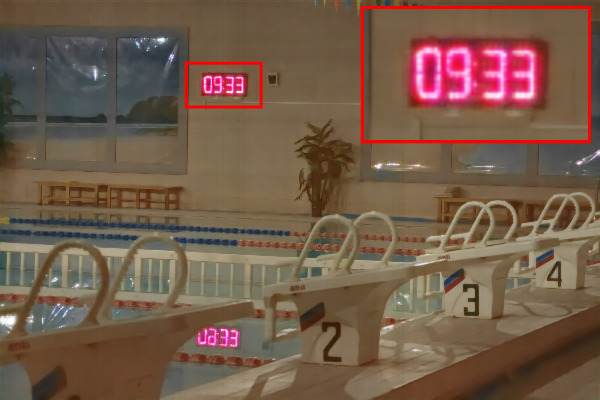} &
            \includegraphics[width=0.36\linewidth]{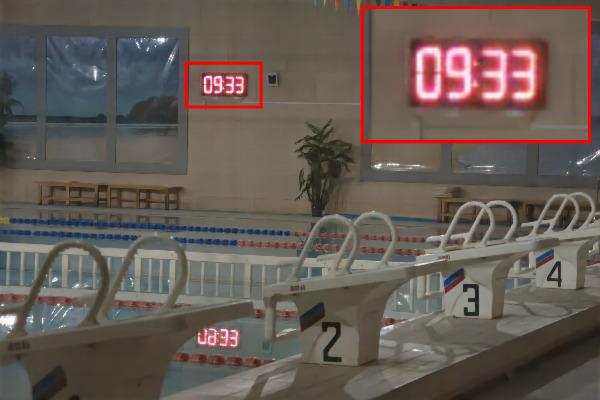} \\
            \scriptsize V2 (current) & \scriptsize V3 & \scriptsize GT \\
            \includegraphics[width=0.36\linewidth]{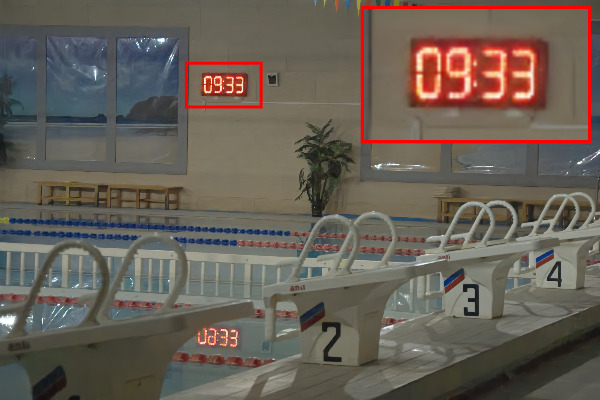} &
            \includegraphics[width=0.36\linewidth]{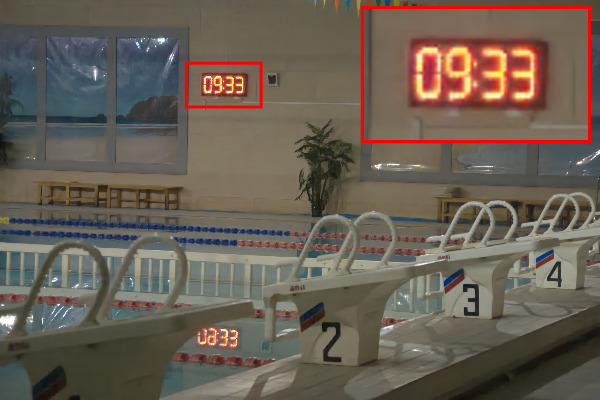} &
            \includegraphics[width=0.36\linewidth]{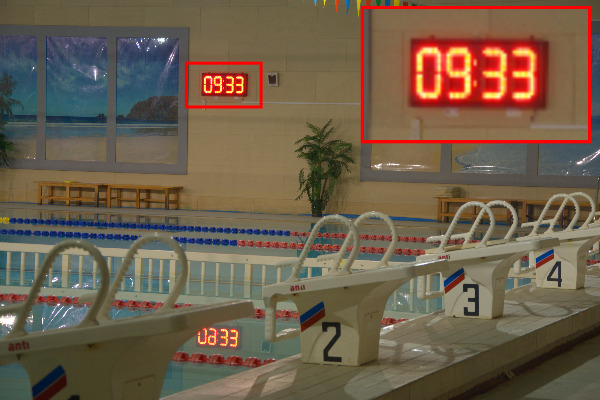} \\
        \end{tabular}
    \end{minipage}
    \hfill
    % ===== Right: LOL-v2-Syn (2x2) =====
    \begin{minipage}[t]{0.38\textwidth}
        \centering
        {\scriptsize \textbf{LOL-v2-Syn}}\par
        \vspace{1mm}
        \begin{tabular}{cc}
            \scriptsize Input & \scriptsize V2 (current) \\
            \includegraphics[width=0.36\linewidth]{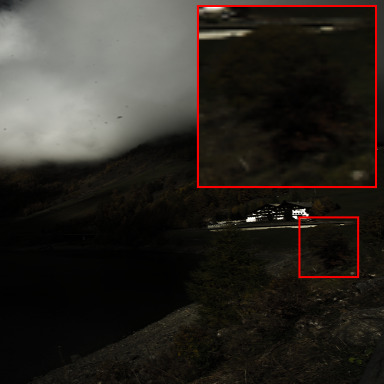} &
            \includegraphics[width=0.36\linewidth]{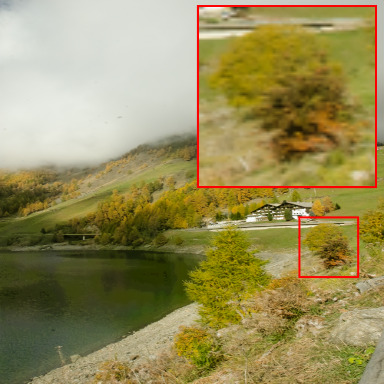} \\
            \scriptsize V3 & \scriptsize GT \\
            \includegraphics[width=0.36\linewidth]{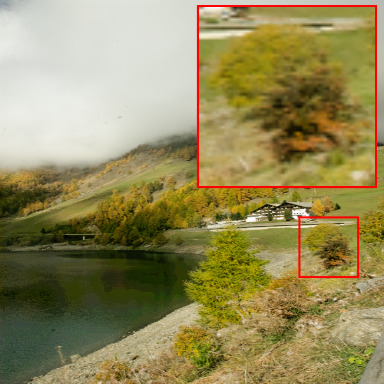} &
            \includegraphics[width=0.36\linewidth]{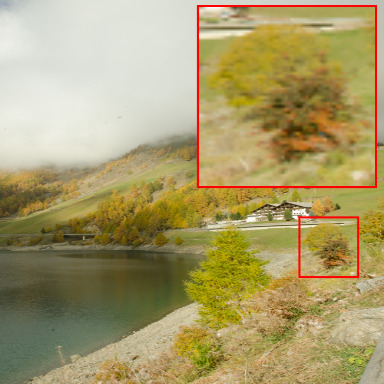} \\
        \end{tabular}
    \end{minipage}

    \caption{Qualitative comparison of different variants on representative samples from LOL-v1 and LOL-v2-Syn.}
    \label{fig:variant_qualitative_combined}
\end{figure*}

\textbf{Qualitative Results Evaluation.}
We visualize representative samples from LOL-v1 and LOL-v2-Syn to compare the perceptual behavior of different variants under challenging low-light conditions. The results of variants (V1a, V1b, V2, and V3) and the input and the corresponding ground truth of LOL-v1 are illustrated on the left of Fig. \ref{fig:variant_qualitative_combined}. The results of variants (V2 and V3) and the input and the corresponding ground truth of LOL-v2-Syn are illustrated on the right of Fig. \ref{fig:variant_qualitative_combined}. Red boxes are used to highlight regions of interest, allowing a clearer visual inspection of character sharpness, highlight diffusion, color fidelity, and structural restoration.

For the LOL-v1 results, V1a exhibits a noticeable color cast, and the bright digital characters show evident highlight spreading. V1b suppresses the highlight spreading a bit when compared with V1a, but there is still obvious color bias in the image. V2 produces the clearest character boundaries among all variants, and the saturation levels that are more in line with human visual preference. V3 is visually closer to the ground truth in enhancement; however, its perceptual appearance is not as favorable as that of V2, and both V3 and the ground truth show slight highlight spreading around the characters and other highlighted regions. 

For the LOL-v2-Syn results, both V2 and V3 significantly improve scene visibility compared with the low light input. Notably, the ground truth itself presents relatively pale color reproduction, while the enhanced results of V2 and V3 exhibit higher saturation and clearer local contrast than the paired ground truth in the displayed examples. Among them, V3 shows relatively stronger color contrast, making local regions appear more vivid and visually prominent. At the same time, V2 provides a more moderate and visually balanced enhancement. The comparison indicates that a result closer to the paired reference is not always equivalent to a more preferred perceptual quality, especially in regions affected by blur or subdued contrast.

\subsection{Non-Reference Perceptual Quality Study}

\begin{table*}[!t]
\centering
\caption{Non-reference image quality assessment results on LOL-v1, LOL-v2-Real, and LOL-v2-Syn. The best and second-best results are highlighted in \textbf{bold} and \underline{underlined}, respectively.}
\label{tab:nr_iqa_comparison}
\setlength{\tabcolsep}{2.5pt}
\renewcommand{\arraystretch}{1.15}
\begin{threeparttable}
\begin{tabular*}{\textwidth}{@{\extracolsep{\fill}}l c ccc ccc ccc@{}}
\toprule
\multirow{2}{*}{Method} &
\multirow{2}{*}{Reference} &
\multicolumn{3}{c}{LOL-v1} &
\multicolumn{3}{c}{LOL-v2-Real} &
\multicolumn{3}{c}{LOL-v2-Syn} \\
\cmidrule(lr){3-5}
\cmidrule(lr){6-8}
\cmidrule(lr){9-11}
& &
MUSIQ$\uparrow$ & LIQE$\uparrow$ & Q-Align$\uparrow$ &
MUSIQ$\uparrow$ & LIQE$\uparrow$ & Q-Align$\uparrow$ &
MUSIQ$\uparrow$ & LIQE$\uparrow$ & Q-Align$\uparrow$ \\
\midrule

KinD \cite{Zhang2019KinD}
& ACM MM'19
& 69.189 & 2.853 & 3.632
& 66.225 & 2.785 & 3.427
& 61.747 & 2.708 & 2.503 \\

MIRNet \cite{zamir2020mirnet}
& ECCV'20
& 67.276 & 3.052 & 3.581
& 63.348 & \textbf{3.103} & 3.457
& 58.519 & 2.898 & 2.551 \\

RUAS \cite{liu2021ruas}
& CVPR'21
& 59.345 & 2.902 & 3.341
& 56.177 & \underline{3.077} & 3.269
& 51.735 & 2.006 & 1.876 \\

Restormer \cite{Restormer}
& CVPR'22
& 64.236 & \textbf{3.266} & 3.390
& 51.711 & 2.897 & 2.782
& 63.545 & 3.817 & 2.857 \\

SNR-Net \cite{xu2022snr}
& CVPR'22
& 65.223 & 2.552 & 3.424
& 62.506 & 2.965 & 3.393
& 66.430 & 3.883 & 3.324 \\

Retinexformer \cite{cai2023retinexformer}
& ICCV'23
& 63.145 & 2.933 & 3.334
& 59.301 & 3.052 & 3.212
& 66.125 & \underline{3.886} & 3.181 \\

BEM \cite{Huang_Yang_Lin_Qi_Bull_Anantrasirichai_2026}
& AAAI'26
& \underline{72.972} & 3.186 & \underline{4.098}
& \underline{68.281} & 3.069 & \textbf{3.884}
& \textbf{67.513} & \textbf{4.005} & \textbf{3.469} \\

GT
& /
& 72.203 & 3.900 & 4.211
& 68.567 & 3.254 & 4.008
& 67.128 & 3.928 & 3.530 \\

\midrule
\textbf{PRIN (Ours)}
& /
& \textbf{73.745} & \underline{3.218} & \textbf{4.120}
& \textbf{69.048} & 2.970 & \underline{3.836}
& \underline{67.091} & 3.867 & \underline{3.360} \\

\bottomrule
\end{tabular*}
\end{threeparttable}
\end{table*}

To better analyze the perceptual naturalness of the enhanced results, we conduct non-reference image quality assessments using MUSIQ, LIQE, and Q-Align. In the implementation, all enhanced images are converted to RGB tensors and evaluated at their original saved resolutions without resizing. The evaluation is implemented using the PyIQA toolbox, where the \texttt{musiq}, \texttt{liqe\_mix}, and \texttt{qalign} models are employed for MUSIQ, LIQE, and Q-Align, respectively. For Q-Align, the quality-assessment task setting is adopted. 

\FloatBarrier

The reported values in Table~\ref{tab:nr_iqa_comparison} are the average scores over all test images in LOL-v1, LOL-v2-Real, and LOL-v2-Syn.

As shown in Table~\ref{tab:nr_iqa_comparison}, PRIN achieves strong no-reference perceptual performance, particularly on MUSIQ and Q-Align benchmarks. On LOL-v1, PRIN obtains the highest MUSIQ and Q-Align scores of 73.745 and 4.120, respectively, while achieving the second-best LIQE score of 3.218. On LOL-v2-Real, PRIN ranks first in MUSIQ with a score of 69.048 and second in Q-Align with a score of 3.836. For LOL-v2-Syn, BEM achieves the highest scores across all three no-reference metrics, whereas PRIN remains competitive by obtaining the second-best MUSIQ and Q-Align scores of 67.091 and 3.360, respectively. The results demonstrate that the proposed PRIN not only preserves reference-based fidelity but also effectively enhances the \replaced[id=rev]{no}{0}-reference perceptual quality in low-light scenes.

\subsection{\revadded{Cross-Dataset Evaluation under Illumination Distribution Shift}}
\added[id=rev]{%
To evaluate the generalization ability of different LLIE methods beyond their paired training domains, we conduct a zero-shot cross-dataset evaluation on five commonly used unpaired real-world benchmarks, including DICM, LIME, MEF, NPE, and VV. We download the original dataset images and re-evaluate all compared methods using their publicly released checkpoints. All methods are evaluated using exactly the same image lists, at the original image resolutions, and under the same PyIQA-based evaluation pipeline with MUSIQ, LIQE, and Q-Align. For supervised methods with multiple checkpoints, the corresponding paired training dataset is indicated in the table. The results are tabulated in Tables~\ref{tab:cross_dataset_a} and~\ref{tab:cross_dataset_b}.
}

\begin{table*}[!t]
\centering
\caption{\added[id=rev]{%
Zero-shot cross-dataset evaluation on DICM, LIME, and MEF.
Higher values indicate better performance for all reported metrics.
The best result in each column is highlighted in bold.
}}
\label{tab:cross_dataset_a}

\scriptsize
\setlength{\tabcolsep}{3.5pt}

\begin{tabular}{
l l
ccc
ccc
ccc
}
\toprule

\multirow{2}{*}{Method}
&
\multirow{2}{*}{Weight}
&
\multicolumn{3}{c}{DICM}
&
\multicolumn{3}{c}{LIME}
&
\multicolumn{3}{c}{MEF}
\\

\cmidrule(lr){3-5}
\cmidrule(lr){6-8}
\cmidrule(lr){9-11}

&
&
MUSIQ$\uparrow$ & LIQE$\uparrow$ & Q-Align$\uparrow$
&
MUSIQ$\uparrow$ & LIQE$\uparrow$ & Q-Align$\uparrow$
&
MUSIQ$\uparrow$ & LIQE$\uparrow$ & Q-Align$\uparrow$
\\

\midrule

Input
& --
& 58.019 & 3.153 & 3.734
& 58.277 & 2.808 & 3.341
& 57.536 & 2.889 & 3.083
\\

\midrule

Zero-DCE \cite{Guo2020}
& Official
& 58.857 & 3.182 & 3.517
& 60.734 & 3.063 & 3.107
& 64.925 & 3.786 & 3.550
\\

RUAS \cite{liu2021ruas}
& Official
& 53.502 & 2.058 & 2.536
& 58.019 & 2.628 & 2.670
& 61.805 & 3.180 & 3.209
\\

SCI \cite{ma2022sci}
& Official
& 53.596 & 2.536 & 2.909
& 59.113 & 2.891 & 2.961
& 62.532 & 3.527 & 3.435
\\

URetinex-Net \cite{Wu2022URetinexNet}
& LOL
& \textbf{61.146} & \textbf{3.329} & 3.679
& \textbf{64.297} & 3.269 & \textbf{3.369}
& \textbf{67.273} & 3.923 & \textbf{3.691}
\\

\midrule

\multirow{2}{*}{Retinexformer \cite{cai2023retinexformer}}
& LOL-v1
& 56.830 & 2.543 & 3.006
& 60.843 & 3.018 & 3.253
& 62.075 & 3.473 & 3.268
\\

& LOL-v2-Syn
& 58.178 & 3.139 & 3.694
& 59.450 & 2.810 & 3.186
& 62.112 & 3.390 & 3.400
\\

\addlinespace[1pt]

\multirow{2}{*}{HVI-CIDNet \cite{Yan2025HVIAN}}
& LOL-v1
& 59.796 & 2.957 & 3.399
& 62.376 & \textbf{3.308} & 3.249
& 66.739 & \textbf{3.980} & 3.633
\\

& LOL-v2-Syn
& 58.932 & 3.244 & \textbf{3.797}
& 59.347 & 2.988 & 3.245
& 62.779 & 3.522 & 3.457
\\

\addlinespace[1pt]

\midrule

\multirow{3}{*}{PRIN (Ours)}
& LOL-v1
& 52.911 & 2.222 & 3.175
& 57.247 & 2.490 & 3.144
& 53.421 & 2.359 & 2.910
\\

& LOL-v2-Syn
& 55.169 & 2.553 & 3.278
& 55.666 & 2.235 & 2.790
& 60.497 & 3.044 & 3.100
\\

\bottomrule
\end{tabular}
\end{table*}

\begin{table*}[!t]
\centering
\caption{\added[id=rev]{%
Zero-shot cross-dataset evaluation on NPE and VV.
Higher values indicate better performance for all reported metrics.
The best result in each column is highlighted in bold.
}}
\label{tab:cross_dataset_b}

\scriptsize
\setlength{\tabcolsep}{6.0pt}

\begin{tabular}{
l l
ccc
ccc
}
\toprule

\multirow{2}{*}{Method}
&
\multirow{2}{*}{Weight}
&
\multicolumn{3}{c}{NPE}
&
\multicolumn{3}{c}{VV}
\\

\cmidrule(lr){3-5}
\cmidrule(lr){6-8}

&
&
MUSIQ$\uparrow$ & LIQE$\uparrow$ & Q-Align$\uparrow$
&
MUSIQ$\uparrow$ & LIQE$\uparrow$ & Q-Align$\uparrow$
\\

\midrule

Input
& --
& 65.131 & 3.853 & \textbf{3.735}
& \textbf{43.600} & 1.468 & 3.485
\\

\midrule

Zero-DCE \cite{Guo2020}
& Official
& \textbf{68.964} & 3.890 & 3.298
& 39.875 & 1.296 & 3.564
\\

RUAS \cite{liu2021ruas}
& Official
& 61.430 & 2.411 & 2.145
& 35.172 & 1.051 & 2.692
\\

SCI \cite{ma2022sci}
& Official
& 62.290 & 3.101 & 2.556
& 32.998 & 1.141 & 3.006
\\

URetinex-Net \cite{Wu2022URetinexNet}
& LOL
& 68.114 & 3.839 & 3.527
& 40.257 & \textbf{1.521} & 3.073
\\

\midrule

\multirow{2}{*}{Retinexformer \cite{cai2023retinexformer}}
& LOL-v1
& 63.339 & 2.946 & 2.790
& 34.876 & 1.157 & 2.666
\\

& LOL-v2-Syn
& 66.069 & 3.734 & 3.461
& 41.230 & 1.373 & 3.392
\\

\addlinespace[1pt]

\multirow{2}{*}{HVI-CIDNet \cite{Yan2025HVIAN}}
& LOL-v1
& 65.925 & 3.444 & 2.947
& 33.702 & 1.180 & 2.809
\\

& LOL-v2-Syn
& 67.003 & \textbf{3.976} & 3.539
& 41.499 & 1.379 & \textbf{3.644}
\\

\addlinespace[1pt]

\midrule

\multirow{3}{*}{PRIN (Ours)}
& LOL-v1
& 58.896 & 2.593 & 3.280
& 35.190 & 1.247 & 3.174
\\

& LOL-v2-Syn
& 61.947 & 2.884 & 3.283
& 37.663 & 1.217 & 3.331
\\

\bottomrule
\end{tabular}
\end{table*}

\added[id=rev]{%
As shown in Tables~\ref{tab:cross_dataset_a} and \ref{tab:cross_dataset_b}, the relative performance of the evaluated methods varies considerably across the five unseen datasets. URetinex-Net performs particularly well on DICM, LIME, and MEF. HVI-CIDNet trained on LOL-v2-Syn achieves strong results on NPE and VV. PRIN also exhibits clear dataset-dependent behavior. For example, with the LOL-v1 checkpoint, its Q-Align scores on NPE (3.280) and VV (3.174) are higher than those of Retinexformer (2.790 and 2.666) and HVI-CIDNet (2.947 and 2.809) trained on the same dataset, while its performance on LIME and MEF is less competitive.
}

\added[id=rev]{%
To investigate whether this variation is specific to PRIN, we further compare the checkpoints trained from LOL-v1 and LOL-v2-Syn. The results illustrates that on NPE and VV, replacing LOL-v1 with LOL-v2-Syn generally improves the results of all three \replaced[id=rev]{supervised}{surprised} methods (Retinexformer, HVI-CIDNet, and PRIN). In contrast, after changing the training dataset, the values of MUSIQ, LIQE, and Q-Align declines on the LIME test dataset. Results on DICM and MEF test dataset show more mixed metric-dependent variations. Since the architecture remains unchanged within each checkpoint pair, these results indicate that the paired training distribution might play an important role in cross-dataset transfer, and that its effect depends on the target domain.
}

\added[id=rev]{%
We found that the behavior may be consistent with the illumination statistics of the training and test data. The low-light inputs of LOL-v1 have an average intensity of only approximately 0.05, while the five unpaired benchmarks cover much more diverse illumination conditions, with dataset-level means of 0.14-0.34 and image-wise means spanning approximately 0.06-0.53. Despite this diversity, the outputs of paired models tend to concentrate within a considerably narrower mean-intensity range of approximately 0.37-0.43. These observations suggest that paired LLIE models partially inherit the exposure distribution of their training targets, making the learned enhancement strength sensitive to a mismatch between the training and test illumination conditions.
}

\begin{figure*}[t]
    \centering
    \setlength{\tabcolsep}{1pt}
    \renewcommand{\arraystretch}{0.96}

    \begin{tabular}{cccccccc}

        %==================== Row 1: P1010676 ====================
        \includegraphics[width=0.121\textwidth]{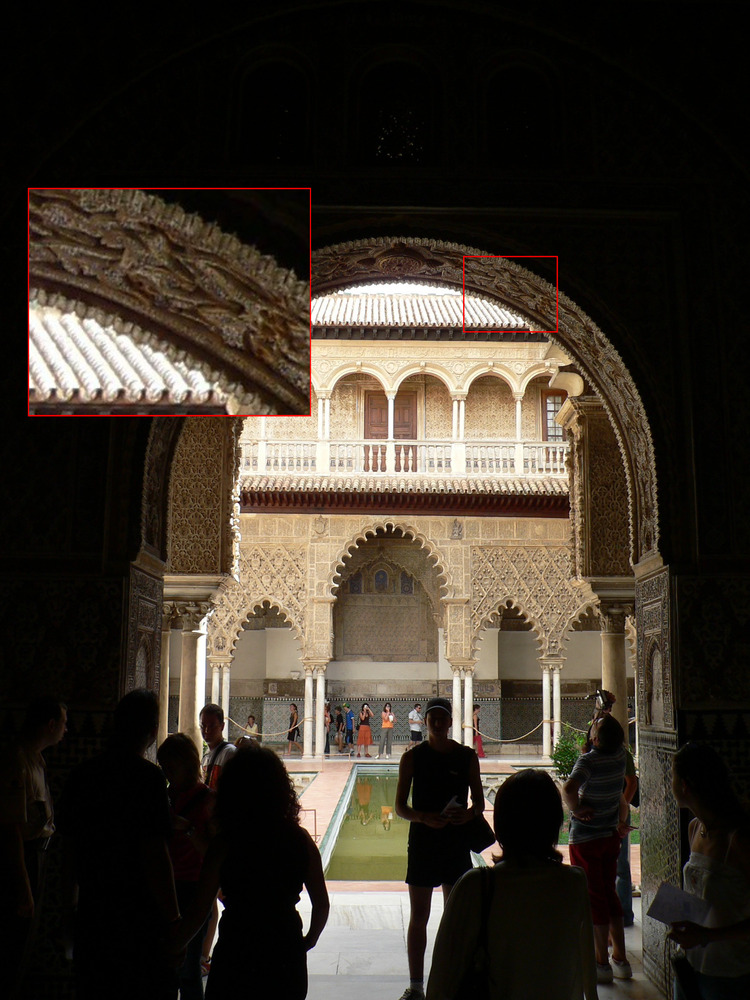} &
        \includegraphics[width=0.121\textwidth]{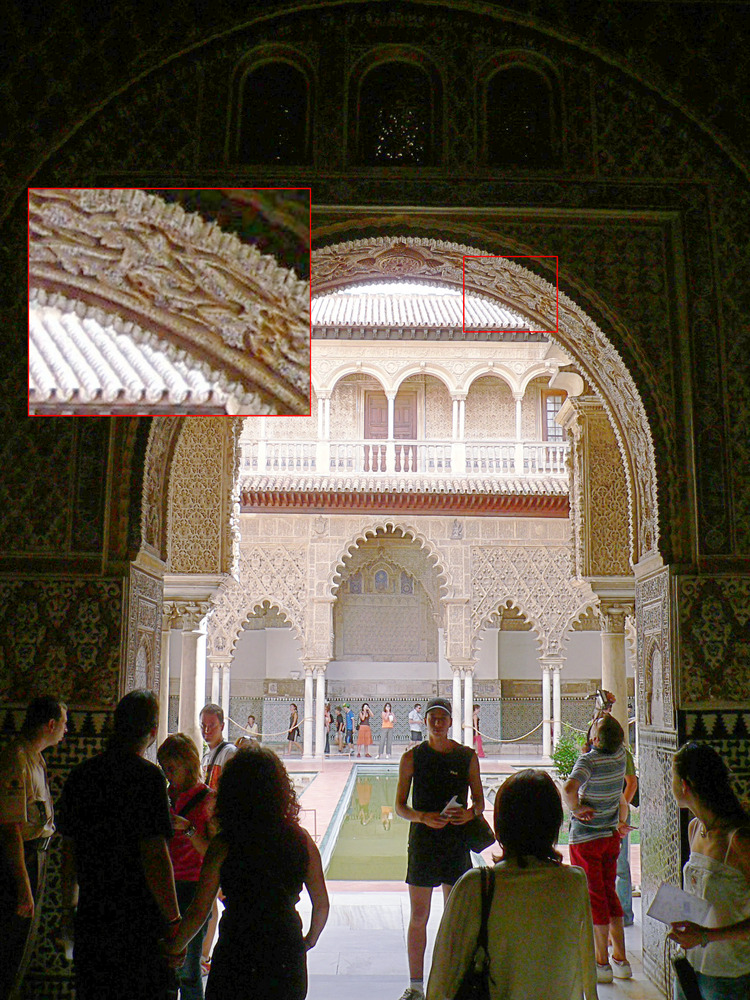} &
        \includegraphics[width=0.121\textwidth]{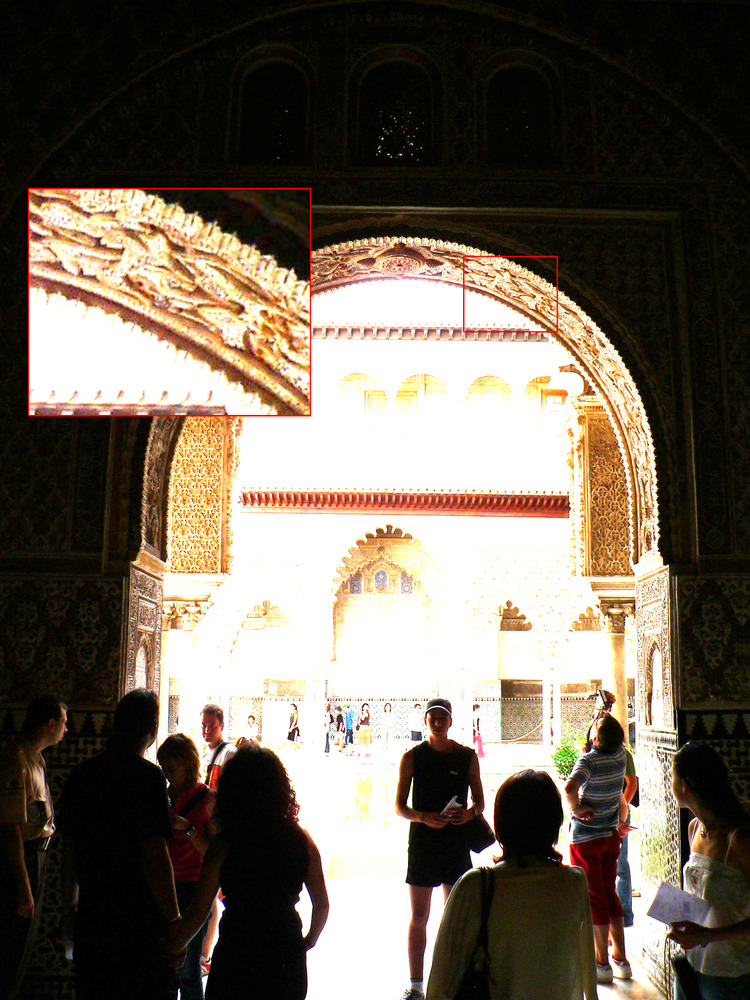} &
        \includegraphics[width=0.121\textwidth]{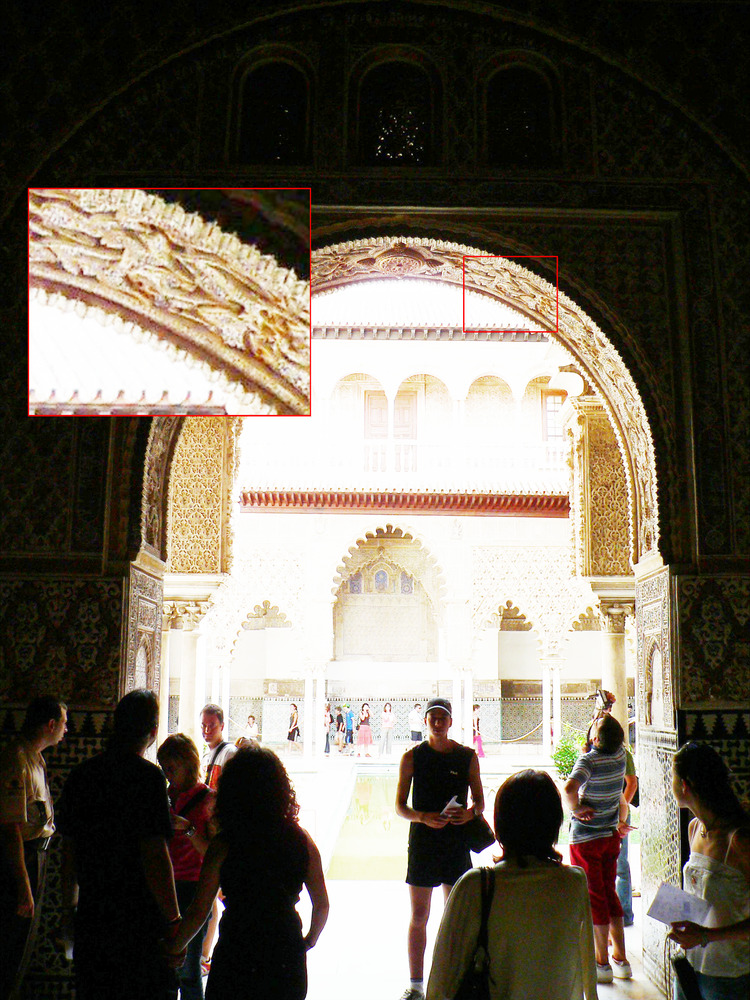} &
        \includegraphics[width=0.121\textwidth]{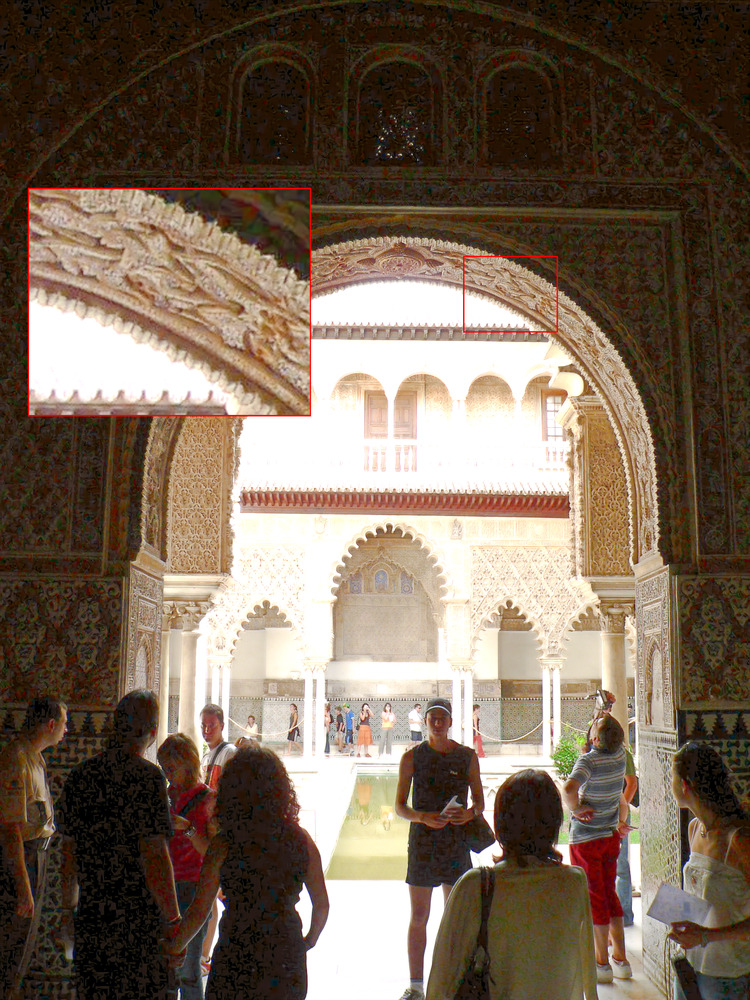} &
        \includegraphics[width=0.121\textwidth]{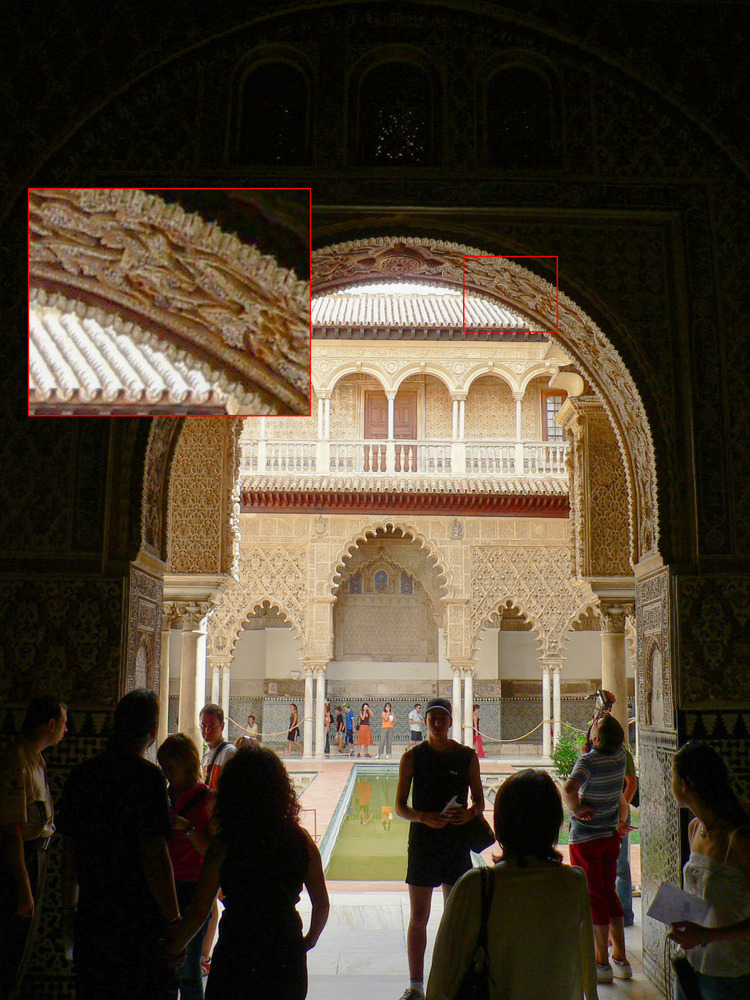} &
        \includegraphics[width=0.121\textwidth]{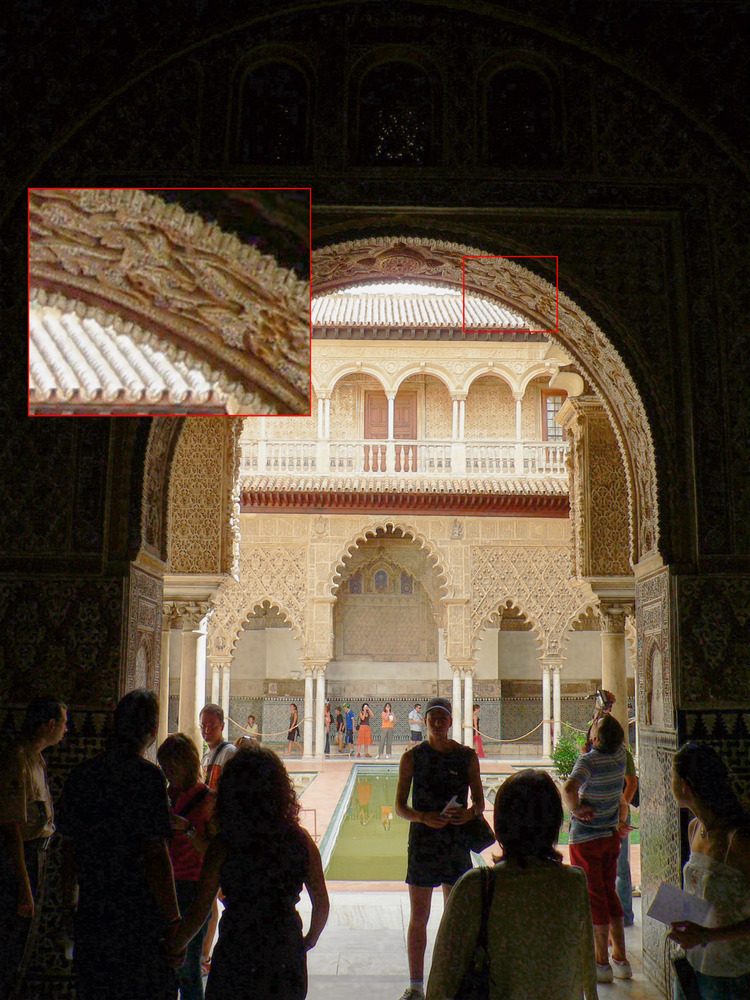} &
        \includegraphics[width=0.121\textwidth]{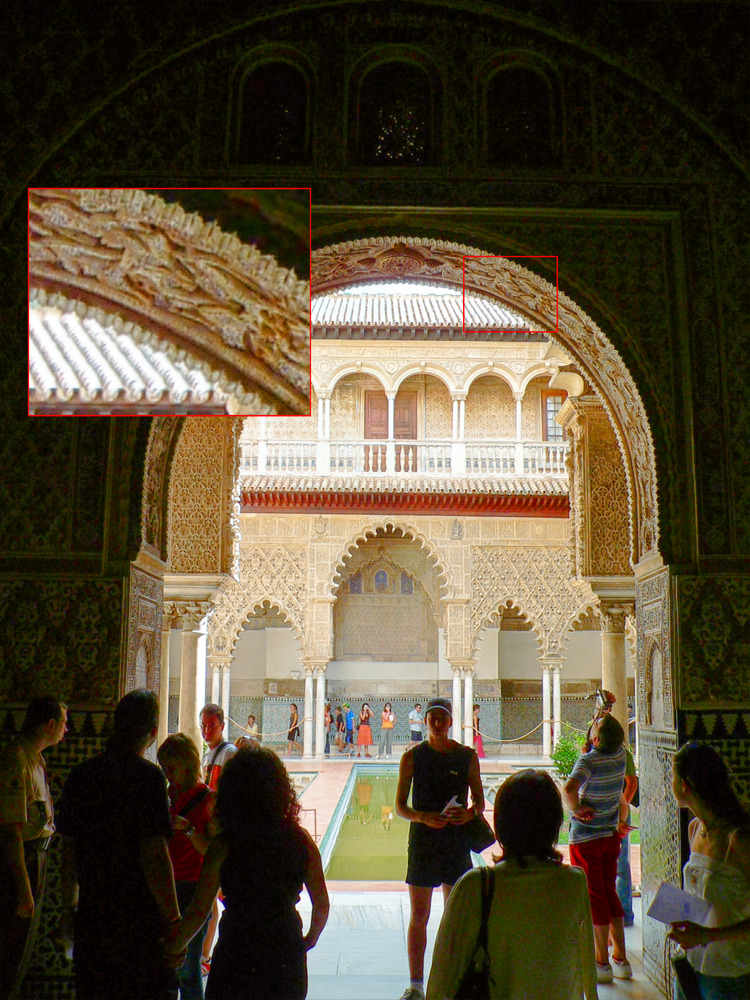}
        \\[1pt]

        %==================== Row 2: P1090815 ====================
        \includegraphics[width=0.121\textwidth]{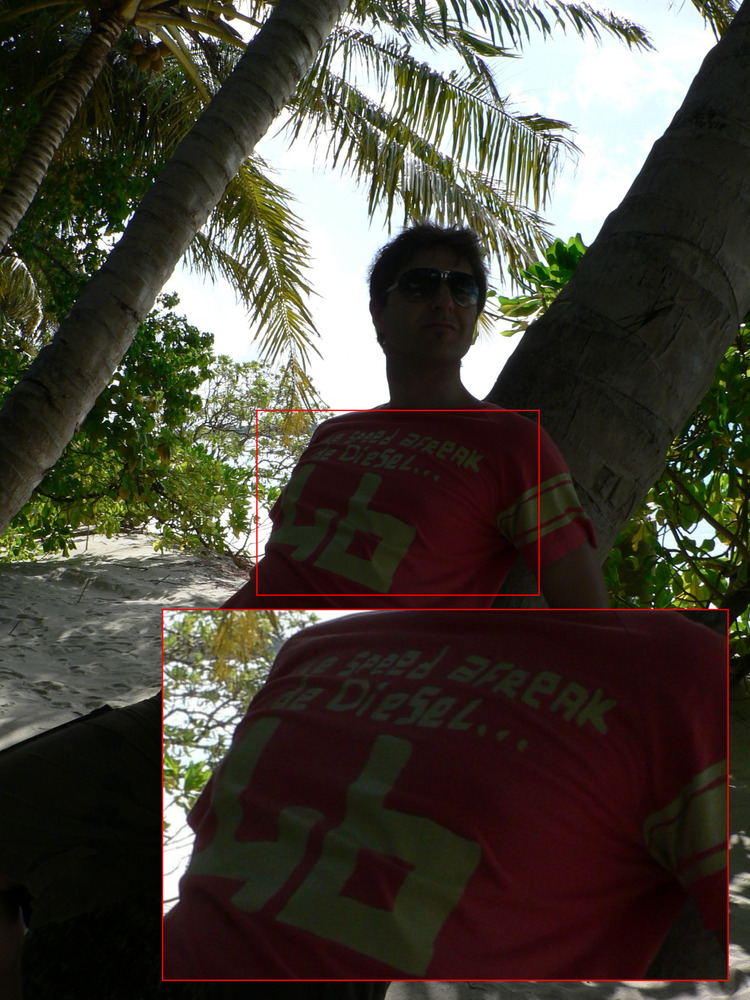} &
        \includegraphics[width=0.121\textwidth]{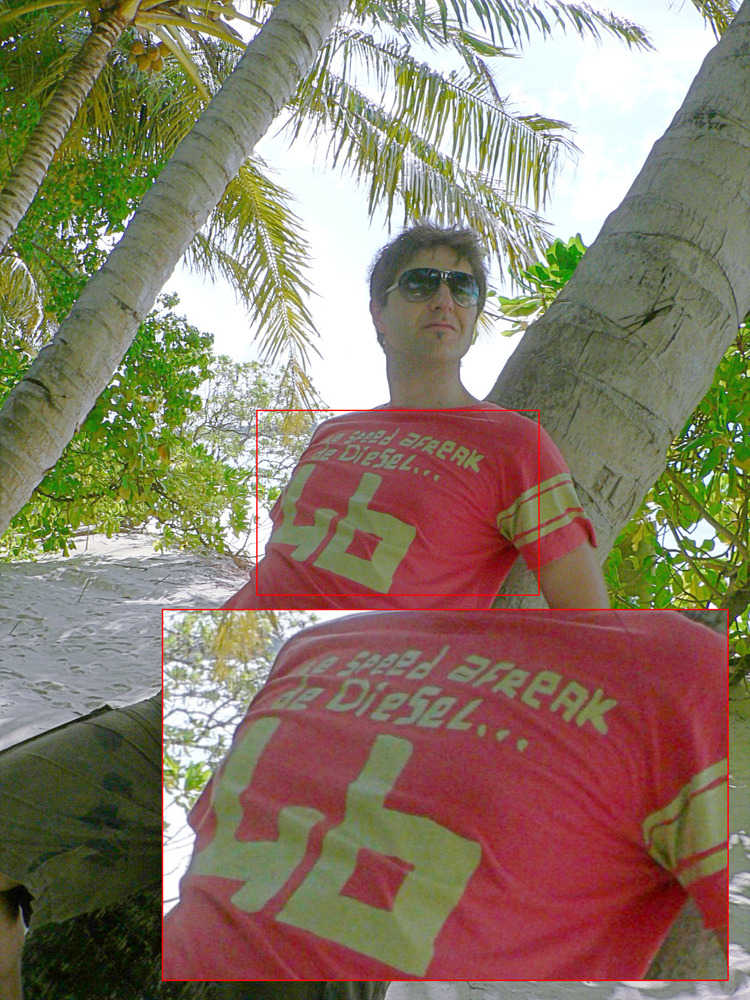} &
        \includegraphics[width=0.121\textwidth]{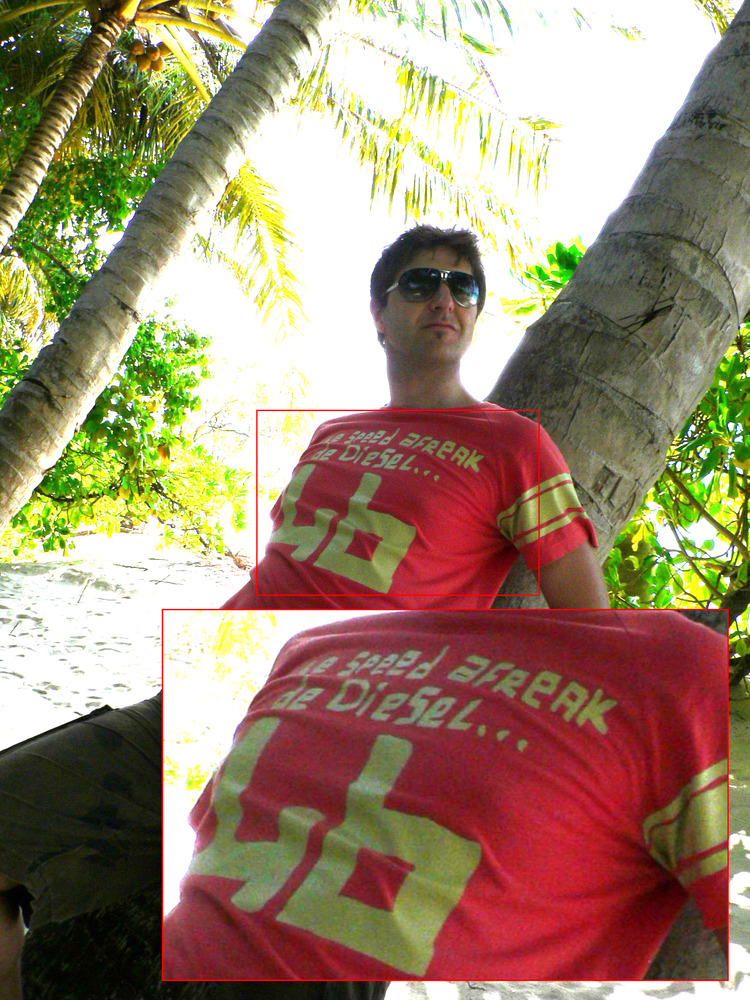} &
        \includegraphics[width=0.121\textwidth]{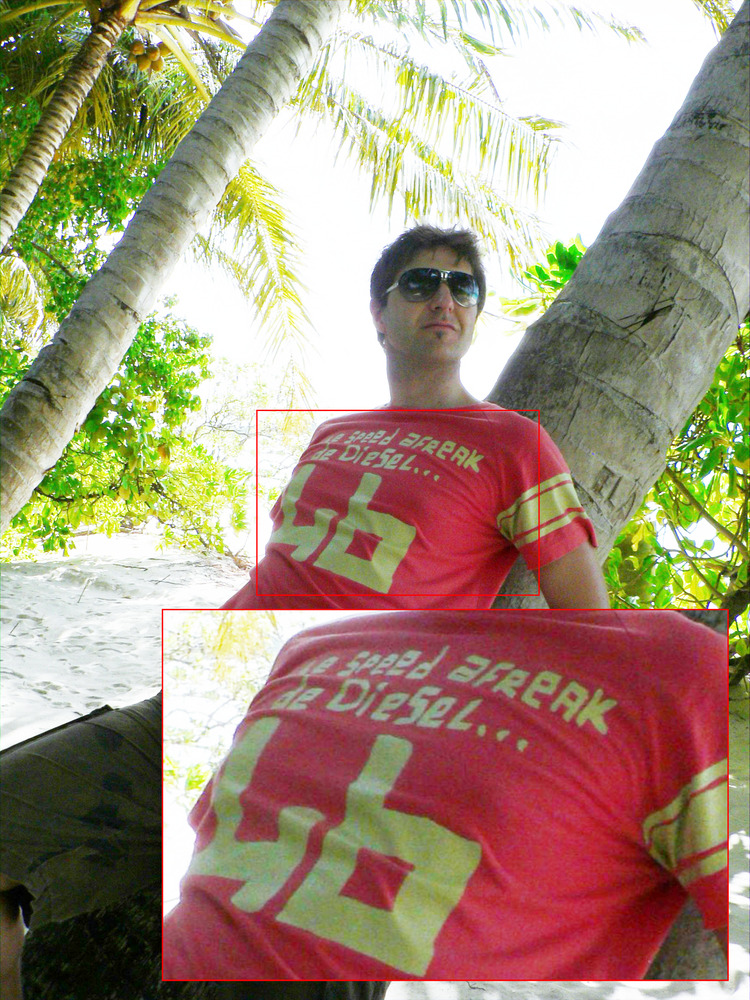} &
        \includegraphics[width=0.121\textwidth]{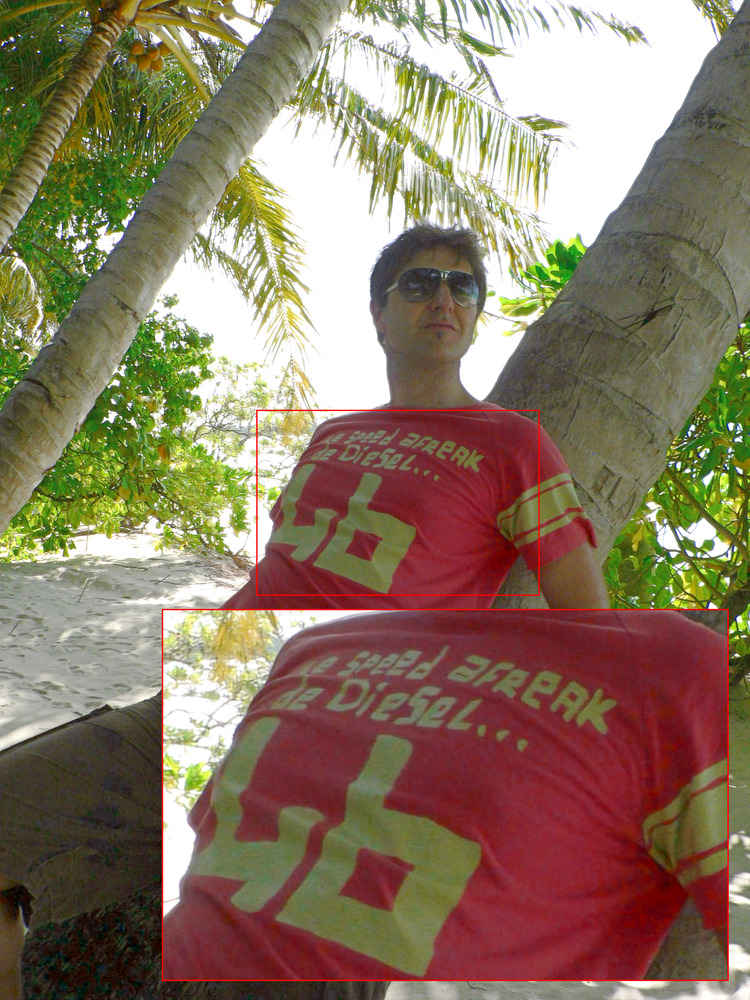} &
        \includegraphics[width=0.121\textwidth]{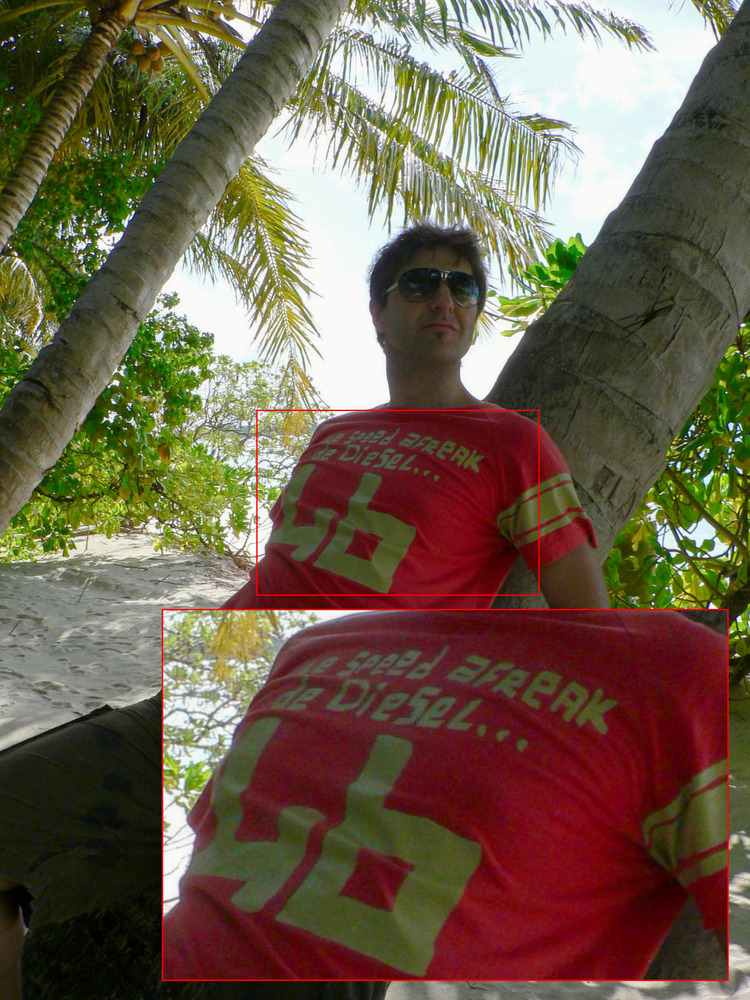} &
        \includegraphics[width=0.121\textwidth]{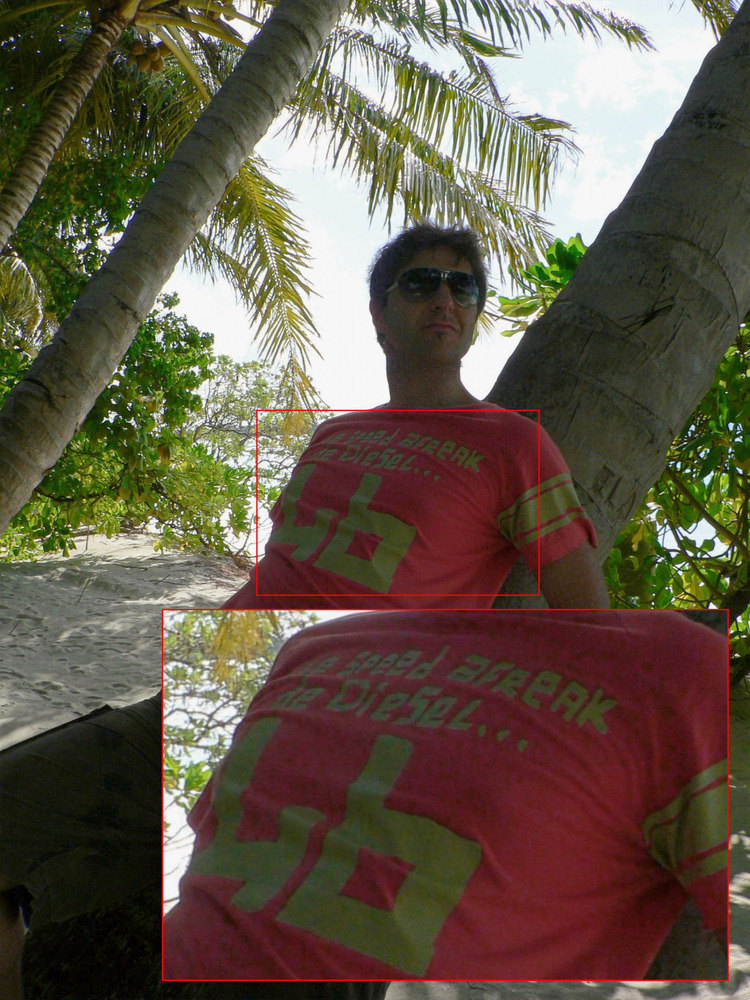} &
        \includegraphics[width=0.121\textwidth]{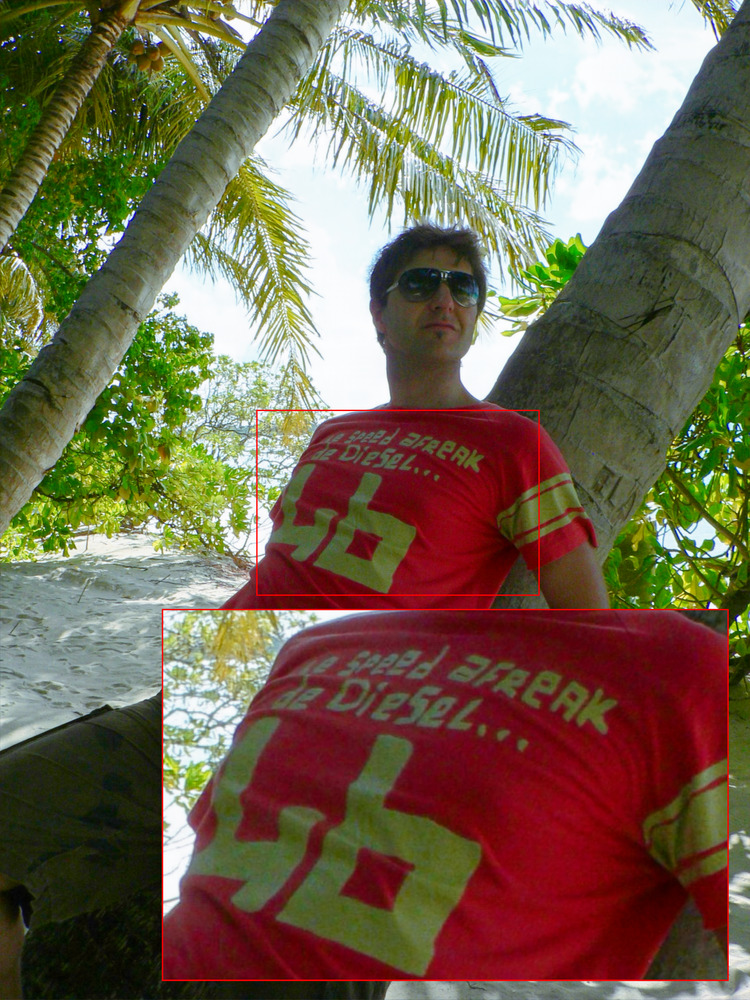}
        \\[1pt]

        %==================== Row 3: road ====================
        \includegraphics[width=0.121\textwidth]{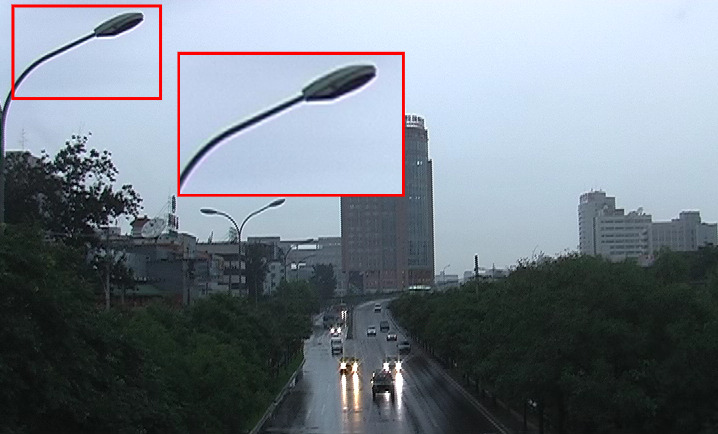} &
        \includegraphics[width=0.121\textwidth]{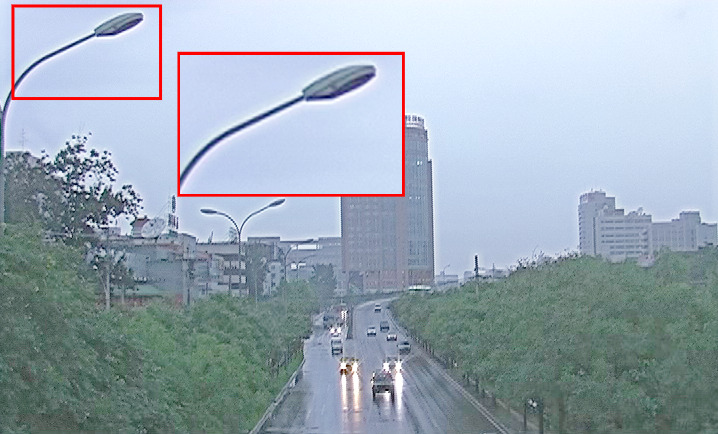} &
        \includegraphics[width=0.121\textwidth]{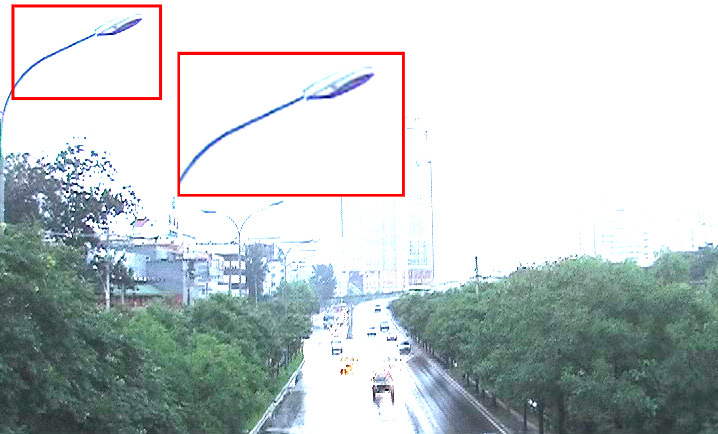} &
        \includegraphics[width=0.121\textwidth]{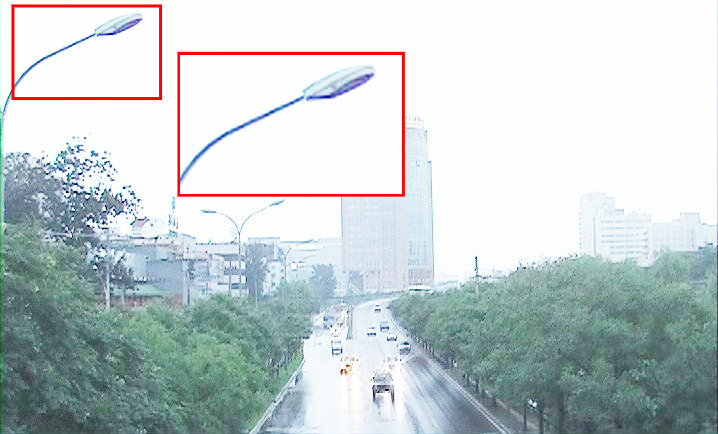} &
        \includegraphics[width=0.121\textwidth]{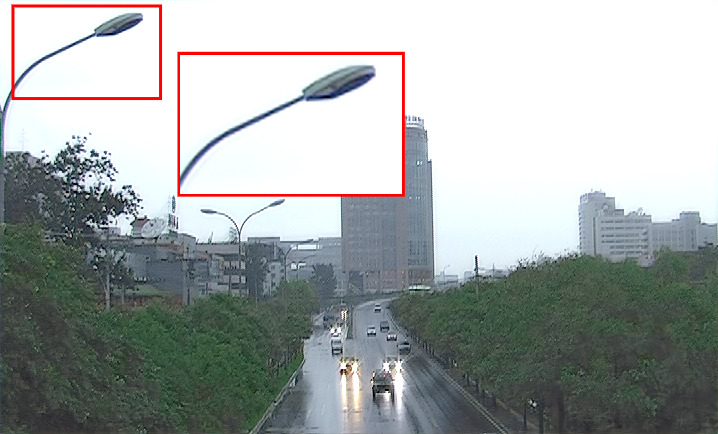} &
        \includegraphics[width=0.121\textwidth]{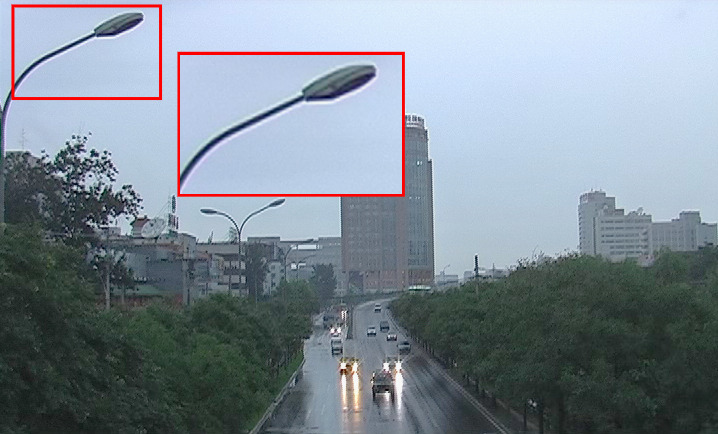} &
        \includegraphics[width=0.121\textwidth]{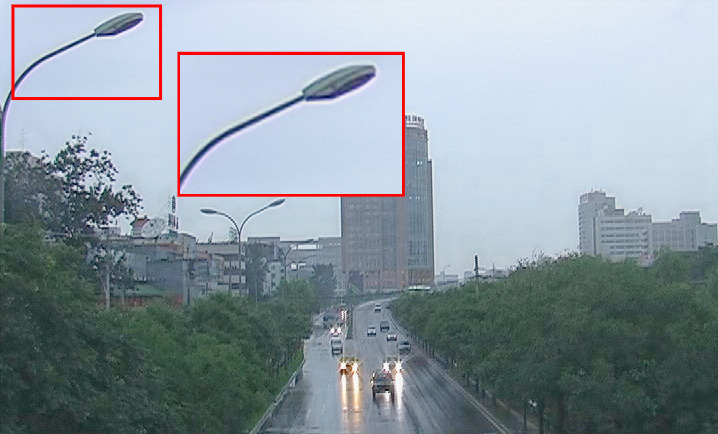} &
        \includegraphics[width=0.121\textwidth]{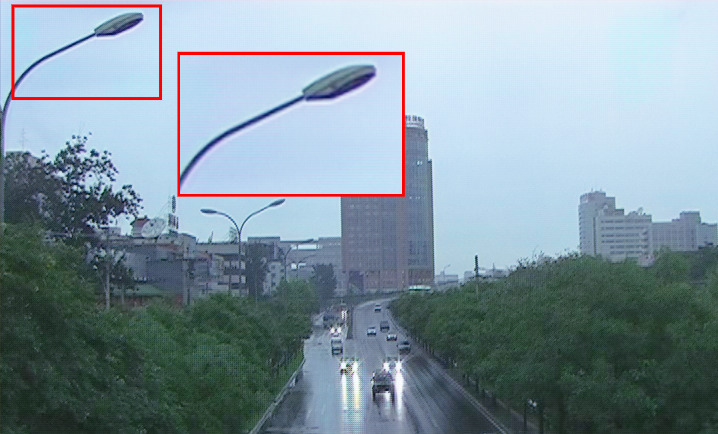}
        \\[2pt]

        %==================== Method labels ====================
        \scriptsize\textbf{Input} &
        \scriptsize\textbf{Zero-DCE} &
        \scriptsize\textbf{RUAS} &
        \scriptsize\textbf{SCI} &
        \scriptsize\textbf{URetinex-Net} &
        \scriptsize\textbf{HVI-CIDNet} &
        \scriptsize\textbf{Retinexformer} &
        \scriptsize\textbf{PRIN (Ours)}
        \\

    \end{tabular}

    \caption{\added[id=rev]{%
    Qualitative comparison on representative images from the unpaired real-world benchmarks. The three rows show scenes with challenging illumination transitions and large differences between shadow and highlight regions.
    }}
    \label{fig:cross_dataset_visual}
\end{figure*}

\added[id=rev]{%
Although PRIN does not achieve the best no-reference scores in the evaluation, the qualitative comparison provides complementary evidence about the type of degradation produced under illumination shift. As shown in Fig.~\ref{fig:cross_dataset_visual}, in the first row, PRIN effectively brightens the dark region above the arch while maintaining a reasonable brightness level in the already well-lit courtyard. It also preserves the fine structural details in the transition region between the extremely dark interior and the bright exterior, as highlighted in the zoomed box. By contrast, Zero-DCE introduces noticeable noise, RUAS, SCI, and URetinex-Net overexpose the bright region, and RUAS, HVI-CIDNet, and Retinexformer fail to provide sufficient enhancement for the dark region above the arch. %
}

\added[id=rev]{%
In the second row, PRIN better preserves the cloud details in the bright sky, while maintaining more natural clothing color and texture, a more realistic trunk color, and smoother color transitions around the human body. Zero-DCE exhibits an overall whitish tone together with excessive sharpening, and the tree trunk is also biased toward white. RUAS, SCI, and URetinex-Net again fail to suppress the overly bright sky and produce overexposed results. Retinexformer yields a relatively dark global tone and noticeable noise in the large red clothing region, whereas HVI-CIDNet produces results that are visually closer to PRIN. %
}

\added[id=rev]{%
In the third row, PRIN effectively suppresses the sky brightness after enhancement, while preserving clear and vivid colors in the foliage. RUAS, SCI, and URetinex-Net are again unable to control the sky brightness and lead to overexposure. HVI-CIDNet shows a relatively whitish tone and more visible noise in the foliage region, while Retinexformer and Zero-DCE also exhibit a whitish overall appearance. Overall, although PRIN does not achieve the highest no-reference scores, it provides a more balanced shadow-highlight trade-off and preserves more visually plausible local details in the selected cross-dataset examples. %
}

\subsection[Module Ablation Study]{\revadded{Module} Ablation Study}
In this section, we conduct ablation experiments on the LOL-v2-Real dataset to verify the effectiveness of the proposed components. We evaluate six configurations, including the baseline model, the baseline with WDGRR only, the baseline with WSAIR only, the baseline with WDGRR and WSAIR, \added[id=rev]{the baseline with the curve-based post-composition enhancement head (DCE) only} and the full model with all proposed components. The quantitative results evaluation on the LOL-v2-Real dataset is tabulated in Table \ref{tab:ablation_main}.

\begin{table}[t]
\centering
\caption{Ablation study of the proposed modules on LOL-v2-Real. $\checkmark$ indicates that the corresponding component is enabled.}
\label{tab:ablation_main}
\small
\setlength{\tabcolsep}{4pt}
\renewcommand{\arraystretch}{1.0}
\begin{tabular}{cccc|ccc}
\toprule
Baseline & WDGRR & WSAIR & DCE & LPIPS$\downarrow$ & SSIM$\uparrow$ & PSNR$\uparrow$ \\
\midrule
\checkmark &  &  &  & 0.269 & 0.809 & 18.870 \\
\checkmark & \checkmark &  &  & \textbf{0.178} & \underline{0.845} & \underline{21.398} \\
\checkmark &  & \checkmark &  & 0.211 & 0.824 & 20.257 \\
\checkmark &  &  & \checkmark & 0.201 & 0.832 & 20.841 \\
\checkmark & \checkmark & \checkmark &  & 0.197 & 0.834 & 20.840 \\
\checkmark & \checkmark & \checkmark & \checkmark & \underline{0.181} & \textbf{0.852} & \textbf{22.428} \\
\bottomrule
\end{tabular}
\end{table}

The results show that each individual module improves the baseline when evaluated independently. For example, the WDGRR module reduces LPIPS from 0.269 to 0.178 while increasing SSIM and PSNR from 0.809 and 18.870 dB to 0.845 and 21.398 dB, respectively. The WSAIR module reduces LPIPS to 0.211 while increasing SSIM and PSNR to 0.824 and 20.257 dB. Moreover, the DCE reduces LPIPS to 0.201 while increasing SSIM and PSNR to 0.832 and 20.841 dB.

It is also observed that directly combining WDGRR and WSAIR without DCE does not further improve the performance over the single-module variant, obtaining 0.197 LPIPS, 0.834 SSIM, and 20.840 dB PSNR. However, after integrating the DCE module, the model achieves 0.181 LPIPS, 0.852 SSIM, and 22.428 dB PSNR, producing the best SSIM and PSNR while maintaining competitive perceptual quality. The results suggest that the DCE can be an effective post-composition calibration module that can help harmonize the interaction of WDGRR and WSAIR and resulting in a better trade-off between perceptual quality, structural preservation, and pixel-level fidelity.

\subsection{\revadded{Loss-Group Ablation}}
\label{sec:loss_ablation}

\added[id=rev]{%
To examine the contribution of the applied loss components, we conduct a loss-group ablation study on the LOL-v1 test set. The primary reconstruction loss $\mathcal{L}_{\mathrm{rec}}$ is retained in all
groups as it provides the fundamental paired supervision for the final enhanced image. The remaining loss terms are organized into four groups based on the functions: $\mathcal{G}_1=\{\mathcal{L}_{\mathrm{ssim}},\mathcal{L}_{\mathrm{edge}}\}$ for structural and edge preservation, $\mathcal{G}_2=\{\mathcal{L}_{\mathrm{wav}}\}$ for frequency-domain fidelity, $\mathcal{G}_3=\{\mathcal{L}_{\mathrm{base}},\mathcal{L}_{\mathrm{illum}}\}$ for decomposition-oriented regularization, and $\mathcal{G}_4=\{\mathcal{L}_{\mathrm{col}}, \mathcal{L}_{\mathrm{oe}}\}$ for color and exposure control.
}

\added[id=rev]{%
It should be noted that the purpose of this experiment is to provide preliminary assessment for loss selection, rather than to establish the fully converged asymptotic performance of each variant. Due to the constraints on computational resources and experimental time, each loss-ablation variant was trained for up to 1,000 epochs. This budget was selected based on our previous training experiments, in which the models typically entered a slow convergence zone at this stage. Although the training objectives could continue to decrease with additional optimization, the subsequent improvements were generally gradual. Thus, we consider 1,000 epochs are sufficient for preliminary assessment of the relative contribution and optimization behavior of the different loss groups. For each variant, the checkpoint with the lowest loss within its 1,000-epoch training is selected. All reported variants were evaluated under the same maximum training budget and otherwise identical architectural, optimization, and data-processing settings.
}

\added[id=rev]{%
As shown in Table~\ref{tab:loss_group_ablation}, removing the structure-edge group $\mathcal{G}_1$ produces obvious degradation. PSNR decreases from 23.36 to 22.68 dB, SSIM decreases from 0.8439 to 0.8065, and LPIPS increases from 0.1293 to 0.1865, although mean absolute error (MAE) is reduced from 0.0712 to 0.0673. The results suggest that without $\mathcal{G}_1$, the MAE is lower, but the structural and perceptual fidelity becomes worse. This indicates that perceptual preservation rather than direct pixel-wise error minimization and reconstruction supervision alone does not fully capture the edge and structure related properties targeted by this group and the $\mathcal{G}_1$ primarily contributes to structural.
}

\added[id=rev]{%
Moreover, after removing the frequency-domain group $\mathcal{G}_2$, the resulting variant obtains a approximately the same PSNR (23.37 versus 23.36 dB) and slightly better SSIM and MAE, while LPIPS increases from 0.1293 to 0.1326.This mixed behavior suggests that the wavelet-domain loss does not improve all output metrics, but instead introduces an additional frequency-aware constraint that slightly favors perceptual similarity. The observed differences are relatively small and should therefore be interpreted as a preliminary indication of its regularizing role rather than evidence of a large independent performance gain.
}

\added[id=rev]{%
Removing the decomposition-oriented group $\mathcal{G}_3$ leads to slight changes that PSNR decreases from
23.36 to 23.23 dB and LPIPS increases from 0.1293 to 0.1315, while the differences in SSIM and MAE are little. However, as shown in Fig.~\ref{fig:g3_component_ablation}, the intermediate reflectance and illumination estimates become less consistent with their Retinex roles that texture and structural information leaks into the illumination component, while smooth exposure variations are partially absorbed by the reflectance component.
}

\par
\medskip
\noindent
\makebox[\columnwidth][c]{%
\begin{minipage}{1.0\columnwidth}
    \centering

    \begin{tabular}{
        @{}
        >{\centering\arraybackslash}p{0.48\linewidth}
        @{\hspace{0.02\linewidth}}
        >{\centering\arraybackslash}p{0.48\linewidth}
        @{}
    }
        \scriptsize Full loss
        &
        \scriptsize w/o regularization
        \\[0.5mm]

        \includegraphics[
            width=\linewidth,
            keepaspectratio
        ]{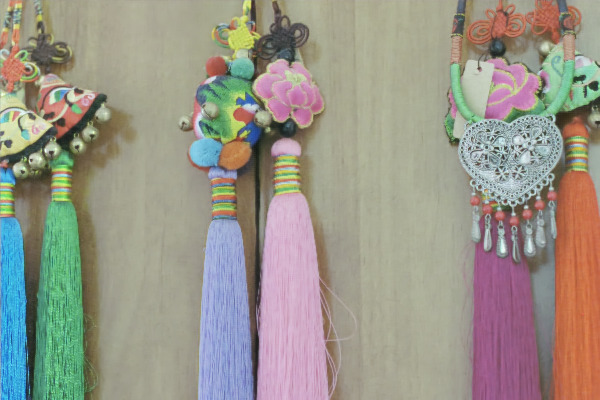}
        &
        \includegraphics[
            width=\linewidth,
            keepaspectratio
        ]{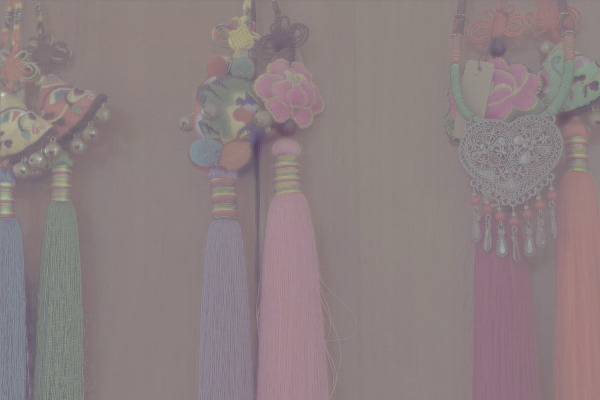}
        \\[-0.5mm]

        \scriptsize Reflectance
        &
        \scriptsize Reflectance
        \\[1.5mm]

        \includegraphics[
            width=\linewidth,
            keepaspectratio
        ]{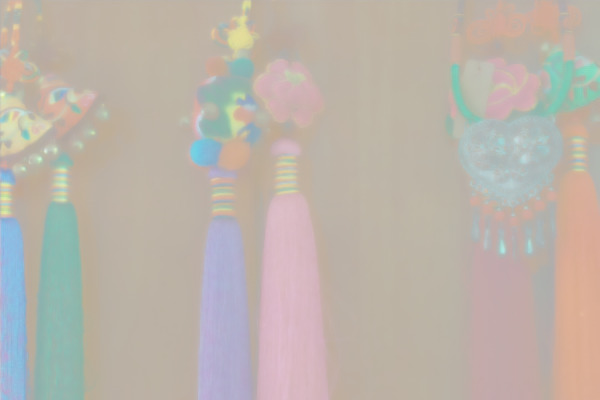}
        &
        \includegraphics[
            width=\linewidth,
            keepaspectratio
        ]{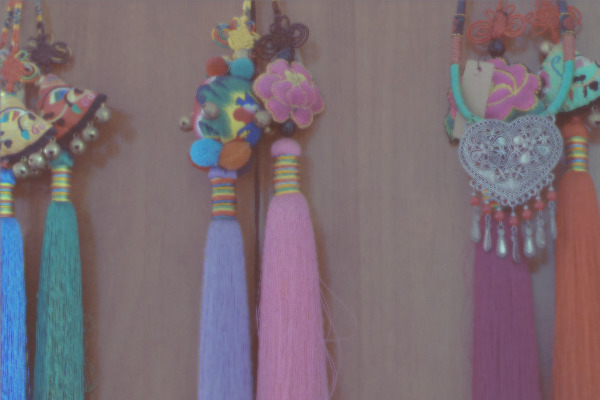}
        \\[-0.5mm]

        \scriptsize Illumination
        &
        \scriptsize Illumination
    \end{tabular}

    \captionof{figure}{\added[id=rev]{%
    Comparison of the intermediate Retinex decomposition obtained using the full loss objective and after removing the decomposition regularization losses.%
    }}
    \label{fig:g3_component_ablation}
\end{minipage}%
}
\par
\medskip

\added[id=rev]{%
A possible explanation is that removing the intermediate reconstruction and illumination regularization terms simplifies the multi-loss optimization, allowing the remaining losses to reach a favorable solution more rapidly within the limited 1,000-epoch training budget. Since the models had not fully converged, the current results do not demonstrate that removing $\mathcal{G}_3$ would bring better performance under a sufficiently long training schedule.
}

\added[id=rev]{%
Finally, removing the color-exposure group $\mathcal{G}_4$ decreases PSNR from 23.36 to 23.26 dB, while slightly improving SSIM, MAE, and LPIPS. Thus, the current quantitative results do not indicate a uniform advantage from including $\mathcal{G}_4$. One possible reason is that
the LOL-v1 dataset provides limited coverage of challenging color and exposure related variations, so the effects targeted by $\mathcal{G}_4$ may not be fully reflected by these general-purpose image-quality metrics. However, the color-consistency and over-exposure terms impose explicit constraints on chromatic and intensity behavior that are not directly measured by metrics. The group loss items are retained as a task-oriented regularizer intended to improve robustness to variations in color and exposure conditions.
}

\begin{table*}[!t]
    \centering
    \caption{\revadded{%
    Loss-group ablation results on the LOL-v1 test set.
    $\mathcal{L}_{\mathrm{rec}}$ is retained in all configurations.}}
    \label{tab:loss_group_ablation}

    \setlength{\tabcolsep}{4.5pt}
    \renewcommand{\arraystretch}{1.0}

    \normalsize
    \begin{tabular}{lcccc}
        \toprule
        Variant
        & PSNR$\uparrow$
        & SSIM$\uparrow$
        & MAE$\downarrow$
        & LPIPS$\downarrow$ \\
        \midrule

        Full objective
        & 23.36 & 0.8439 & 0.0712 & 0.1293 \\

        w/o $\mathcal{G}_1$ (structure-edge)
        & 22.68 & 0.8065 & 0.0673 & 0.1865 \\

        w/o $\mathcal{G}_2$ (frequency-domain)
        & 23.37 & 0.8507 & 0.0684 & 0.1326 \\

        w/o $\mathcal{G}_3$ (decomposition regularization)
        & 23.23 & 0.8446 & 0.0707 & 0.1315 \\

        w/o $\mathcal{G}_4$ (color-exposure)
        & 23.26 & 0.8519 & 0.0688 & 0.1259 \\

        \bottomrule
    \end{tabular}
\end{table*}

\subsection{\revadded{Role-Aware Behavior and Coefficient Sensitivity}}
\subsubsection{\added{The branches play their assigned roles}}
\added{%
Section 4.7 shows qualitatively that removing the decomposition regularization causes the illumination map to absorb scene texture. We quantify the branch separation in the full model on the 15 LOL-v1 test images using three measurements: the normalized gradient energy of $\hat R$ and of $\hat L$,  the Pearson correlation between each branch's gradient-magnitude map and that of the ground-truth image, and illumination invariance, obtained by passing the low-light input and its normal-light reference through the same trained network and correlating the two resulting reflectance maps and separately the two illumination maps. %
}

\begin{table}[!t]
    \centering

    % Only the table number and caption are blue
    \captionsetup{
        labelfont={color=blue},
        textfont={color=blue}
    }
    \caption{\revadded{%
    Quantitative evidence of behavioral separation between the reflectance and illumination branches.}}
    \label{tab:role_behavior}

    \setlength{\tabcolsep}{4pt}
    \renewcommand{\arraystretch}{1.15}
    \footnotesize

    \begin{tabularx}{\columnwidth}{
        @{}
        >{\raggedright\arraybackslash}X
        >{\centering\arraybackslash}p{1.15cm}
        >{\centering\arraybackslash}p{1.15cm}
        >{\centering\arraybackslash}p{1.55cm}
        @{}
    }
        \toprule
        \textbf{Measurement} &
        \textbf{$\hat{R}$} &
        \textbf{$\hat{L}$} &
        \textbf{$p$} \\
        \midrule

        Normalized gradient energy &
        0.0254 &
        0.0096 &
        $6\times10^{-5}$ \\

        Correlation with ground-truth edges &
        0.875 &
        0.431 &
        $6\times10^{-5}$ \\

        Illumination invariance
        $\rho(X_{\mathrm{low}},X_{\mathrm{ref}})$ &
        0.977 &
        0.743 &
        $6\times10^{-5}$ \\

        \bottomrule
    \end{tabularx}
\end{table}

\added{%
The illumination map has (2.6$\times$) lower gradient energy than the reflectance map, while reflectance edges align with the true scene edges twice as well as illumination edges do and reflectance is nearly invariant to a change of illumination while illumination is not. Every separation holds on all 15 images (two-sided paired Wilcoxon). Thus, the results show that the two branches exhibit distinct restoration behavior consistent with their assigned roles. Notably, none of these three quantities is supervised by any term of the training objective: they are properties the decomposition acquires from the structural asymmetry itself, which is as the definition in Section~3. %
}

\subsubsection{\added{Sensitivity to the refinement coefficients}}
\added{%
We then examine whether this behavior depends on a particular choice of the three residual-scale coefficients: the reflectance refinement scale ($s_R$), illumination refinement scale ($s_L$), and edge-injection scale ($s_E$). Seven configurations covering changes from (0.5$\times$) to (2$\times$) are trained from scratch using the same 1000-epoch schedule and evaluated using the final-epoch weights. %
}

\begin{table*}[!t]
    \centering
    \caption{\added{Sensitivity of PRIN to the residual-scale coefficients on LOL-v1. $s_R$, $s_L$, and $s_E$ denote the reflectance-refinement, illumination-refinement, and edge-injection scales, respectively. Values in parentheses report two-sided paired Wilcoxon $p$-values against C0 over the 15 test images.  $^{*}p<0.05$ and $^{**}p<0.01$.}}
    \label{tab:coefficient_sensitivity}

    \setlength{\tabcolsep}{4pt}
    \renewcommand{\arraystretch}{1.15}
    \footnotesize

    \begin{tabularx}{\textwidth}{
        @{}p{3.15cm}
        p{2.55cm}
        Y
        Y
        Y@{}
    }
        \toprule
        \textbf{Config} &
        \textbf{$(s_R,s_L,s_E)$} &
        \textbf{PSNR$\uparrow$} &
        \textbf{SSIM$\uparrow$} &
        \textbf{LPIPS$\downarrow$} \\
        \midrule

        C0 default &
        $(0.10,0.05,0.10)$ &
        22.620 &
        0.8441 &
        0.1388 \\
        \rowrule

        C1 ratio 1:1 &
        $(0.10,0.10,0.10)$ &
        \textbf{23.377} (0.107) &
        \textbf{0.8520} (0.022$^{*}$) &
        0.1252 (0.0006$^{**}$) \\
        \rowrule

        C2 ratio 1:2 &
        $(0.05,0.10,0.10)$ &
        23.056 (0.083) &
        0.8497 (0.010$^{*}$) &
        0.1243 (0.0003$^{**}$) \\
        \rowrule

        C3 magnitude $0.5\times$ &
        $(0.05,0.025,0.10)$ &
        23.017 (0.454) &
        0.8456 (0.252) &
        0.1294 (0.0026$^{**}$) \\
        \rowrule

        C4 magnitude $2\times$ &
        $(0.20,0.10,0.10)$ &
        22.814 (0.454) &
        0.8451 (0.303) &
        0.1269 (0.0004$^{**}$) \\
        \rowrule

        C5 edge $0.5\times$ &
        $(0.10,0.05,0.05)$ &
        22.729 (0.561) &
        0.8432 (0.600) &
        0.1276 (0.0067$^{**}$) \\
        \rowrule

        C6 edge $2\times$ &
        $(0.10,0.05,0.20)$ &
        22.959 (0.804) &
        0.8458 (0.389) &
        \textbf{0.1238} (0.0054$^{**}$) \\
        \midrule

        \textit{Spread across C0--C6} &
        -- &
        \textit{0.757} &
        \textit{0.0088} &
        \textit{0.0150} \\

        \textit{Checkpoint-selection variance, single run} &
        -- &
        \textit{0.784} &
        \textit{0.0045} &
        \textit{0.0015} \\

        \bottomrule
    \end{tabularx}
\end{table*}

\added{Table \ref{tab:coefficient_sensitivity} illustrates that the coefficients do affect quantitative performance, but no single setting is optimal across all metrics. PSNR varies by 0.757 dB across the seven configurations, with no configuration significantly different from the default (p>0.08). SSIM varies within 0.0088, while C1 and C2 obtain small but statistically significant improvements. LPIPS is more sensitive: all six alternative configurations improve over the default, with C6 achieving the lowest value of 0.1238.
}

\added{These results indicate that the default coefficients represent one reasonable operating point under the adopted training objective, rather than a unique optimum. Varying the refinement scales shifts the recovered solution but causes only moderate performance variation within the tested range, while different metrics favor different configurations: C1 achieves the best PSNR and SSIM, whereas C6 gives the best LPIPS. The coefficients therefore regulate the strength of the refinement operations rather than define their restoration roles; the latter arise from the branch-specific architectural and objective constraints.}

\subsubsection{\added{Stability of the observable under decomposition ambiguity}}

\added{The limited variation in the enhanced output does not imply that the coefficients have no effect on the internal decomposition. Because the Retinex decomposition is ill-posed, different pairs ($\hat{R}$,$\hat{L}$) may produce similar restored images. We therefore compare the enhanced output and the two recovered factors across all 21 pairs of coefficient configurations.}

\begin{table}[!t]
    \centering
    \caption{\added{Cross-configuration agreement of the enhanced output and the recovered Retinex factors.}}
    \label{tab:cross_configuration_agreement}

    \setlength{\tabcolsep}{3pt}
    \renewcommand{\arraystretch}{1.12}
    \footnotesize

    \begin{tabularx}{\columnwidth}{
        @{}
        >{\raggedright\arraybackslash}X
        >{\centering\arraybackslash}p{1.65cm}
        >{\centering\arraybackslash}p{2.15cm}
        @{}
    }
        \toprule
        \textbf{Quantity} &
        \textbf{Agreement} &
        \textbf{Relative disagreement} \\
        \midrule

        Enhanced output $\hat{I}$ &
        0.9825 &
        --- \\

        Reflectance $\hat{R}$ &
        0.9710 &
        $1.7\times$ \\

        Illumination $\hat{L}$ &
        0.7797 &
        $\mathbf{12.6\times}$ \\

        \bottomrule
    \end{tabularx}
\end{table}

\added{As shown in Table \ref{tab:cross_configuration_agreement},the cross-configuration agreement is 0.9825 for the enhanced output and 0.9710 for reflectance, but only 0.7797 for illumination. Thus, different coefficient settings preserve the restored image and the reflectance estimate to a much greater extent than the illumination estimate. The decomposition ambiguity is therefore concentrated primarily in the illumination factor, while reflectance remains comparatively stable under coefficient perturbations. This observation is also consistent with the strong illumination invariance of the reflectance branch reported in Section 4.8.1.}

\subsubsection{\added{Scope}}

\added{The above experiments support three conclusions: the two branches exhibit measurable behavioral separation; this separation is not tied to a unique residual-scale setting; and coefficient-induced variation in the decomposition is concentrated mainly in the illumination factor. They do not establish that each asymmetric mechanism is individually necessary, nor that the coefficients have no influence on quantitative performance. Accordingly, we regard the default coefficients as a representative operating point rather than a metric-wise optimum.}

\subsection{GT-mean Study}
To evaluate the model's capability in structural preservation, textural consistency, and high-level perceptual representation, we employ a GT-mean correction to eliminate the metric penalties induced by global luminance mismatch. In this study, a post GT-mean correction was applied to the generated enhanced images by using the paired ground-truth mean as the target brightness statistic and then re-evaluating the corrected results with the same full-reference metrics. The results are tabulated in Table \ref{tab:gtmean_effect}.

\begin{table*}[!t]
\centering
\caption{Effect of GT-mean correction on quantitative results.}
\label{tab:gtmean_effect}
\setlength{\tabcolsep}{4.5pt}
\begin{tabular}{l|ccc|ccc|ccc}
\toprule
\multirow{2}{*}{Dataset} &
\multicolumn{3}{c|}{PSNR$\uparrow$} &
\multicolumn{3}{c|}{SSIM$\uparrow$} &
\multicolumn{3}{c}{LPIPS$\downarrow$} \\
& w/o & w/ & $\Delta$ & w/o & w/ & $\Delta$ & w/o & w/ & $\Delta$ \\
\midrule
LOL-v1      & 23.390 & \textbf{27.055} & \textbf{+3.666} & 0.853 & \textbf{0.872} & \textbf{+0.018} & 0.119 & \textbf{0.114} & \textbf{-0.005} \\
LOL-v2-Real & 22.428 & \textbf{27.019} & \textbf{+4.591} & 0.852 & \textbf{0.876} & \textbf{+0.023} & 0.181 & \textbf{0.170} & \textbf{-0.011} \\
LOL-v2-Syn  & 24.596 & \textbf{28.610} & \textbf{+4.014} & 0.931 & \textbf{0.942} & \textbf{+0.011} & 0.056 & \textbf{0.054} & \textbf{-0.002} \\
\bottomrule
\end{tabular}
\end{table*}

The quantitative results show that after GT-mean correction, PSNR increases significantly, while SSIM and LPIPS change only slightly. On LOL-v1, PSNR improves from 23.390 to 27.055, with a gain of 3.666 dB, while SSIM increases from 0.853 to 0.872 and LPIPS decreases from 0.119 to 0.114. On LOL-v2-Real, PSNR rises from 22.428 to 27.019, corresponding to the gain of 4.591 dB, whereas SSIM changes from 0.852 to 0.876 and LPIPS decreases from 0.181 to 0.170. On LOL-v2-Syn, PSNR improves from 24.596 to 28.610, with a gain of 4.014 dB, while SSIM increases from 0.931 to 0.942 and LPIPS decreases only slightly from 0.056 to 0.054.

The results suggest that the proposed method already preserves structural information and perceptual similarity reasonably well, as shown by the small changes in SSIM and LPIPS after GT-mean correction. However, the main remaining gap to the ground truth is associated with global luminance or tone mismatch. In other words, the current model still has room for improvement in global brightness calibration, while the structural restoration ability and perceptual similarity restoration ability are comparatively stable.

\subsection{\added[id=rev]{Efficiency Evaluation}}
\added[id=rev]{%
To evaluate computational complexity and practical inference efficiency, we report the number of parameters, computational cost, single-image inference latency, and frames per second (FPS). The computational cost is measured using THOP for source-image resolutions of $256\times256$ and $600\times400$, while latency and FPS are evaluated using a $600\times400$ image with a batch size of one. Official pretrained weights, model configurations, and inference pipelines are used for all compared methods whenever available, including their required padding, resizing, and iterative sampling procedures. Runtime is measured on a single NVIDIA GeForce RTX 5080 laptop GPU after three warm-up iterations and averaged over 15 runs, with CUDA synchronization applied for accurate timing. Image loading, checkpoint loading, and output saving are excluded from the reported latency. The final results are tabulated in Table \ref{tab:efficiency}%
}

\begin{table*}[t]
    \centering
    \caption{\added[id=rev]{%
    Comparison of computational complexity and practical inference
    efficiency.%
    }}
    \label{tab:efficiency}

    \resizebox{\textwidth}{!}{%
    \begin{tabular}{lccccc}
        \toprule
        Method
        & Params (M)$\downarrow$
        & FLOPs@$256\times256$ (G)$\downarrow$
        & FLOPs@$600\times400$ (G)$\downarrow$
        & Latency (ms)$\downarrow$
        & FPS$\uparrow$ \\
        \midrule

        MIRNet \cite{zamir2020mirnet}
        & 31.79
        & 511.75
        & 1874.08
        & 294.2
        & 3.4 \\

        SNR-Net \cite{xu2022snr}
        & 39.12
        & 23.96
        & 88.91
        & 18.8
        & 53.3 \\

        Uformer-T \cite{wang2022uformer}
        & 5.20
        & 9.98
        & 62.37
        & 79.3
        & 12.6 \\

        Restormer \cite{Restormer}
        & 26.10
        & 140.99
        & 516.32
        & 269.0
        & 3.7 \\

        LLFormer \cite{wang2023llformer}
        & 24.52
        & 22.03
        & 81.77
        & 371.9
        & 2.7 \\

        Retinexformer \cite{cai2023retinexformer}
        & 1.61
        & 15.76
        & 57.72
        & 36.0
        & 27.8 \\

        DiffLL \cite{jiang2023diffll}
        & 22.06
        & 87.83
        & 338.94
        & 153.7
        & 6.5 \\

        QuadPrior \cite{wang2024quadprior}
        & 1313.39
        & 11444.98
        & 17145.47
        & 3666.1
        & 0.27 \\

        HVI-CIDNet \cite{Yan2025HVIAN}
        & 1.97
        & 8.13
        & 29.78
        & 40.9
        & 24.4 \\

        \midrule
        \textbf{PRIN (Ours)}
        & \textbf{24.41}
        & \textbf{96.60}
        & \textbf{372.79}
        & \textbf{79.2}
        & \textbf{12.6} \\

        \bottomrule
    \end{tabular}%
    }

    \vspace{1mm}
    \begin{minipage}{0.98\textwidth}
        \footnotesize
        \added[id=rev]{%
        For a $600\times400$ source image, PRIN and DiffLL process an effective resolution of $608\times416$; SNR-Net and LLFormer process $608\times400$; Uformer-T pads the input to $640\times640$; and QuadPrior resizes the short side to 512, resulting in an effective resolution of $768\times512$. DiffLL and QuadPrior include their complete default sampling procedures. QuadPrior runtime is evaluated using its official FP16 setting, while the remaining reported runtimes use FP32.%
        }
    \end{minipage}
\end{table*}

\added[id=rev]{%
As Table \ref{tab:efficiency} illustrates, PRIN contains 24.41M parameters, which is comparable to DiffLL (22.06M), LLFormer (24.52M), and Restormer (26.10M), while being smaller than MIRNet and SNR-Net. However, it is substantially larger than compact Retinex-oriented methods such as Retinexformer, and HVI-CIDNet.%
}

\added[id=rev]{%
For a $256\times256$ source image, PRIN requires 96.60G module-level multiply-accumulate operations. When processing a $600\times400$ source image, the input is internally padded to $608\times416$, resulting in 372.79G operations. The computational cost is mainly associated with the asymmetric dual decoders, branch-specific wavelet refinement modules, and the post-composition curve refinement head. Although this cost is higher than that of compact Transformer- and Retinex-based methods, it remains lower than that of MIRNet, Restormer, and QuadPrior.%
}

\added[id=rev]{%
At the $600\times400$ source resolution, PRIN achieves an average inference latency of 79.2~ms, corresponding to 12.6~FPS. PRIN is slower than SNR-Net, Retinexformer, and HVI-CIDNet, but faster than MIRNet, Restormer, LLFormer, DiffLL, and QuadPrior. Its measured latency is also nearly identical to that of Uformer-T. In particular, although PRIN and LLFormer have similar parameter counts, PRIN exhibits approximately $4.7\times$ lower measured latency under the current hardware and implementation environment.
}

\added[id=rev]{%
Overall, the efficiency analysis shows that PRIN is not a lightweight edge-oriented model, but rather a quality-oriented architecture that accepts a moderate computational overhead in exchange for stronger perceptual and structural restoration. Despite its relatively high operation count, PRIN and maintains a practical inference latency of 79.2~ms at the $600\times400$ source resolution \added[id=rev]{and} outperforms several computationally intensive baselines in runtime while remaining competitive in model size. When considered together with its consistently favorable LPIPS and SSIM results, these findings indicate that PRIN provides a reasonable quality-efficiency trade-off for desktop-GPU low-light image enhancement.
}

\subsection{Perceptual-Structural Performance Analysis}
To further analyze the perceptual-structural balance of different low-light enhancement methods, we plot the SSIM-LPIPS distribution on LOL-v1 in Fig. \ref{fig:ssim_lpips_lolv1}. In the figure, the horizontal axis denotes LPIPS (smaller is better for perceptual similarity), and the vertical axis denotes SSIM (larger is better for structural fidelity). Therefore, methods closer to the upper-left region achieve a better balance. As the results show, our proposed PRIN is positioned in the extreme upper-left region, which means it achieves a relatively good balance between the LPIPS and SSIM among all compared methods.

\begin{center}
    \includegraphics[width=0.85\linewidth]{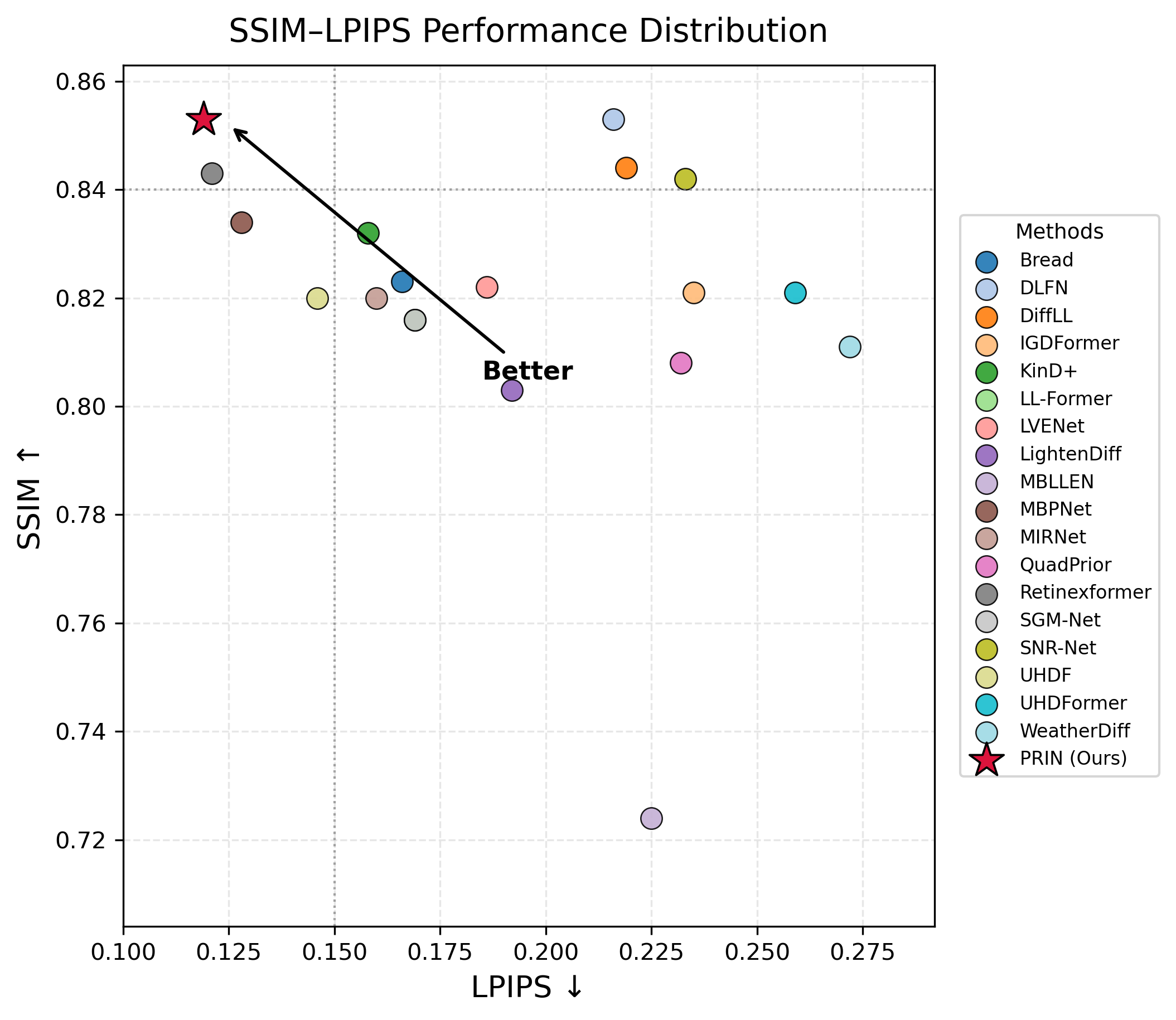}
    \captionof{figure}{SSIM-LPIPS performance distribution on LOL-v1.}
    \label{fig:ssim_lpips_lolv1}
\end{center}

\section{Limitations and Future Work}
Although PRIN achieves strong perceptual and structural performance on the evaluated benchmarks, several limitations remain to be addressed.

First, the GT-mean study reveals that global luminance and tone calibration remain an important limitation of the current model. After applying ground truth mean correction, the PSNR values increase substantially by 3.666 dB, 4.591 dB, and 4.014 dB on LOL-v1, LOL-v2-Real, and LOL-v2-Syn, respectively, while the corresponding changes in SSIM and LPIPS are relatively limited. The observation suggests that PRIN already preserves local structures and perceptual features reasonably well, but its global exposure statistics are not yet fully aligned with the reference images. Future work will focus on scene-adaptive tone regulation and exposure calibration strategies that improve global brightness consistency without requiring reference-derived statistics during inference.

Second, the perceptual advantages of PRIN are not completely uniform across different evaluation settings. \replaced[id=rev]{Although PRIN achieves the best or second-lowest LPIPS scores among the compared methods on the three full-reference benchmarks and performs competitively on several non-reference quality measures, its non-reference scores are still dataset-dependent.}{Although PRIN achieves the lowest LPIPS scores among the compared methods on all three full-reference benchmarks and performs competitively on several non-reference quality measures, its non-reference scores are still dataset-dependent.} Especially on LOL-v2-Syn, competing methods achieve higher scores on some non-reference metrics. Since the current model is mainly trained and evaluated on paired datasets, its robustness under unknown camera conditions, severe sensor noise, motion blur, and complex mixed-illumination conditions remain to be further examined. Future studies will extend the evaluation to more diverse real-captured datasets and explore domain adaptation or self-supervised regularization strategies.

Third, PRIN introduces a dual-branch architecture, branch-specific wavelet refiners, attention modules, and a curve-based post-composition head to improve restoration quality. Although these components contribute to enhanced perceptual-structural performance, the resulting model contains 24.412M parameters, which is larger than the lightweight variants considered in this work. Future work will increase computational efficiency and explore compact attention mechanisms, model pruning, and knowledge distillation for edge applications.

\added[id=rev]{%
In addition, we did not experimentally compare the Haar DWT with other frequency-domain decomposition methods. Therefore, we do not claim that Haar is the optimal choice for PRIN, and other frequency representations or learned decompositions may potentially provide better performance. In this work, Haar DWT is adopted primarily for its structural compatibility with the proposed refiners, while a systematic investigation of alternative frequency decomposition methods is left for future work.}

Finally, the current framework is designed for single-image enhancement and does not explicitly consider temporal coherence. Extending the proposed restoration-role-aware Retinex formulation to low-light video enhancement, with temporal constraints and cross-frame feature propagation, will be a meaningful direction for future applications such as nighttime surveillance and autonomous navigation.

\section{Conclusion}
In this paper, we proposed PRIN, a restoration-role-aware Retinex-inspired network for low-light image enhancement. PRIN explicitly distinguishes the enhancement objectives of reflectance and illumination through asymmetric decoding branches. The proposed WDGRR and WSAIR modules further perform branch-specific wavelet-domain refinement, allowing detail-sensitive reflectance restoration and spatially coherent illumination adjustment to be addressed separately. A curve-based post-composition enhancement head is then employed to calibrate residual tone and exposure imbalance after Retinex recomposition.

Experiments on LOL-v1, LOL-v2-Syn, and LOL-v2-Real indicate that PRIN achieves a favorable balance between perceptual quality (LPIPS) and structural fidelity (SSIM). \replaced[id=rev]{%
In particular, PRIN ties for the best SSIM and obtains the
second-lowest LPIPS on LOL-v1, achieves the best SSIM and LPIPS on
LOL-v2-Syn, and yields the highest PSNR and the second-lowest LPIPS
on LOL-v2-Real.%
}{%
In particular, PRIN obtains the lowest LPIPS values among the compared
methods across all three benchmarks, achieves the best or tied-best
SSIM performance on LOL-v1 and LOL-v2-Syn, and yields the highest
PSNR on LOL-v2-Real.%
} The non-reference perceptual quality study \replaced[id=rev]{shows}{show} competitive perceptual naturalness results. In addition, the ablation results further indicate that each proposed module contributes to the network, and the curve-based post-composition module can effectively coordinate the branch-specific refiners to achieve better results.

The GT-mean study further shows that the remaining performance gap is mainly associated with global luminance and tone mismatch rather than severe degradation of structural or perceptual representations. Overall, PRIN provides an interpretable and effective framework for low-light image enhancement by combining Retinex-motivated component restoration with branch-specific frequency refinement. Future work will focus on more adaptive tone calibration, stronger robustness in real-world applications, and more efficient model designs.

%\appendix
%\section{My Appendix}
%Appendix sections are coded under \verb+\appendix+.

%\verb+\printcredits+ command is used after appendix sections to list 
%author credit taxonomy contribution roles tagged using \verb+\credit+ 
%in frontmatter.

\printcredits

\section*{Declaration of Competing Interest}
The authors declare that they have no known competing financial interests or personal relationships that could have appeared to influence the work reported in this paper.

\section*{Funding}
This work is supported by the Xiamen University Malaysia Research Fund (XMUMRF) scheme under the grant of XMUMRF/2025-C16/IECE/0070.

%% Loading bibliography style file
%\bibliographystyle{model1-num-names}
% \bibliographystyle{cas-model2-names}
\bibliographystyle{elsarticle-num}

% Loading bibliography database
\bibliography{cas-refs}

\newpage
%\vskip3pt
\bio{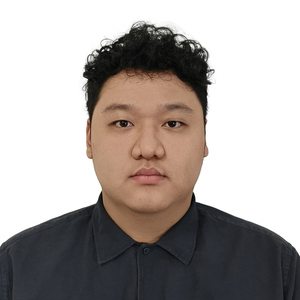}
Qiyi Xie received the B.Eng. (Honours) degree in Electrical and Electronic Engineering from Xiamen University Malaysia, Sepang, Malaysia, in 2025. He is currently pursuing the M.S.E. degree in Electrical Engineering with a concentration in Artificial Intelligence at the University of Pennsylvania, Philadelphia, PA. His current research interests include computer vision, computational photography, and signal processing.
\endbio

\bio{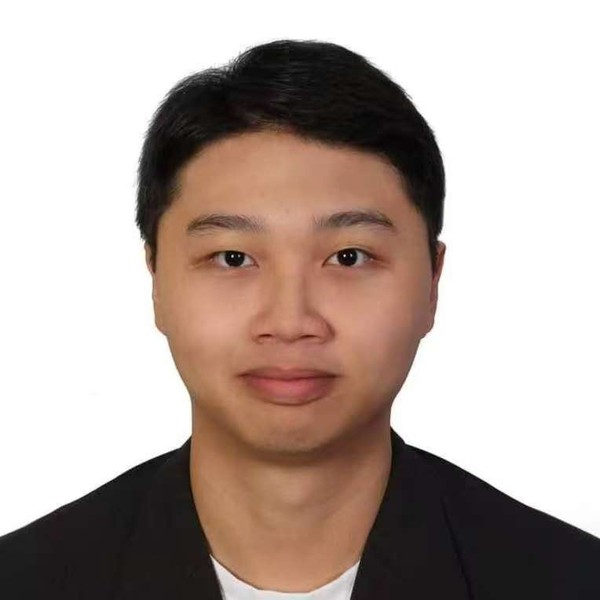}
Wei Chen Ng received the B.Eng. (Honours) degree in Artificial Intelligence from Xiamen University Malaysia, Sepang, Malaysia, in 2025. He is currently pursuing the M.S. degree in Computer Science and Technology at Xiamen University Malaysia, School of Computing and Data Science. His current research interests include deep learning and computer vision.
\endbio

\bio{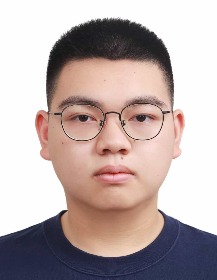}
Ziyao Li is an Electrical and Electronic Engineering student at Xiamen University Malaysia. His work centers on embedded systems, FPGA-based computing, and artificial intelligence, with a focus on computer vision and intelligent perception for embedded platforms. His research interests include hardware–software co-design and efficient computing architectures for intelligent electronic systems.
\endbio

\bio{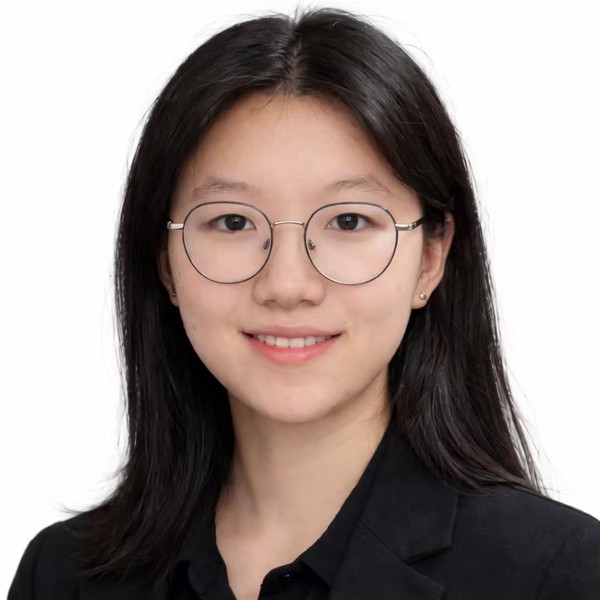}
Yuning Wu is an Electrical and Electronic Engineering student at Xiamen University Malaysia. Her research focuses on embedded artificial intelligence and computer vision for intelligent robotic and edge systems. She is particularly interested in deploying perception algorithms on resource-constrained platforms, including microcontrollers, FPGA-based accelerators, and system-on-chip architectures. Her work explores hardware–software co-design and efficient computing architectures to enable real-time intelligent perception in embedded and robotic systems.
\endbio

\bio{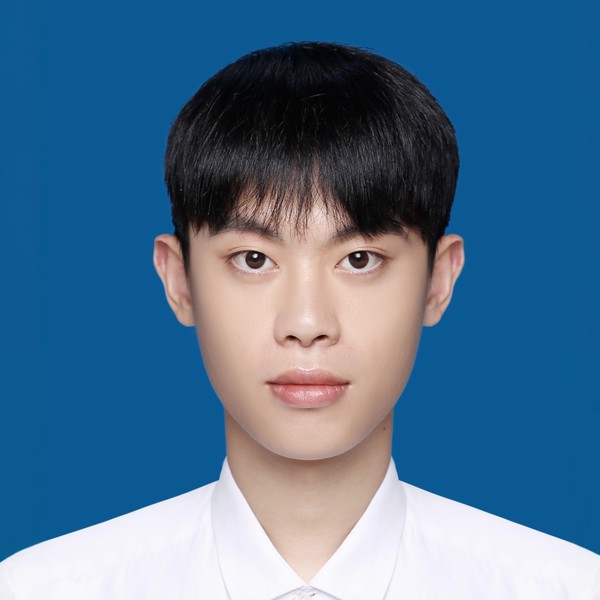}
Huasen Wu received the B.Eng. degree in Robotics and Electrical Engineering from the University of Sussex, U.K., and Zhejiang Gongshang University, Hangzhou, China, in 2025. He is currently pursuing the M.S.E. degree in Electrical Engineering with a concentration in Artificial Intelligence at the University of Pennsylvania, Philadelphia, PA, USA. His current research interests include Deep learning and Signal Processing.
\endbio

\bio{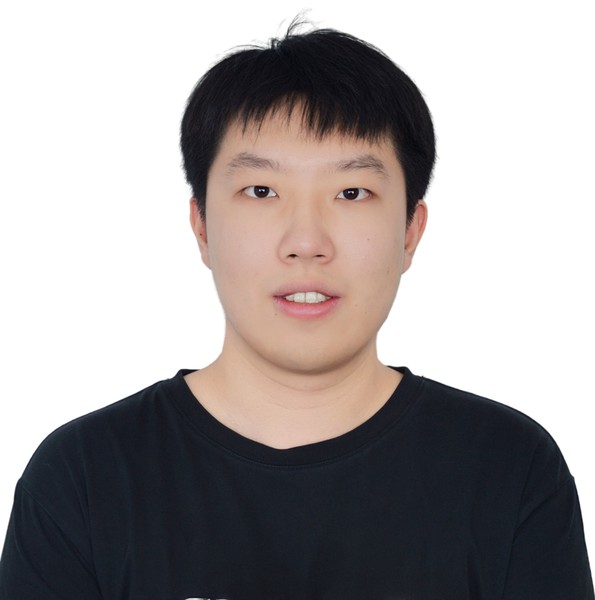}
Tianshi Miao received the B.S. degree in Information Management and Information Systems from Southwest Jiaotong University in 2024. He is currently pursuing the M.S. degree in Computer and Information Science and Systems Engineering at the University of Pennsylvania. His current research and professional interests include distributed systems, backend engineering, and cloud infrastructure.
\endbio

\bio{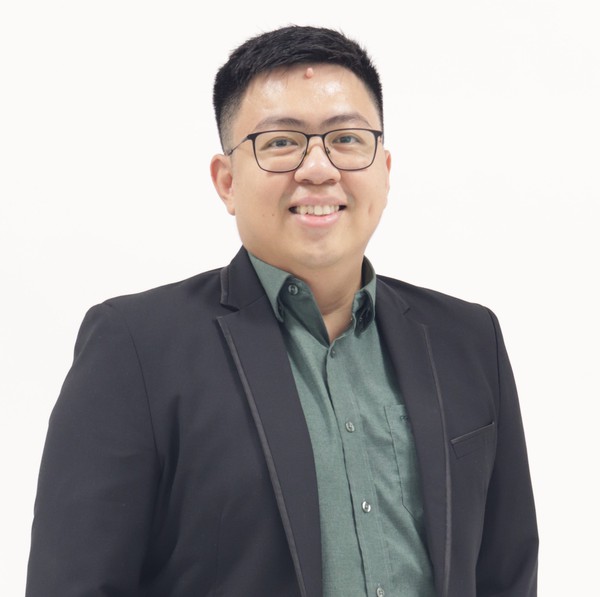}
Dr. Tan Jian Ding is an Associate Professor and Head of Robotics and Automation Engineering at the School of Artificial Intelligence and Robotics, Xiamen University Malaysia, and serves as the Director of the S.M.A.R.T. Nexus Center of Excellence at the same institution. He is a registered member of the Board of Engineers Malaysia (BEM) and the Institution of Engineers, Malaysia (IEM) in the field of electrical and electronic engineering. With over 15 years of academic and research experience, Dr. Tan has been actively engaged in both governmental and industrial consultancies, contributing his expertise in areas such as artificial intelligence, control engineering, robotic systems, industrial automation, embedded systems, microcontrollers, renewable energy, and Industry 4.0 technologies. He received his PhD in Engineering from the University of Malaya, Malaysia. His research interests encompass artificial intelligence, robotics, industrial automation, optimization, heuristic search, Internet of Things (IoT), renewable energy, and embedded control systems.
\endbio

\end{document}